\documentclass[twocolumn]{aastex631}
\usepackage{amsmath}
\usepackage{graphicx}
\usepackage{subfigure}
\usepackage{etoolbox}
\usepackage{ragged2e}
\makeatletter
\patchcmd{\@floatboxreset}{\@parboxrestore}{}{}{}
\makeatother

\begin{document}

\title{The Effect of Radiation and Supernovae Feedback on LyC Escape in Local Star-forming Galaxies}

\author[0000-0003-4166-2855]{Cody A. Carr}
 \affiliation{Center for Cosmology and Computational Astrophysics, Institute for Advanced Study in Physics \\ Zhejiang University, Hangzhou 310058,  China}
\affiliation{Institute of Astronomy, School of Physics, Zhejiang University, Hangzhou 310058,  China}

\author[0000-0001-8531-9536]{Renyue Cen}
 \affiliation{Center for Cosmology and Computational Astrophysics, Institute for Advanced Study in Physics \\ Zhejiang University, Hangzhou 310058,  China}
\affiliation{Institute of Astronomy, School of Physics, Zhejiang University, Hangzhou 310058,  China}

\author[0000-0002-9136-8876]{Claudia Scarlata}
 \affiliation{Minnesota Institute for Astrophysics, School of
   Physics and Astronomy, University of Minnesota \\ 316 Church str
 SE, Minneapolis, MN 55455,USA}

 \author[0000-0002-9217-7051]{Xinfeng Xu}
 \affiliation{Department of Physics and Astronomy, Northwestern University \\ 2145 Sheridan Road, Evanston, IL, 60208, USA.}

 \affiliation{Center for Interdisciplinary Exploration and Research in Astrophysics (CIERA), Northwestern University \\ 1800 Sherman Avenue, Evanston, IL, 60201, USA.}

\author[0000-0002-6586-4446]{Alaina Henry}
 \affiliation{Space Telescope Science Institute, 3700 San Martin Drive, Baltimore, MD 21218, USA}
 \affiliation{Center for Astrophysical Sciences, Department of Physics \& Astronomy, Johns Hopkins University, Baltimore, MD 21218, USA}

  \author[0000-0001-8442-1846]{Rui Marques-Chaves}
  \affiliation{Observatoire de Genève, Université de Genève, 51 Ch. des Maillettes, 1290 Versoix, Switzerland}

 \author[0000-0001-7144-7182]{Daniel Schaerer}
  \affiliation{Observatoire de Genève, Université de Genève, 51 Ch. des Maillettes, 1290 Versoix, Switzerland}

\author[0000-0001-5758-1000]{Ricardo O. Amor\'{i}n}
\affiliation{Instituto de Astrof\'{i}sica de Andaluc\'{i}a (CSIC), Apartado 3004, 18080 Granada, Spain}

 \author[0000-0002-5808-1320]{M. S. Oey}
\affiliation{Astronomy Department, University of Michigan, Ann Arbor, MI 48109, USA}

  \author[0000-0002-5235-7971]{Lena Komarova}
 \affiliation{Astronomy Department, University of Michigan, Ann Arbor, MI 48109, USA}

    \author[0000-0002-0159-2613]{Sophia Flury}
 \affiliation{Department of Astronomy, University of Massachusetts, Amherst,MA 01003, USA}
 \affiliation{Institute for Astronomy, University of Edinburgh, Royal Observatory, Edinburgh, EH9 3HJ, UK}

\author[0000-0002-6790-5125]{Anne Jaskot}
 \affiliation{Department of Astronomy, Williams College, Williamstown, MA 01267, USA}

 \author[0000-0001-8419-3062]{Alberto Saldana-Lopez}
\affiliation{Department of Astronomy, Oskar Klein Centre, Stockholm University, 106 91 Stockholm, Sweden}

   \author[0000-0001-7673-2257]{Zhiyuan Ji}
\affiliation{Steward Observatory, University of Arizona, 933 N. Cherry Avenue, Tucson, AZ 85721, USA}

 \author[0009-0002-9932-4461]{Mason Huberty}
 \affiliation{Minnesota Institute for Astrophysics, School of
   Physics and Astronomy, University of Minnesota \\ 316 Church str
 SE, Minneapolis, MN 55455,USA}

\author[0000-0003-1127-7497]{Timothy Heckman}

\affiliation{Center for Astrophysical Sciences, Department of Physics \& Astronomy, Johns Hopkins University, Baltimore, MD 21218, USA}

\author[0000-0002-3005-1349]{Göran Östlin}
\affiliation{Department of Astronomy, Oskar Klein Centre, Stockholm University, 106 91 Stockholm, Sweden}

\author[0000-0003-2722-8841]{Omkar Bait}
  \affiliation{Observatoire de Genève, Université de Genève, 51 Ch. des Maillettes, 1290 Versoix, Switzerland}
\affiliation{SKA Organization, Jodrell Bank, Lower Whitington, Macclesfield, SK11 9FT, UK}

 \author[0000-0001-8587-218X]{Matthew James Hayes}
\affiliation{Department of Astronomy, Oskar Klein Centre, Stockholm University, 106 91 Stockholm, Sweden}

\author[0000-0001-5331-2030]{Trinh Thuan}
\affiliation{Astronomy Department, University of Virginia, Charlottesville, VA 22904, USA}

   \author[0000-0002-4153-053X]{Danielle A. Berg}
\affiliation{Department of Astronomy, The University of Texas at Austin, 2515 Speedway, Stop C1400, Austin, TX 78712, USA}

\author[0000-0002-7831-8751]{Mauro Giavalisco}
\affiliation{Department of Astronomy, University of Massachusetts, Amherst,MA 01003, USA}

\author[0000-0002-2724-8298]{Sanchayeeta Borthakur}
\affiliation{School of Earth \& Space Exploration, Arizona State University, Tempe, AZ 85287, USA}

\author[0000-0002-0302-2577]{John Chisholm}
\affiliation{Department of Astronomy, The University of Texas at Austin, 2515 Speedway, Stop C1400, Austin, TX 78712-1205, USA}

\author[0000-0001-7113-2738]{Harry C. Ferguson}
\affiliation{Space Telescope Science Institute, 3700 San Martin Drive, Baltimore, MD 21218, USA}

\author{Leo Michel-Dansac}
\affiliation{Univ Lyon, Univ Lyon1, ENS de Lyon, CNRS, Centre de Recherche Astrophysique de Lyon UMR5574 \\ 69230 Saint-Genis-Laval, France}

\author[0000-0002-2201-1865]{Anne Verhamme}
\affiliation{Observatoire de Genève, Université de Genève, 51 Ch. des Maillettes, 1290 Versoix, Switzerland}

\author[0000-0003-0960-3580]{Gábor Worseck}
\affiliation{Institut für Physik und Astronomie, Universität Potsdam, Karl-Liebknecht-Str. 24/25, D-14476 Potsdam, Germany}

\begin{abstract}
Feedback is widely recognized as an essential condition for Lyman continuum (LyC) escape in star-forming galaxies. However, the mechanisms by which galactic outflows clear neutral gas and dust remain unclear. In this paper, we model the Mg II 2796\AA, 2804\AA\ absorption + emission lines in 29 galaxies taken from the Low-z LyC Survey (LzLCS) to investigate the impact of (radiation + mechanical) feedback on LyC escape. Using constraints on $\rm Mg^+$ and photoionization models, we map the outflows’ neutral hydrogen content and predict $f_{esc}^{LyC}$ with a multiphase wind model.  We measure mass, momentum, and energy loading factors for the neutral winds, which carry up to 10\% of the momentum and 1\% of the energy in SFR-based deposition rates.  We use SED template fitting to determine the relative ages of stellar populations, allowing us to identify radiation feedback dominant systems.  We then examine feedback related properties (stellar age, loading factors, etc.) under conditions that optimize feedback efficiency, specifically high star formation rate surface density and compact UV half-light radii.  Our findings indicate that the strongest leakers are radiation feedback dominant, lack Mg II outflows, but have extended broad components in higher ionization lines like [O III] 5007Å, as observed by \cite{Amorin2024}.  In contrast, galaxies experiencing supernovae feedback typically exhibit weaker $f_{esc}^{LyC}$ and show evidence of outflows in both Mg II and higher ionization lines.  We attribute these findings to rapid or “catastrophic” cooling in the radiation-dominant systems, which, given the low metallicities in our sample, are likely experiencing delayed supernovae.  
\end{abstract}


\section{Introduction} \label{sec:intro}

As the James Webb Space Telescope (JWST) era unfolds, the next two decades should see rapid progress in understanding the physics connecting galaxy formation to the Reionization of the early Universe.  Critical to this challenge is the spatial and temporal mapping of the ionizing emissivity, $\dot{n}_{\rm UV} = \rho_{\rm UV}\xi_{\rm UV} f_{esc}^{LyC}$ [photon $\rm s^{-1}$ $\rm Mpc^{-3}$], where $\rho_{\rm UV}$ [erg $\rm s^{-1}$ $\rm Hz^{-1}$ $\rm Mpc^{-3}$] is the co-moving total UV luminosity density, $\xi_{\rm UV}$ [photon $\rm erg^{-1}$ $\rm Hz$] is the ionizing, or Lyman continuum (LyC) photon production efficiency, and $f_{esc}^{LyC}$ is the fraction of LyC photons able to escape a galaxy and its surroundings to ionize intergalactic space \citep{Robertson2022}.  Early results already indicate a larger than expected population of UV bright ($M_{\rm UV}=-20$) objects at redshifts $z>10$ \citep{Finkelstein2023}, and high ionization rates at redshifts $4 \lesssim z \lesssim 9$ \citep{Sanders2023,Llerena2024,Simmonds2024}. 

Constraining $f_{esc}^{LyC}$ is still a challenge, however, since attenuation by the neutral intergalactic medium (IGM) precludes measurements at $z>4$ \citep{Inoue2014}.  Consequently, astronomers must rely on indirect diagnostics of $f_{esc}^{LyC}$ developed at local and intermediate redshifts.  To date, the largest of these studies has been the Low-z Lyman Continuum Survey (LzLCS, \citealt{Flury2022a}), which studied 89 galaxies spanning a large range of values for potential $f_{esc}^{LyC}$ indicators such as $[\rm O\ III]\ 5007\text{\AA}/[\rm O\ II]\ 3726\text{\AA},\ 3729\text{\AA}$ ratios ($\rm O_{32}$, \citealt{Jaskot2013,Nakajima2014,Izotov2017,Izotov2018b,Izotov2020}), star formation rate surface density ($\Sigma_{\rm SFR}$, \citealt{Heckman2001,Clarke2002,Sharma2017,Naidu2020}), and the slope of the non-ionizing UV continuum ($\beta_{UV}$, \citealt{Zackrisson2013,Zackrisson2017,Chisholm2022}).  While no perfect indicator was found \citep{Flury2022b}, it is clear that LyC emitters are sensitive to the column density of neutral hydrogen and dust \citep{Gazagnes2018,Gazagnes2020,Chisholm2018b,Chisholm2022,Saldana-Lopez2022}, ionization state \citep{Izotov2021,Flury2022b}, and age of the starburst \citep{Hayes2023a,Hayes2023b}.  These factors underscore the crucial, yet intricate, relationship between feedback and the geometry of neutral gas and dust in the interstellar medium (ISM) and circumgalactic medium (CGM) in facilitating LyC escape.

Indeed, followup studies of the LzLCS galaxies have hinted at the important role of feedback in facilitating LyC escape.  For example, \cite{Amorin2024} found a strong correlation between the width of the broad line component in optical emission lines and $f_{esc}^{LyC}$, indicating that ionized winds driven by radiation and/or mechanical feedback play a role in shaping the geometry of the ISM/CGM.  Further evidence from radio observations by \cite{Bait2023} suggest an age dependence.  In particular, they found that the slope of the radio spectrum was flatter in galaxies with higher $f_{esc}^{LyC}$, indicating young stars and a lack of supernovae.  This result may indicate an evolution in $f_{esc}^{LyC}$ dependent on how different types of feedback (radiation + mechanical) influence the geometry of the ISM/CGM.

The neutral ISM/CGM geometries of LyC leakers are typically categorized into two scenarios.  In the density-bounded scenario, a Strömgren sphere fails to form, allowing LyC radiation to escape directly through the ISM/CGM \citep{Zackrisson2013,Nakajima2014}.  This condition is often linked to intense radiation feedback and high $\rm O_{32}$ ratios in the strongest LyC leakers \citep{Izotov2018b,Gazagnes2020}.  Alternatively, LyC photons may escape through low-density channels or small holes in a “picket fence” type geometry \citep{Heckman2011,Gazagnes2020,Saldana-Lopez2022}.  Both radiation and mechanical feedback could contribute to this geometry.  In simulations of H II regions, \cite{Kakiichi2021} demonstrated that turbulence can create low-density channels, which are then cleared by thermal pressure from the ionization front to facilitate LyC escape.  Alternatively, supernovae blastwaves could lift cool clouds to large distances, clearing paths for LyC escape \citep{Heckman2011,Cen2020,Komarova2021}. The picket fence geometry has also been associated with catastrophic cooling \citep{Silich2003,Silich2004,Silich2018,Gray2019,Danehkar2021,Danehkar2022}, where rapid cooling leads to cloud fragmentation on small scales, forming a picket fence type structure \citep{Jaskot2017,Jaskot2019}.  

Combinations are also likely.  For example, in the two-stage burst scenario, mechanical feedback from supernovae clears the ISM/CGM of dense clouds to create low-density channels, while radiation feedback simultaneously ionizes them \citep{Micheva2017,Enders2023}.  Such conditions have been proposed to explain some of the strongest leakers in LzLCS (\citealt{Flury2022b}, Flury et al. in prep). 

\begin{table*}[htbp]
\centering
\caption{LzLCS galaxy properties.}
\resizebox{\textwidth}{!}{$%
\begin{tabular}{ccccccccccc}
\hline\hline
Galaxy&$z$&$\text{E(B-V})_{\text{UV}}$&$12+\log{\left(\frac{\text{O}}{\text{H}}\right)}$&$r_{1/2}(\rm UV)$&$\log{M_{\star}}$&$\log\text{SFR},\text{H}_{\beta}$&$\log{\Sigma_{\text{SFR},\text{H}_{\beta}}}$&$\rm \log{\text{O}_{32}}$&$\text{significance}$&$f_{esc}^{LyC}$\\[.5 ex]
& &&&$\text{kpc}$&$M_{\odot}$&$M_{\odot}\ \text{yr}^{-1}$&$M_{\odot}\ \text{yr}^{-1} \text{kpc}^{-2}$&&&$\%$\\[.5 ex]
\hline
J003601+003307 &    0.3477 & $0.013\substack{0.033\\0.033}$ & $7.8\substack{0.037\\0.037}$ &   $0.45\substack{0.03\\0.03}$ &  $8.8\substack{0.44\\0.43}$ &  $1.2\substack{0.024\\0.024}$ &      $1.1\substack{0.15\\0.15}$ &   $1.1\substack{0.039\\0.039}$ &          1.50 &  $2.9\substack{-99\\-99}$ \\
J004743+015440 &    0.3533 &    $0.13\substack{0.04\\0.04}$ & $8.0\substack{0.037\\0.037}$ & $0.62\substack{0.029\\0.029}$ &  $9.2\substack{0.44\\0.43}$ &  $1.3\substack{0.024\\0.024}$ &       $0.93\substack{0.1\\0.1}$ &  $0.66\substack{0.026\\0.026}$ &          5.10 &          $1.3\substack{2.1\\0.3}$ \\
J011309+000223 &    0.3059 &  $0.14\substack{0.049\\0.049}$ &   $8.3\substack{0.11\\0.11}$ &   $0.63\substack{0.03\\0.03}$ &  $9.1\substack{0.44\\0.43}$ & $0.64\substack{0.076\\0.076}$ &     $0.25\substack{0.12\\0.12}$ &  $0.36\substack{0.086\\0.086}$ &          4.60 &          $2.2\substack{1.6\\1.2}$ \\
J012217+052044 &    0.3651 &   $0.075\substack{0.04\\0.04}$ & $7.8\substack{0.064\\0.064}$ &   $0.71\substack{0.03\\0.03}$ &  $8.8\substack{0.45\\0.42}$ & $0.93\substack{0.041\\0.041}$ &       $0.43\substack{0.1\\0.1}$ &  $0.88\substack{0.047\\0.047}$ &          5.00 &          $3.8\substack{4.6\\1.6}$ \\
J012910+145935 &    0.2799 &  $0.16\substack{0.036\\0.036}$ & $8.4\substack{0.044\\0.044}$ &   $0.64\substack{0.03\\0.03}$ &  $9.2\substack{0.58\\0.31}$ &  $1.1\substack{0.026\\0.026}$ &   $0.71\substack{0.091\\0.091}$ &  $0.34\substack{0.031\\0.031}$ &          0.62 & $0.69\substack{-99\\-99}$ \\
J072326+414608 &    0.2966 &  $0.18\substack{0.034\\0.034}$ & $8.2\substack{0.044\\0.044}$ & $0.44\substack{0.029\\0.029}$ &  $9.5\substack{0.48\\0.38}$ &   $0.93\substack{0.03\\0.03}$ &     $0.85\substack{0.13\\0.13}$ &  $0.65\substack{0.034\\0.034}$ &          0.71 & $0.37\substack{-99\\-99}$ \\
J081112+414146 &    0.3329 &  $0.18\substack{0.022\\0.022}$ &   $7.9\substack{0.09\\0.09}$ & $0.68\substack{0.032\\0.032}$ &  $8.4\substack{0.45\\0.42}$ & $0.56\substack{0.062\\0.062}$ &    $0.094\substack{0.12\\0.12}$ &   $1.1\substack{0.071\\0.071}$ &          6.80 &         $2.0\substack{1.2\\0.49}$ \\
J081409+211459 &    0.2268 &  $0.26\substack{0.016\\0.016}$ & $8.1\substack{0.032\\0.032}$ &  $1.4\substack{0.034\\0.034}$ &  $9.6\substack{0.44\\0.43}$ &    $1.2\substack{0.02\\0.02}$ &  $0.086\substack{0.042\\0.042}$ &   $0.2\substack{0.021\\0.021}$ &          1.20 & $0.75\substack{-99\\-99}$ \\
J082652+182052 &    0.2972 &  $0.19\substack{0.049\\0.049}$ & $8.3\substack{0.051\\0.051}$ &   $0.56\substack{0.03\\0.03}$ &  $8.5\substack{0.45\\0.42}$ & $0.82\substack{0.032\\0.032}$ &     $0.52\substack{0.11\\0.11}$ &  $0.74\substack{0.037\\0.037}$ &          0.00 & $0.89\substack{-99\\-99}$ \\
J091113+183108 &    0.2620 &  $0.24\substack{0.025\\0.025}$ & $8.1\substack{0.037\\0.037}$ & $0.44\substack{0.029\\0.029}$ &  $10.0\substack{0.5\\0.37}$ &  $1.4\substack{0.022\\0.022}$ &      $1.3\substack{0.12\\0.12}$ &  $0.38\substack{0.026\\0.026}$ &          8.20 &         $2.3\substack{1.8\\0.71}$ \\
J091208+505009 &    0.3275 &    $0.16\substack{0.03\\0.03}$ & $8.2\substack{0.039\\0.039}$ &    $1.2\substack{0.03\\0.03}$ &  $8.8\substack{0.66\\0.36}$ &  $1.1\substack{0.025\\0.025}$ &    $0.2\substack{0.059\\0.059}$ &   $0.6\substack{0.029\\0.029}$ &          0.46 & $0.25\substack{-99\\-99}$ \\
J091703+315221 &    0.3003 & $0.077\substack{0.024\\0.024}$ & $8.5\substack{0.035\\0.035}$ & $0.41\substack{0.029\\0.029}$ &  $9.3\substack{0.44\\0.43}$ &    $1.3\substack{0.02\\0.02}$ &      $1.3\substack{0.14\\0.14}$ &  $0.42\substack{0.023\\0.023}$ &          8.20 &         $16.0\substack{7.3\\5.5}$ \\
J092552+395714 &    0.3141 &  $0.18\substack{0.072\\0.072}$ & $8.2\substack{0.049\\0.049}$ &    $1.1\substack{0.03\\0.03}$ &  $8.8\substack{0.55\\0.39}$ & $0.77\substack{0.031\\0.031}$ & $-0.088\substack{0.065\\0.065}$ &  $0.46\substack{0.033\\0.033}$ &          0.00 & $0.45\substack{-99\\-99}$ \\
J095700+235709 &    0.2444 &  $0.37\substack{0.023\\0.023}$ & $8.4\substack{0.035\\0.035}$ &  $1.3\substack{0.037\\0.037}$ & $11.0\substack{0.44\\0.43}$ &  $1.4\substack{0.019\\0.019}$ &     $0.37\substack{0.05\\0.05}$ & $-0.19\substack{0.021\\0.021}$ &          0.00 & $0.08\substack{-99\\-99}$ \\
J095838+202508 &    0.3013 &  $0.15\substack{0.058\\0.058}$ & $7.8\substack{0.037\\0.037}$ & $0.49\substack{0.029\\0.029}$ &  $8.7\substack{0.45\\0.42}$ &  $1.2\substack{0.021\\0.021}$ &      $1.0\substack{0.12\\0.12}$ &  $0.91\substack{0.031\\0.031}$ &          2.20 &          $1.9\substack{2.8\\1.2}$ \\
J103344+635317 &    0.3465 & $0.068\substack{0.026\\0.026}$ & $8.2\substack{0.039\\0.039}$ & $0.54\substack{0.029\\0.029}$ &  $9.1\substack{0.44\\0.43}$ &  $1.4\substack{0.025\\0.025}$ &      $1.1\substack{0.12\\0.12}$ &    $0.66\substack{0.03\\0.03}$ &          8.20 &       $31.0\substack{15.0\\12.0}$ \\
J103816+452718 &    0.3256 &  $0.21\substack{0.019\\0.019}$ & $8.4\substack{0.032\\0.032}$ &   $0.63\substack{0.03\\0.03}$ & $10.0\substack{0.45\\0.42}$ &  $1.6\substack{0.019\\0.019}$ &    $1.2\substack{0.098\\0.098}$ &  $0.27\substack{0.022\\0.022}$ &          4.60 &        $0.7\substack{0.25\\0.22}$ \\
J112933+493525 &    0.3448 & $0.037\substack{0.037\\0.037}$ & $8.3\substack{0.061\\0.061}$ & $0.44\substack{0.029\\0.029}$ &  $8.4\substack{0.47\\0.41}$ &   $0.69\substack{0.04\\0.04}$ &     $0.61\substack{0.15\\0.15}$ &  $0.66\substack{0.044\\0.044}$ &          0.00 &  $2.1\substack{-99\\-99}$ \\
J113304+651341 &    0.2414 &  $0.14\substack{0.023\\0.023}$ & $8.0\substack{0.039\\0.039}$ &  $0.7\substack{0.031\\0.031}$ &  $9.6\substack{0.48\\0.38}$ & $0.87\substack{0.024\\0.024}$ &   $0.38\substack{0.077\\0.077}$ &   $0.7\substack{0.027\\0.027}$ &          2.50 &         $2.2\substack{2.2\\0.87}$ \\
J115855+312559 &    0.2430 &  $0.18\substack{0.018\\0.018}$ & $8.4\substack{0.031\\0.031}$ & $0.55\substack{0.029\\0.029}$ &  $9.7\substack{0.52\\0.35}$ &  $1.3\substack{0.021\\0.021}$ &    $1.1\substack{0.092\\0.092}$ &  $0.37\substack{0.023\\0.023}$ &          8.20 &          $6.6\substack{3.0\\1.5}$ \\
J123519+063556 &    0.3326 &    $0.08\substack{0.02\\0.02}$ & $8.4\substack{0.053\\0.053}$ & $0.45\substack{0.029\\0.029}$ &  $8.9\substack{0.44\\0.43}$ &  $1.2\substack{0.032\\0.032}$ &      $1.1\substack{0.14\\0.14}$ &  $0.75\substack{0.037\\0.037}$ &          6.20 &          $5.2\substack{2.8\\1.4}$ \\
J124619+444902 &    0.3220 &   $0.2\substack{0.056\\0.056}$ & $8.0\substack{0.041\\0.041}$ &  $1.0\substack{0.032\\0.032}$ &  $8.9\substack{0.44\\0.43}$ &  $1.5\substack{0.019\\0.019}$ &   $0.67\substack{0.067\\0.067}$ &   $0.7\substack{0.031\\0.031}$ &          2.30 &        $0.44\substack{0.93\\0.2}$ \\
J124835+123403 &    0.2629 & $0.099\substack{0.026\\0.026}$ & $8.2\substack{0.036\\0.036}$ & $0.33\substack{0.037\\0.037}$ &  $8.7\substack{0.44\\0.43}$ &  $1.2\substack{0.023\\0.023}$ &        $1.4\substack{0.2\\0.2}$ &  $0.66\substack{0.026\\0.026}$ &          2.20 &          $4.7\substack{4.3\\2.6}$ \\
J130128+510451 &    0.3476 & $0.097\substack{0.031\\0.031}$ & $8.3\substack{0.045\\0.045}$ & $0.61\substack{0.029\\0.029}$ &  $9.1\substack{0.44\\0.43}$ &  $1.4\substack{0.027\\0.027}$ &      $1.0\substack{0.11\\0.11}$ &  $0.64\substack{0.032\\0.032}$ &          5.90 &          $2.7\substack{1.3\\1.4}$ \\
J131037+214817 &    0.2830 &  $0.19\substack{0.038\\0.038}$ & $8.4\substack{0.044\\0.044}$ & $0.42\substack{0.029\\0.029}$ &  $9.0\substack{0.48\\0.42}$ &  $1.1\substack{0.024\\0.024}$ &      $1.1\substack{0.13\\0.13}$ &  $0.31\substack{0.027\\0.027}$ &          5.70 &         $1.6\substack{2.0\\0.58}$ \\
J131419+104739 &    0.2960 &  $0.26\substack{0.029\\0.029}$ & $8.3\substack{0.036\\0.036}$ & $0.97\substack{0.057\\0.057}$ & $10.0\substack{0.45\\0.42}$ &  $1.4\substack{0.022\\0.022}$ &     $0.58\substack{0.11\\0.11}$ &  $0.17\substack{0.024\\0.024}$ &          0.00 &  $0.1\substack{-99\\-99}$ \\
J132633+421824 &    0.3176 &   $0.055\substack{0.08\\0.08}$ & $8.2\substack{0.032\\0.032}$ & $0.61\substack{0.029\\0.029}$ &  $8.7\substack{0.44\\0.43}$ &  $1.3\substack{0.023\\0.023}$ &   $0.97\substack{0.099\\0.099}$ &  $0.63\substack{0.025\\0.025}$ &          5.00 &        $12.0\substack{14.0\\8.4}$ \\
J134559+112848 &    0.2371 &   $0.3\substack{0.022\\0.022}$ & $8.3\substack{0.055\\0.055}$ &  $1.3\substack{0.034\\0.034}$ & $10.0\substack{0.45\\0.41}$ &  $1.2\substack{0.023\\0.023}$ &   $0.17\substack{0.047\\0.047}$ &  $0.17\substack{0.024\\0.024}$ &          0.00 &  $0.2\substack{-99\\-99}$ \\
J141013+434435 &    0.3557 & $0.072\substack{0.056\\0.056}$ & $8.0\substack{0.035\\0.035}$ & $0.51\substack{0.029\\0.029}$ &  $8.7\substack{0.44\\0.43}$ &  $1.3\substack{0.023\\0.023}$ &      $1.0\substack{0.13\\0.13}$ &   $1.1\substack{0.029\\0.029}$ &          7.70 &         $11.0\substack{7.1\\7.0}$ \\
\hline
\end{tabular} $}
\justifying
{\\ \\ From left to right: (1) Object (2) redshift (3) UV extinction (4) gas metallicity (5) NUV COS 1/2 light radius (6) log of stellar mass (7) log of SFR, derived from $H_{\beta}$ (8) log of SFR surface density, using $H_{\beta}$ derived SFR (9) LyC detection significance (10) LyC escape fraction, derived from UV SED fitting. Redshift values were obtained by \cite{Xu2023} using Balmer lines and by \cite{Marques-Chaves2022a} using non-resonant H II lines.  UV extinction values were taken from \cite{Saldana-Lopez2022}.  All other values were taken from \cite{Flury2022a}.}

\label{tab:lzlcs_data}
\end{table*}

Understanding the optimal geometry for LyC escape and its timing relative to star formation remains an active area of research.  For instance, \cite{Choustikov2023} studied LyC emitters in SPHINX simulations and postulated a framework for escape which emphasized the need for copious LyC production, supernovae feedback, and a precise timeline (3.5-10 Myr) to allow feedback enough time to clear the neutral ISM/CGM before massive stars reach the end of their life cycles and LyC production wanes (see also \citealt{Hayes2019,Rosdahl2022}).  They proposed a linear model to predict $f_{esc}^{LyC}$ based on several observable indicators to ensure the above conditions, which they used on real observations.  \cite{Jaskot2024a} took a similar multi-variate approach to predicting $f_{esc}^{LyC}$ from several LyC escape indicators using a survival analysis model, but their model was trained on the LzLCS data instead of simulations.  They made predictions of $f_{esc}^{LyC}$ in the same high-z galaxies as \cite{Choustikov2023}, but got different values.  \cite{Jaskot2024b} traced the discrepancies to a subset of compact galaxies with high $\rm O_{32}$ and $f_{esc}^{LyC}$ values in LzLCS that were absent in SPHINX, suggesting that \cite{Choustikov2023} might underestimate the role of radiation feedback in compact galaxies with young massive stars. Ultimately, these tests highlight the complexity of the relationship between feedback, ISM/CGM geometry, and LyC escape, emphasizing the need for a deeper understanding of their interplay.  

In this work, we aim to reveal the relationship between feedback, the geometry of the ISM/CGM, and LyC escape in star-forming galaxies by performing radiation transport modeling of the Mg II 2796\AA, 2804\AA\ lines observed in higher resolution follow-up spectra of 29 confirmed LyC leakers and non-leakers from LzLCS.  Mg II has emerged as a promising probe of the neutral gas ($\rm H^0$) content of galaxies, complementary to studies of the ionized outflows, and as a successful proxy of LyC escape \citep{Henry2018}.  For example, \cite{Chisholm2020} was able to infer the column density of $\rm H^0$ from optically thin Mg II lines to predict $f_{esc}^{LyC}$ in 10 local star forming galaxies.  Later, \cite{Xu2022b,Xu2023} expanded on the work of \cite{Chisholm2020}, and extended their models to include LyC emitters with optically thick and thin partial coverings using an empirical based correction to quantify the covering fraction of optically thick gas.  They were able to successfully predict $f_{esc}^{LyC}$ values for a sample of LzLCS galaxies, many of which are in our data set. 

We build on these works by performing a comprehensive radiation transport modeling using the semi-analytical line transfer (SALT) model presented in \cite{Carr2023}. SALT is designed to handle both absorption-dominated and absorption+emission-dominated systems, the latter of which \cite{Katz2022} found to be common in simulations of high-redshift galaxies. By incorporating photoionization models, we will map the $\rm H^0$ content of the winds using our model constraints on $\rm Mg^+$.  Additionally, by comparing outflow properties with light fractions—the fraction of total starlight from stellar populations within a specified age range (Flury et al., in prep)—we aim to connect outflows to their driving mechanisms and ultimately establish the link between feedback, ISM/CGM geometry, and LyC escape.  

The remainder of this paper is organized as follows: In Section 2, we describe our dataset and summarize the general galaxy properties measured by \cite{Flury2022a}. Section 3 provides an overview of the SALT model and details our fitting procedure. In Section 4, we use photoionization modeling to determine the $\rm H^0$ content of the winds and model $f_{esc}^{LyC}$. Section 5 explores the general properties of the winds traced by $\rm H^0$. In Section 6, we examine the relationship between $f_{esc}^{LyC}$ and outflows. Our discussion in Section 7 interprets the results and develops a framework for understanding how feedback influences LyC escape. Finally, we present our conclusions in Section 8.

Throughout this work we assume a solar metallicity value of 12 + $\log(\text{O}/\text{H})_{\odot} = 8.69$ \citep{Asplund2009}.

\section{Data} \label{sec:data}

Our data set consists of two samples of higher resolution follow up observations of objects in LzLCS.  The smaller sample consists of eight objects observed with the VLT X-Shooter spectrograph (3000–5600\AA) by \cite{Marques-Chaves2022a}, and the larger sample consists of 23 objects observed with the MMT blue spectrograph by \cite{Henry2018,Xu2023}.  There are two objects with observations from both instruments.  This places the total unique galaxies in our sample at 29.   The approximate spectral resolution, near the Mg II $2796$\AA, $2803$\AA\ doublet, is $55 \ \rm km \ s^{-1}$ for X-Shooter and $90 \ \rm km \ s^{-1}$ for the MMT.  In what follows, we provide a brief summary of the data reduction and refer the reader to the references above for further details.

The X-Shooter observations were obtained as part of the ESO program ID 106.215K.001. Observations were carried out between October 2020 and April 2021 under seeing of $\simeq 0.7^{\prime \prime}-1.1^{\prime \prime}$ (FWHM) and grey moon conditions. We used 0.9$^{\prime \prime}$ , 1.0$^{\prime \prime}$, and 1.0$^{\prime \prime}$ slits in the UVB, VIS and NIR arms providing power resolution of $R \sim 5400$, 8900, and 5600, from 3000–5600\AA, 5500–10200\AA, and 10200–24000\AA,  respectively. Slits were oriented with the parallactic angle.  Observations were performed in nodding-on-slit mode with a standard ABBA sequence and total on-source exposure times of 46~min and 92~min, depending on the brightness of each source.  

The X-Shooter spectra were reduced in a standard manner using the ESO Reflex reduction pipeline (version 2.11.5; \citealt{Freudling2013}). For each observing block (OB), the pipeline performs bias (UVB and VIS) or dark (NIR) subtraction, flat-field correction, wavelength calibration, and 2D mapping. Sky subtraction is done by taking the difference of the 2D spectra from nodding positions A and B. Finally, the pipeline produces flux-calibrated 2D spectra using standard stars. 1D spectra are extracted and co-added if more than one OB was observed for a specific source. We use the extinction curve of \cite{Fitzpatrick1999} and the extinction map of \cite{Schlafly2011} to correct for the reddening effect in the Galaxy.  Finally, the flux of the spectrum is matched to that obtained from the SDSS photometry in the g-band to account for slit losses. 

The MMT observations were conducted on six nights in three different semesters (2017, 2019A, 2020A, and 2021A) with exposure times ranging from 30 to 180 minutes at 3100-4100\AA.  The slit was reset to the parallactic angle for each exposure.  The data were reduced using standard IDL+IRAF routines following the methodology of \cite{Henry2018}.  The wavelength calibration is applied from the HeArHgCd arc lamps.  By matching arc lines, the rms of the residuals was found to be $< 0.1$\AA\ ($\sim 10\ \rm km^{-1}$ around the Mg II spectral regions). 

In both data sets, the spectra were normalized by performing a linear fit to adjacent flux values near emission and absorption features, making sure to avoid the features themselves, to establish the continuum for normalization.  

We have listed all relevant galaxy properties studied in this work in Table~\ref{tab:lzlcs_data}.

\section{Modeling} \label{sec:model}

\begin{figure}
	\centering
	\includegraphics[scale=0.48]{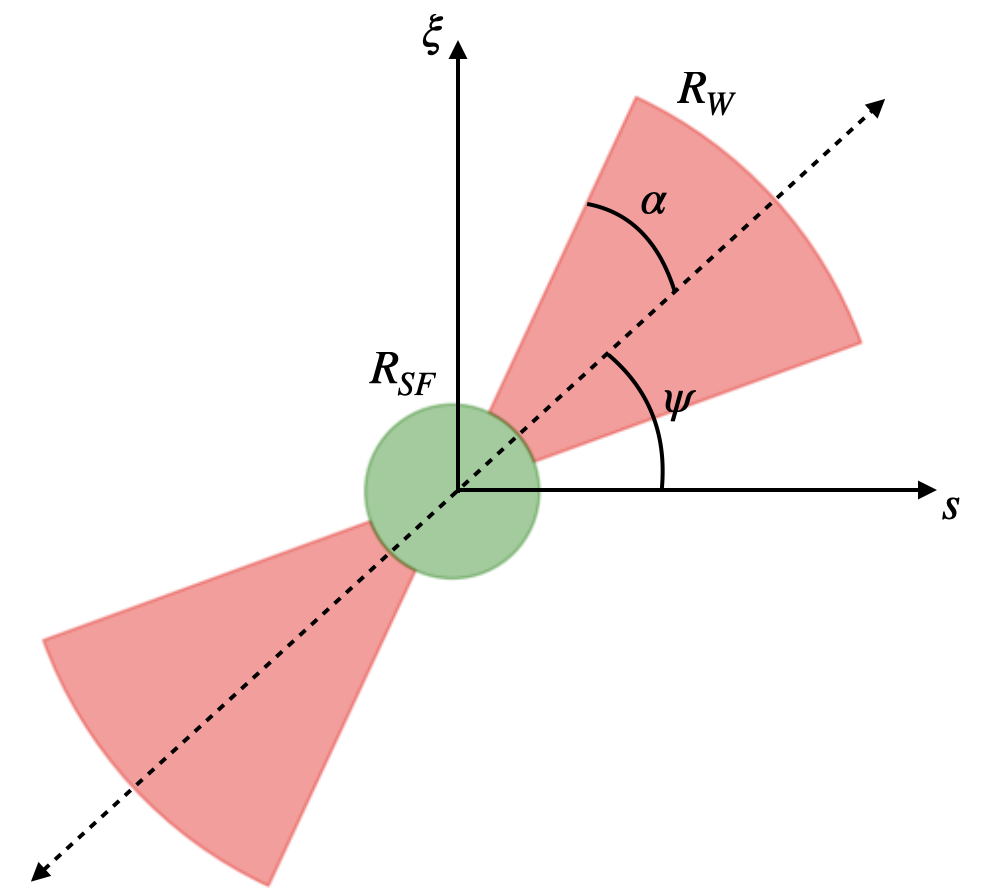}
	\caption{shows a crosssectional view of the idealized outflow underlying the SALT calculations. The model consists of a spherical source ($r=R_{\rm SF}$) of isotropic radiation surrounded by a bi-conical outflow of half-opening angle, $\alpha$, and orientation angle, $\psi$, which extends to a terminal radius, $R_{\rm W}$.  The $s$ and $\xi$-axes point in directions parallel and perpendicular to the observer, respectively, and are measured in units of distance normalized by $R_{SF}$.}
	\label{fig:bicone}
\end{figure} 

\begin{table*}[ht]
\centering
\caption{Atomic Data for Mg II ion.}
\begin{tabular}{c|c|c|c|c|c|c|c}
\hline\hline
Ion & Vac. Wavelength & $A_{ul}$ & $f_{ul}$ & $E_{l}-E_{u}$ & $g_l-g_u$ & Lower Level & Upper Level \\
& \AA & $s^{-1}$ & & $eV$ & & Conf.,Term,J & Conf.,Term,J \\ [.5ex]
\hline
Mg II & $2796.352$ & $2.60\times 10^{8}$ & $0.608$ & $0.0000-35 760.88$ & $2-4$ & $2p^63s, {}^2S, 1/2$ & $2p^63p, {}^2P^0, 3/2$ \\
& $2803.530$ & $2.57\times 10^{8}$ & $0.303$ & $0.0000-35 669.31$ & $2-2$ & $2p^63s, {}^2S, 1/2$ & $2p^63p, {}^2P^0, 1/2$ \\
\hline
\end{tabular}

\vspace{0.2cm} 

\begin{minipage}{\textwidth}
\justifying
\small
\noindent From left to right: (1) ion name (2) wavelength (3) Einstein coefficient (4) oscillator strength (5) energy difference (6) atomic weight (7) lower state configuration (8) upper state configuration. Data taken from the NIST Atomic Spectra Database (\url{http://www.nist.gov/pml/data/asd.cfm})
\end{minipage}

\label{tab:atomicdata}
\end{table*}

To model the Mg II 2796\AA, 2804\AA\ spectral lines in LzLCS, we use the semi-analytical line transfer (SALT) model of \cite{Carr2023}.  SALT is a forward model of galactic outflows to spectral line predictions and represents a fully consistent picture of radiation transport in a 3-dimensional idealized outflow.  It is an attempt at solving the radiation transport equation exactly, while assuming the Sobolev approximation \citep{Lamers1999}, for the case of bound-bound interactions.  SALT has been rigorously tested against numerical simulations to identify degeneracy \citep{Carr2018,Carr2023} and the validity/meaning of its idealized parameter space (\citealt{Carr2023}; Carr et al., in prep).  The goal of this section is to introduce the reader to the SALT formalism and its limitations.  Since SALT is well documented in the literature (see \citealt{Scarlata2015,Carr2018,Carr2021a,Carr2023}), we provide only the essential details and refer the reader to the literature for the more nuanced information.  We conclude this section by discussing our model fitting procedure.


\subsection{SALT Model}
The SALT calculations are performed on an idealized model of a galactic outflow.  The base model features a spherical source of isotropic radiation with a radius of $R_{\rm SF}$, surrounded by a bi-conical outflow that extends to a terminal radius of $R_{\rm W}$.  Physically, the source is meant to embody the star forming regions of a galaxy from which an outflow is expelled.  In this work, we set $R_{\rm SF} = r_{1/2}$, where $r_{1/2}$ is the UV half-light radius measured from COS acquisition images \citep{Carr2021a,Huberty2024}.  The outflow geometry is characterized by a half-opening angle, $\alpha$, and orientation angle, $\psi$, subtended by the axis of the bi-cone and line of sight.  A diagram is provided in Figure~\ref{fig:bicone}. 

The outflow is further characterized by a density field of power law, 
\begin{eqnarray}
n = n_0\left(\frac{R_{SF}}{r}\right)^{\delta},
\end{eqnarray}
where $n_0$ is the number density at $R_{SF}$ and $\delta$ is the power law index.  Likewise, the velocity field of the outflow assumes a power law of the form,
\begin{equation}
\begin{aligned}
v &= v_0\left(\frac{r}{R_{\text{SF}}}\right)^{\gamma} &&\text{for}\ r < R_{W} \\[1em]
v &= v_{\infty}  &&\text{for} \ r \geq R_{W}, 
\end{aligned}
\label{Velocity_Equation}
\end{equation}  
\noindent where $v_0$ is the launch velocity at $R_{SF}$, $v_{\infty}$ is the terminal velocity at $R_W$, and $\gamma$ is the power law index.  The porosity of the outflow is described by the parameter, $f_c$, which represents the fraction of a bi-conical shell covered by material. $f_c$ is assumed to be constant with radius (see \citealt{Carr2018,Carr2021a}).  Note that $f_c$ is not generally equal to the covering fraction, $C_f$, typically inferred from spectra.  The latter quantity also encodes information about the large scale geometry of the bi-cone relative to the line of sight. 

SALT models both the absorption and re-emission (resonant + fluorescent) of photons by the outflow.  Absorption is controlled by the free parameter, $\tau$, which is related to the optical depth, $\tau_0$, through the relation, $\tau = \tau_0/f_{ul}\lambda_{ul}$, where $f_{ul}$ and $\lambda_{ul}$ are the oscillator strength and wavelength from the relevant transitions, respectively.  $n_0$ can be recovered from $\tau$ through the relation,
\begin{eqnarray}
    \tau = \frac{\pi e^2}{mc}n_0\frac{R_{SF}}{v_0},  
\end{eqnarray}
where $e$ is the electron charge, $m$ is the electron mass, and $c$ is the speed of light.  All relevant atomic information for the Mg II 2796\AA, 2804\AA\ doublet is provided in Table~\ref{tab:atomicdata}. 

SALT can account for observational effects such as a limiting circular observing aperture with projected radius, $R_{AP}$, on the plane of the sky.  $v_{ap}$ represents the value of the velocity field at $R_{AP}$ and designates the free parameter in SALT controlling the aperture.  A failure of the aperture to capture the full extent of an outflow will result in reduced emission (see \citealt{Scarlata2015}).  SALT can account for dust extinction in the CGM assuming a fixed dust-to-gas ratio as well.  This parameter is controlled by $\kappa$ which is related to the dust opacity (see \citealt{Carr2021a}).

All of our galaxies show Mg II emission emanating from H II regions within the star forming regions of the galaxy in addition to the continuum.  To account for this, we include the addition of Gaussian-shaped line profiles, centered on each resonant line, to be propagated through the outflow in SALT along with the otherwise flat continuum distribution (see \citealt{Carr2023} for a more detailed description).  Each Gaussian is controlled by two free parameters: the height, $A$, and the standard deviation or width, $\sigma$.  A complete list of the SALT and ISM free parameters along with their definitions is provided in Table~\ref{tab:free_parameters}.

\subsection{Derived Quantities: Mass Outflow Rate, Column Density, etc.}

In the SALT formalism, the mass outflow rate, $\dot{M}$, can be computed directly from the free parameters as
\begin{eqnarray}
\dot{M}(v) &=& 4\pi(1-\cos{\alpha)}R_\text{SF}^2 \nonumber\\
&& \times m_{\rm ion}n_{0}v_0 \left( \frac{r}{R_{\rm SF}} \right)^{2+\gamma - \delta},
\label{eq:SALT_MOR}
\end{eqnarray}
where $m_{\rm ion}$ is the mass of the relevant ion.  Furthermore, the momentum outflow rate, $\dot{P}$, can be computed as 
\begin{eqnarray}\label{eq:SALT_POR}
    \dot{P}(v)=\dot{M} v,
\end{eqnarray}
and the energy outflow rate, $\dot{E}$, as 
\begin{eqnarray}\label{eq:SALT_EOR}
    \dot{E}(v)=\frac{1}{2}\dot{M}v^2.
\end{eqnarray}

The column density can be computed directly from the free parameter space as well.  However, because SALT assumes an extended source, care must be taken to account for variations in the density field along different lines of sight.  To do this, we average the column density over different paths after randomly sampling over the surface of the source.  The column density, $N_{l\xi}$, along an arbitrary sight line can be computed as
\begin{eqnarray}
N_{l\xi} = \int_{L}^{U}\Phi_{\rm cone}n_{0,\rm ion}\left(\frac{l^2+{s^{\prime}}^2+\xi^2}{R_\text{SF}^2}\right)^{-\delta/2}\text{d}s^{\prime}
\label{eq:Nse}
\end{eqnarray}
where,
\begin{eqnarray}
U = (R_W^2-l^2-\xi^2)^{1/2},
\end{eqnarray}
\begin{eqnarray}
L = (R_\text{SF}^2-l^2-\xi^2)^{1/2},
\end{eqnarray}
and
\begin{gather} 
\Phi_{\rm cone} \equiv 
\begin{cases}
1 &\ \rm{if} \  (s^{\prime},l,\xi) \in \text{bi-cone} \\[1em]
0 & \ \rm{otherwise}. \\[1em]
\end{cases}
\end{gather}
The indicator function, $\Phi_{\rm cone}$, accounts for the bi-conical geometry of the outflow, and $s$, $\xi$, $l$ denote locations in the $s\xi l$-space.  Note that $l$ is perpendicular to the $s\xi$-plane (see Figure~\ref{fig:bicone}).  A more detailed description with a diagram is provided in the Appendix of \cite{Huberty2024}.  To achieve the final column density, $N$, responsible for the line profile observed in SALT, we average $N_{l\xi}$ over multiple lines of sight - that is, 
\begin{equation}
    N = \frac{\sum_i^I N_{l(i)\xi(i)}}{I},
    \label{eq:average_N}
\end{equation}
where the sum occurs over $I$ lines of sight drawn uniformly over the surface of the source.  We find that the values of $N$ start to converge to roughly two significant figures by $I = 10^4$.

We can adapt this same Monte Carlo technique to easily compute the covering fraction, $C_f$, of the source by the outflow.  Note that, in general, $C_f$ is a complicated function which depends on parameters $f_c$, $\alpha$, and $\psi$.  When uniformly sampling over the source, we register a “hit” at coordinate $(l,\xi)$ if $N_{l\xi} > 0$.  $C_f$ can then be calculated from the total number of hits, $H$, as\footnote{Note that this definition is highly idealized and suited for the SALT formalism.  A more general expression, for instance, may exclude lines of sight which fall below a minimal column density greater than one.} 
\begin{eqnarray}
    C_f = f_c H/I.
\end{eqnarray}

\begin{table}
\centering
\caption{SALT Model Parameters}
\resizebox{\columnwidth}{!}{$\begin{tabular}{cccc}
\hline\hline
Parameter & Description & Prior&ICs \\[.5 ex]
\hline
\multicolumn{3}{l}{Outflow}\\
\hline
$\alpha$ & half-opening angle [rad] & $[0,\pi/2]$& $[0,\pi/2]$ \\
$\psi$ & orientation angle [rad] & $[0,\pi/2]$& $[0,\pi/2]$ \\
$v_0$ & launch velocity [$\mathrm{km \ s^{-1}}$] & $[0,150]$ & $[0,150]$\\
$v_{\infty}$ & terminal velocity [$\mathrm{km \ s^{-1}}$] & $[150,1200], \ > v_0$ & $[200,800]$ \\
$v_{ap}$ & velocity at $R_{AP}$ [$\mathrm{km \ s^{-1}}$] & $[200,1200]$ & $[200,800]$\\
$\gamma$ & velocity field power law index & $[0.5,1]$ &$[0.5,1]$\\
$\delta$ & density field power law index & $0.5<\delta-\gamma<3.5$& $[-1.5,1.5]$ \\
$\tau$ & optical depth divided by $f_{ul}\lambda_{ul}$ [$\text{\AA}^{-1}$] & $-3<\log{\tau}<1 $& $[-2,1]$\\
$f_c$ & covering fraction & $[0,1]$ & $[0,1]$\\
$\kappa$ & dust opacity multiplied by $R_\text{SF}n_{0,\text{dust}}$ & -- & -- \\
\hline
\multicolumn{3}{l}{ISM}\\
\hline
$\sigma_{2796}$ & 2796\AA\ intrinsic emission width [$\mathrm{km \ s^{-1}}$] & $[0,300]$ & $[0,300]$ \\
$A_{2796}$ & 2796\AA\ intrinsic emission height & $[0,12]$ & $[0,12]$ \\
$\sigma_{2804}$ & 2804\AA\ intrinsic emission width [$\mathrm{km \ s^{-1}}$] & $\sigma_{2796}=\sigma_{2804}$&-- \\
$A_{2804}$ &2804\AA\ intrinsic emission height& $1<A_{2796}/A_{2804}<2.5$ & $[0,12]$ \\
\hline
\end{tabular} $}
\justifying
From left to right: parameter (1) name (2) description (3) Prior (4) Initial Conditions.  See \cite{Carr2023} for additional definitions. Dashed lines refer to constrained quantities.  
\label{tab:free_parameters}
\end{table}

\subsection{Constraints}

We use the Calzetti attenuation law, $k(2800\text{\AA})$ \citep{Calzetti2000}, to constrain the dust optical depth, $\tau_{\rm dust} = k(2800\text{\AA}) \rm{E(B-V)}_{\rm UV}/1.086$.  We note, however, that \cite{Saldana-Lopez2022} used the attenuation law of \cite{Reddy2016} to obtain the $\rm{E(B-V)}_{\rm UV}$ values used by \cite{Flury2022a}.  We changed laws because $2800\text{\AA}$ falls outside the domain of the Reddy Law.  \cite{Reddy2016} note, however, that the Calzetti attenuation law and their relation tend to converge at longer wavelengths.  $\tau_{\rm dust}$ can then be used to compute, $\kappa$, with the relations, 
\begin{equation}
  \tau_{\rm dust} =
    \begin{cases}
      \frac{k}{\gamma}\ln{(\frac{v_{\infty}}{v_0})}& \text{if} \ \delta=1\\
      \frac{k}{1-\delta}\left[\left(\frac{v_{\infty}}{v_0}\right)^{(1-\delta)/\gamma}-1\right] & \text{otherwise.}\\
    \end{cases}       
\end{equation}
We estimate the uncertainty on $\kappa$ using distributions in the same manner described in the next subsection.  To generate distributions: we assume the uncertainty in $\rm{E(B-V)}_{\rm UV}$ to follow a normal distribution, then we sample the corresponding Gaussian $10^4$ times along with the SALT parameter chains obtained from model fitting.


In the absence of radiative transfer effects, the line ratio governing the intrinsic nebular emission from the central H II regions is determined solely by the ratio of the wavelength-oscillator strength products, which for the Mg II 2796\AA, 2804\AA\ doublet is approximately two. However, we do expect some scattering within the ISM, which is currently unaccounted for in the SALT formalism. Based on observational constraints from \cite{Chisholm2020} for optically thin systems, and theoretical studies by \cite{Katz2022} and \cite{Chang2024}, we constrain the amplitude ratio from the emission lines emerging from the ISM to lie within the range
\begin{eqnarray}
    1 < \frac{A_{2796}}{A_{2804}} < 2.5.
\end{eqnarray}
We note that radiation transfer effects occurring on the scale of H II regions have been proposed as a potential means to explain the broadening of Ly$\alpha$ lines.  For example, this was proposed by \cite{Orlitova2018} who required much broader intrinsic Ly$\alpha$ profiles than expected from H$\beta$ to explain the observed profiles in 10 Green Pea galaxies using (spherical) shell models of galactic outflows.  Lastly, we have chosen to set $\sigma_{2796} = \sigma_{2804}$ and fix the position of each Gaussian to line center.  

During our initial modeling efforts, the relatively low spectral resolution of the observations limited our ability to constrain the upper bound on $\tau$ for most galaxies. To address this, we turned to the FIRE-2 simulations \citep{Hopkins2018}.  To find a reasonable range for $\tau$, we sampled $\tau$ values for magnesium from simulaiton m11c, an intermediate-mass simulation ($\log{M_{\star}} \approx 8.9$, $Z/Z_{\odot} = 0.22$) take from the “core” FIRE-2 suite \citep{Pandya2021}.  Specifically, we extracted $\tau$ values from 35 snapshots covering a redshift range of $0.008 < z < 0.07$, which spans approximately three outflow episodes or baryonic cycles. Measurements of $\tau$ were taken at various radii, ranging from one to two times the stellar half-mass radius, with the maximum value being on the order of unity. The FIRE snapshots spanned a range of $0.00029 < \rm SFR < 0.53$, or approximately a factor of two lower than the SFRs measured in our galaxies.  Considering the lower SFRs and redshifts in the simulations, as well as the fact that $\tau$ corresponds to the total amount of magnesium rather than just $\rm Mg^+$, we chose a conservative upper bound for the prior of $\tau = 10\ \rm \text{\AA}^{-1}$. 

Following \cite{Huberty2024}, we constrain $0.5<\delta-\gamma<3.5$.  The provided range allows for deviations from a constant outflow rate (i.e., $\delta = \gamma +2$) up to $\pm 1.5$.   

Given that our spectra were obtained from long slit spectrographs (i.e., non-circular apertures), we have chosen to leave $v_{ap}$ as a free parameter (cf. \citealt{Carr2021a}).  We suspect the limiting aperture to be the dominant source of uncertainty in this study (see below).  We intend to generalize SALT to account for non-circular apertures (e.g., in the application of integral field unit spectroscopy) in future works.      

In total, our model contains 12 free parameters. 
\begin{table*}
\caption{SALT model parameter constraints for $\rm{Mg}^{+}$ gas}
\resizebox{\textwidth}{!}{$%
\begin{tabular}{cccccccccccccc}
\hline\hline
Galaxy&$\alpha$&$\psi$&$f_c$&$v_0$&$v_{\infty}$&$\gamma$&$\tau$&$\delta$&$\kappa$&$v_{ap}$&$A_{2796}$&$A_{2804}$&$\sigma_{2796,2804}$\\[.5 ex]
& deg &deg&&$\rm km \ s^{-1}$& $\rm km \ s^{-1}$&&$\text{\AA}^{-1}$&&&$\rm km \ s^{-1}$&&&$\rm km \ s^{-1}$\\[.5 ex]
\hline
J012910+145935m & $30.0\substack{35.0\\13.0}$ &  $18.0\substack{23.0\\9.2}$ &     $0.68\substack{0.15\\0.3}$ &  $99.0\substack{26.0\\36.0}$ & $790.0\substack{190.0\\330.0}$ &  $0.74\substack{0.55\\0.14}$ &    $0.016\substack{0.5\\0.011}$ &  $2.3\substack{1.1\\0.38}$ &   $1.6\substack{1.4\\0.41}$ & $400.0\substack{360.0\\130.0}$ &    $8.1\substack{2.1\\2.7}$ &   $4.0\substack{1.7\\0.97}$ &    $27.0\substack{14.0\\7.0}$  \\
J072326+414608m &   $8.6\substack{44.0\\2.5}$ &   $6.3\substack{43.0\\3.6}$ &   $0.84\substack{0.073\\0.44}$ &  $68.0\substack{43.0\\43.0}$ & $680.0\substack{240.0\\230.0}$ &   $1.1\substack{0.47\\0.32}$ &      $0.15\substack{1.6\\0.12}$ & $2.5\substack{0.92\\0.32}$ &   $2.1\substack{1.4\\0.47}$ & $>680.0$ &   $3.7\substack{3.0\\0.98}$ &   $1.6\substack{2.8\\0.29}$ &   $69.0\substack{49.0\\39.0}$  \\
J081112+414146m &  $80.0\substack{4.0\\32.0}$ &  $13.0\substack{29.0\\5.5}$ &      $0.6\substack{0.14\\0.2}$ &  $140.0\substack{3.6\\63.0}$ & $870.0\substack{140.0\\220.0}$ &   $1.8\substack{0.12\\0.56}$ &    $0.14\substack{0.85\\0.099}$ &  $4.4\substack{0.35\\1.1}$ &    $4.7\substack{0.5\\1.6}$ & $570.0\substack{240.0\\190.0}$ &    $7.3\substack{1.5\\1.6}$ &   $3.3\substack{1.3\\0.38}$ &   $82.0\substack{26.0\\25.0}$ \\
J082652+182052m & -- & -- & -- & -- & -- & -- & -- & -- & -- & -- &  $0.68\substack{4.6\\0.25}$ &  $0.29\substack{4.9\\0.08}$ & $150.0\substack{30.0\\140.0}$ \\
J091208+505009m &  $40.0\substack{25.0\\8.0}$ &  $15.0\substack{17.0\\7.4}$ &   $0.99\substack{0.0042\\0.1}$ &    $5.7\substack{19.0\\1.5}$ & $770.0\substack{180.0\\370.0}$ & $0.66\substack{0.63\\0.087}$ &      $0.45\substack{1.5\\0.35}$ &  $1.7\substack{1.4\\0.18}$ &  $0.82\substack{1.6\\0.18}$ &   $210.0\substack{410.0\\3.5}$ &   $3.5\substack{0.43\\1.4}$ &   $2.5\substack{6.2\\0.88}$ &   $91.0\substack{24.0\\14.0}$ \\
J091703+315221m & $57.0\substack{20.0\\17.0}$ & $48.0\substack{13.0\\25.0}$ &    $0.53\substack{0.17\\0.15}$ & $120.0\substack{12.0\\54.0}$ &  $490.0\substack{210.0\\68.0}$ &  $0.72\substack{0.57\\0.11}$ &   $0.036\substack{0.65\\0.023}$ &  $2.5\substack{1.0\\0.36}$ & $0.94\substack{0.63\\0.27}$ & $>490.0$ &    $2.0\substack{4.4\\1.1}$ &   $1.6\substack{3.8\\0.77}$ &   $30.0\substack{57.0\\20.0}$  \\
J092552+395714m &  $37.0\substack{29.0\\9.7}$ &   $7.4\substack{26.0\\2.3}$ &    $0.79\substack{0.11\\0.24}$ &  $140.0\substack{3.0\\54.0}$ &  $330.0\substack{240.0\\31.0}$ &  $0.74\substack{0.71\\0.12}$ &      $0.52\substack{2.1\\0.46}$ &  $2.2\substack{1.4\\0.16}$ &   $2.1\substack{1.7\\0.71}$ & $>330.0$ &    $7.9\substack{2.0\\2.4}$ &    $4.8\substack{1.6\\1.3}$ &     $21.0\substack{7.1\\5.3}$  \\
J103344+635317m & -- & -- & -- & -- & -- & -- & -- & -- & -- & -- &    $7.1\substack{2.6\\2.5}$ &   $3.5\substack{2.4\\0.95}$ &   $32.0\substack{17.0\\10.0}$  \\
J103816+452718m &   $6.9\substack{36.0\\3.1}$ &  $1.6\substack{45.0\\0.86}$ &    $0.53\substack{0.24\\0.33}$ &  $93.0\substack{31.0\\51.0}$ & $670.0\substack{210.0\\300.0}$ &  $0.94\substack{0.51\\0.24}$ &        $2.1\substack{2.4\\2.0}$ &  $1.9\substack{1.4\\0.21}$ &   $1.6\substack{2.1\\0.27}$ & $670.0\substack{230.0\\250.0}$ &   $1.9\substack{3.1\\0.28}$ & $0.85\substack{2.7\\0.044}$ &  $75.0\substack{130.0\\53.0}$  \\
J112933+493525m & $57.0\substack{16.0\\18.0}$ &   $9.7\substack{22.0\\5.1}$ &   $0.86\substack{0.063\\0.26}$ &  $140.0\substack{2.9\\42.0}$ & $770.0\substack{220.0\\110.0}$ &   $1.5\substack{0.26\\0.52}$ & $0.0067\substack{0.14\\0.0025}$ &  $2.2\substack{1.2\\0.38}$ & $0.43\substack{0.51\\0.33}$ & $660.0\substack{260.0\\200.0}$ &    $5.6\substack{2.1\\1.1}$ &   $2.7\substack{1.5\\0.46}$ &   $55.0\substack{17.0\\21.0}$  \\
J113304+651341m & -- & -- & -- & -- & -- & -- & -- & -- & -- & --  &   $12.0\substack{0.0\\2.2}$ &   $6.7\substack{0.49\\1.3}$ &    $22.0\substack{5.0\\0.95}$  \\
J115855+312559m & -- & -- & -- & -- & -- & -- & -- & -- & -- & -- &    $5.0\substack{3.0\\1.3}$ &   $2.1\substack{1.7\\0.44}$ &    $35.0\substack{9.1\\16.0}$  \\
J123519+063556m & -- & -- & -- & -- & -- & -- & -- & -- & -- & --  &  $11.0\substack{0.69\\6.0}$ &    $5.0\substack{2.5\\3.0}$ &     $7.5\substack{11.0\\1.0}$  \\
J124619+444902m & $56.0\substack{18.0\\36.0}$ &  $80.0\substack{4.7\\25.0}$ &   $0.92\substack{0.065\\0.56}$ &    $5.7\substack{58.0\\1.8}$ &  $230.0\substack{380.0\\16.0}$ &   $1.1\substack{0.41\\0.32}$ &        $1.8\substack{2.5\\1.7}$ &  $1.7\substack{1.9\\0.13}$ &   $1.1\substack{2.6\\0.24}$ &  $340.0\substack{280.0\\68.0}$ & $12.0\substack{0.13\\0.88}$ &   $6.3\substack{1.0\\0.35}$ &    $36.0\substack{9.7\\33.0}$ \\
J130128+510451m &  $10.0\substack{30.0\\5.4}$ &   $8.6\substack{23.0\\4.7}$ &    $0.74\substack{0.15\\0.42}$ & $100.0\substack{20.0\\56.0}$ & $610.0\substack{210.0\\200.0}$ &    $1.0\substack{0.47\\0.2}$ &        $4.0\substack{2.4\\3.9}$ &  $2.3\substack{1.0\\0.26}$ &  $1.1\substack{0.76\\0.31}$ & $460.0\substack{310.0\\160.0}$ &   $1.7\substack{4.2\\0.37}$ &  $0.75\substack{3.5\\0.14}$ & $110.0\substack{180.0\\91.0}$\\
J131419+104739m &  $46.0\substack{25.0\\9.2}$ & $28.0\substack{15.0\\14.0}$ &  $0.77\substack{0.088\\0.069}$ & $100.0\substack{25.0\\33.0}$ &   $700.0\substack{50.0\\33.0}$ &  $0.73\substack{0.59\\0.12}$ &        $2.3\substack{2.9\\1.9}$ &  $2.7\substack{0.97\\0.6}$ &    $3.3\substack{1.8\\1.1}$ & $700.0\substack{230.0\\240.0}$ &   $3.1\substack{4.3\\0.87}$ &   $2.1\substack{3.0\\0.48}$ &   $42.0\substack{19.0\\27.0}$ \\
J132633+421824m & $51.0\substack{18.0\\25.0}$ &  $18.0\substack{29.0\\8.6}$ &    $0.78\substack{0.11\\0.36}$ &  $65.0\substack{39.0\\29.0}$ & $840.0\substack{170.0\\370.0}$ &   $1.6\substack{0.19\\0.55}$ &   $0.063\substack{0.68\\0.049}$ &  $5.1\substack{0.16\\1.9}$ &   $1.7\substack{0.88\\1.3}$ & $540.0\substack{300.0\\200.0}$ &    $7.3\substack{2.3\\1.9}$ &   $4.2\substack{1.6\\0.94}$ &   $40.0\substack{16.0\\13.0}$\\
J134559+112848m &  $69.0\substack{8.6\\22.0}$ &  $11.0\substack{19.0\\5.6}$ &  $0.97\substack{0.017\\0.055}$ &  $63.0\substack{20.0\\34.0}$ &   $530.0\substack{24.0\\16.0}$ &   $1.0\substack{0.42\\0.32}$ &      $0.37\substack{3.1\\0.24}$ & $1.6\substack{0.76\\0.22}$ &   $1.9\substack{1.8\\0.51}$ &  $250.0\substack{440.0\\26.0}$ &    $1.7\substack{1.2\\0.3}$ &  $1.1\substack{0.99\\0.16}$ &  $100.0\substack{22.0\\38.0}$ \\
J141013+434435m & -- & -- & -- & -- & -- & -- & -- & -- & -- & --  &   $11.0\substack{0.3\\4.7}$ &    $7.2\substack{2.6\\2.8}$ &    $25.0\substack{18.0\\3.7}$ \\
J003601+003307x &  $33.0\substack{26.0\\4.6}$ &   $5.2\substack{17.0\\1.7}$ &    $0.96\substack{0.02\\0.11}$ &   $2.7\substack{19.0\\0.32}$ & $700.0\substack{240.0\\280.0}$ &    $1.6\substack{0.19\\0.6}$ &        $7.0\substack{1.4\\6.3}$ & $3.0\substack{0.67\\0.67}$ &  $0.2\substack{0.33\\0.27}$ & $>700.0$ &    $5.3\substack{1.7\\2.4}$ &    $3.0\substack{1.7\\1.3}$ &   $58.0\substack{26.0\\13.0}$  \\
J004743+015440x & $51.0\substack{22.0\\13.0}$ &  $36.0\substack{5.7\\22.0}$ &     $0.46\substack{0.1\\0.04}$ &   $12.0\substack{21.0\\2.6}$ &   $560.0\substack{90.0\\31.0}$ &  $0.62\substack{0.57\\0.07}$ &        $1.6\substack{3.0\\1.0}$ &  $1.5\substack{1.2\\0.24}$ &  $0.53\substack{1.2\\0.16}$ & $>560.0$ &  $2.4\substack{0.21\\0.31}$ &  $1.6\substack{0.21\\0.23}$ &    $84.0\substack{11.0\\7.4}$  \\
J011309+000223x &  $80.0\substack{5.2\\18.0}$ &  $69.0\substack{6.9\\47.0}$ &    $0.82\substack{0.07\\0.15}$ &    $7.8\substack{26.0\\1.2}$ & $750.0\substack{190.0\\390.0}$ &   $1.5\substack{0.22\\0.46}$ &       $0.4\substack{1.3\\0.32}$ &  $2.5\substack{1.0\\0.37}$ &   $1.5\substack{1.2\\0.48}$ & $>750.0$ &  $2.7\substack{0.85\\0.54}$ &  $1.4\substack{0.56\\0.15}$ &    $86.0\substack{11.0\\8.8}$  \\
J012217+052044x &  $50.0\substack{9.7\\11.0}$ & $18.0\substack{14.0\\11.0}$ &    $0.94\substack{0.04\\0.07}$ &    $6.2\substack{12.0\\1.7}$ & $750.0\substack{170.0\\270.0}$ &    $1.1\substack{0.4\\0.29}$ &      $0.56\substack{1.4\\0.42}$ & $2.2\substack{0.83\\0.38}$ & $0.66\substack{0.63\\0.29}$ & $>750.0$ &  $5.1\substack{0.45\\0.99}$ &   $3.1\substack{1.1\\0.61}$ &    $65.0\substack{11.0\\4.0}$ \\
J081409+211459x &  $29.0\substack{30.0\\7.4}$ & $29.0\substack{16.0\\12.0}$ &     $0.67\substack{0.18\\0.3}$ &   $47.0\substack{51.0\\9.9}$ &  $470.0\substack{160.0\\34.0}$ &   $2.0\substack{0.01\\0.52}$ &        $2.4\substack{2.6\\2.3}$ & $4.1\substack{0.49\\0.82}$ &   $6.1\substack{0.97\\1.6}$ &  $350.0\substack{330.0\\91.0}$ &    $9.4\substack{1.3\\5.4}$ &    $3.8\substack{1.8\\2.0}$ &     $3.5\substack{5.0\\0.79}$\\
J091113+183108x & $57.0\substack{14.0\\36.0}$ & $42.0\substack{22.0\\21.0}$ &     $0.31\substack{0.3\\0.11}$ &  $140.0\substack{3.1\\62.0}$ & $700.0\substack{200.0\\170.0}$ &  $0.85\substack{0.53\\0.15}$ &     $0.01\substack{0.39\\0.01}$ &   $1.8\substack{1.4\\0.2}$ &   $1.8\substack{2.2\\0.31}$ & $530.0\substack{250.0\\200.0}$ &  $2.0\substack{0.94\\0.21}$ &  $1.2\substack{0.77\\0.14}$ &   $68.0\substack{26.0\\26.0}$\\
J095838+202508x & $52.0\substack{20.0\\25.0}$ & $42.0\substack{25.0\\20.0}$ &    $0.97\substack{0.02\\0.47}$ &  $28.0\substack{45.0\\12.0}$ &  $320.0\substack{320.0\\57.0}$ &   $1.2\substack{0.37\\0.32}$ &      $0.49\substack{1.4\\0.47}$ &  $4.1\substack{0.45\\1.1}$ &    $3.5\substack{1.0\\1.5}$ & $>320.0$ &    $5.1\substack{3.2\\1.5}$ &   $2.1\substack{2.4\\0.59}$ &   $44.0\substack{18.0\\24.0}$ \\
\hline
\end{tabular} $}   
\justifying
{\\ \\ Column definitions are listed in Table~\ref{tab:free_parameters}.  Only ISM constraints are provided for galaxies without outflows.  The \emph{m} and \emph{x} in the Galaxy ID refer to data observed with MMT and X-Shooter, respectively.}     
\label{tab:best_fits_MgII}
\end{table*}

\subsection{Model Fitting}
We use the same Bayesian based fitting procedure as \cite{Carr2021a} to fit the SALT model to the Mg II 2796\AA, 2804\AA\ spectral lines of each galaxy.  The method utilizes the Python code \texttt{emcee} \citep{Foreman-Mackey2013}, which implements Goodman and Weare’s Affine Invariant Markov Chain Monte Carlo (MCMC) Ensemble sampler in Python \citep{Goodman2010}.  We sample a Gaussian likelihood function assuming uniform distributions for the priors and initial conditions.  The exact parameter ranges have been specified in Table~\ref{tab:free_parameters}.  Each model was smoothed to the appropriate instrumental resolution using a convolution with SciPy's Gaussian filter before fitting.  

To ensure convergence of the posteriors, we explored each parameter space with twice as many walkers as free parameters (i.e., 24 walkers) for 10000 steps.  We removed the first 4000 steps or so called “burn-in period” from our analysis to ensure our estimates of the uncertainties reflect only the converged portion of the chains.  

We utilized two statistics to define the “best-fit”  For dependent quantities, e.g., $\gamma$, $\delta$, etc., we use the maximum likelihood estimate (MLE) from the parameter chains.  For dependent quantities, e.g., mass outflow rates, column densities, etc., we assemble histograms from the parameter chains and take the median of the resulting distribution to define the best fit \cite{Huberty2024}.  Carr et al. (in prep) found the latter to be a better choice in tests aimed at recovering the column densities and mass outflow rates from simulations. 

Error bars were drawn from the posterior distributions or histograms.  When considering the median, the lower (upper) error bars refer to the 16(84) percentiles.  In regards to the MLE, the lower (upper) error bars represent the median value of the deviation from below (above) the MLE value \citep{Carr2021a}.  The errors represent how well we can constrain the SALT parameter space and represent a measure of degeneracy.             

\begin{figure*}
	\centering
	\includegraphics[width=\textwidth]{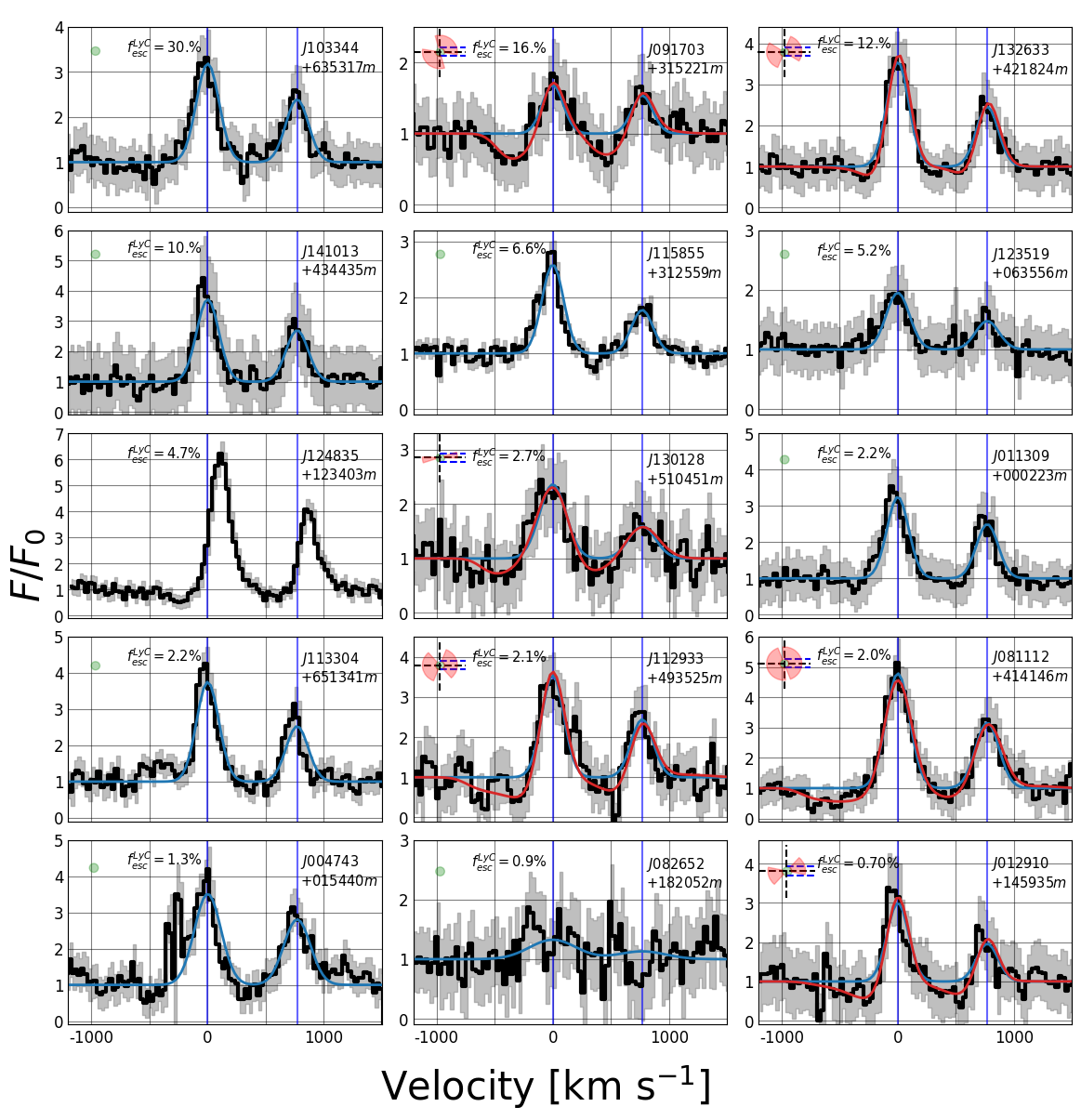}
	\caption{shows the Mg II 2796\AA, 2804\AA\ spectra lines observed with the MMT blue spectrograph ($\rm Res \sim 90 \rm \ km \ s^{-1}$).  The SALT models fits are shown in red, the nebular emission component in blue, the spectra in black, and the errors on the data are shaded in grey.   Galaxies are ordered from left to right, top to bottom with decreasing $f_{esc}^{LyC}$.  The emblems in the upper left corners represent the constrained outflow geometries and are to be viewed from the right. Isolated green circles represent objects which did not register an outflow.  Galaxy $\rm J124835+123403$ was not fit, as it shows highly redshifted emission lines which lie outside the SALT domain.}
	\label{fig:MgII_fits_MMT}
\end{figure*} 

\begin{figure*}
\addtocounter{figure}{-1}
	\centering
	\includegraphics[width=\textwidth]{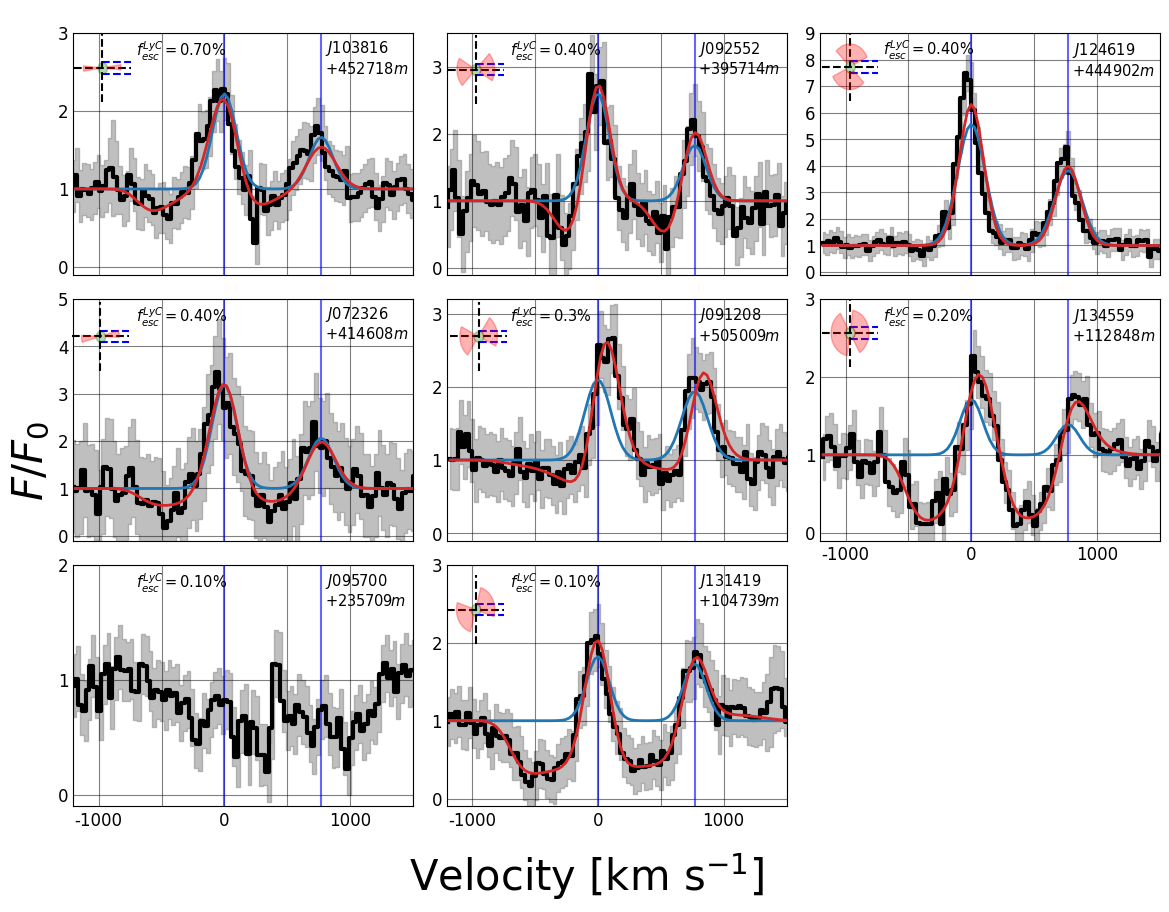}
	\caption{\emph{continued}.  Galaxy J095700+235709 was not fit because it shows significant redshifted absorption, characteristic of galactic inflows. }
\end{figure*}

\begin{figure*}
	\centering
	\includegraphics[width=\textwidth]{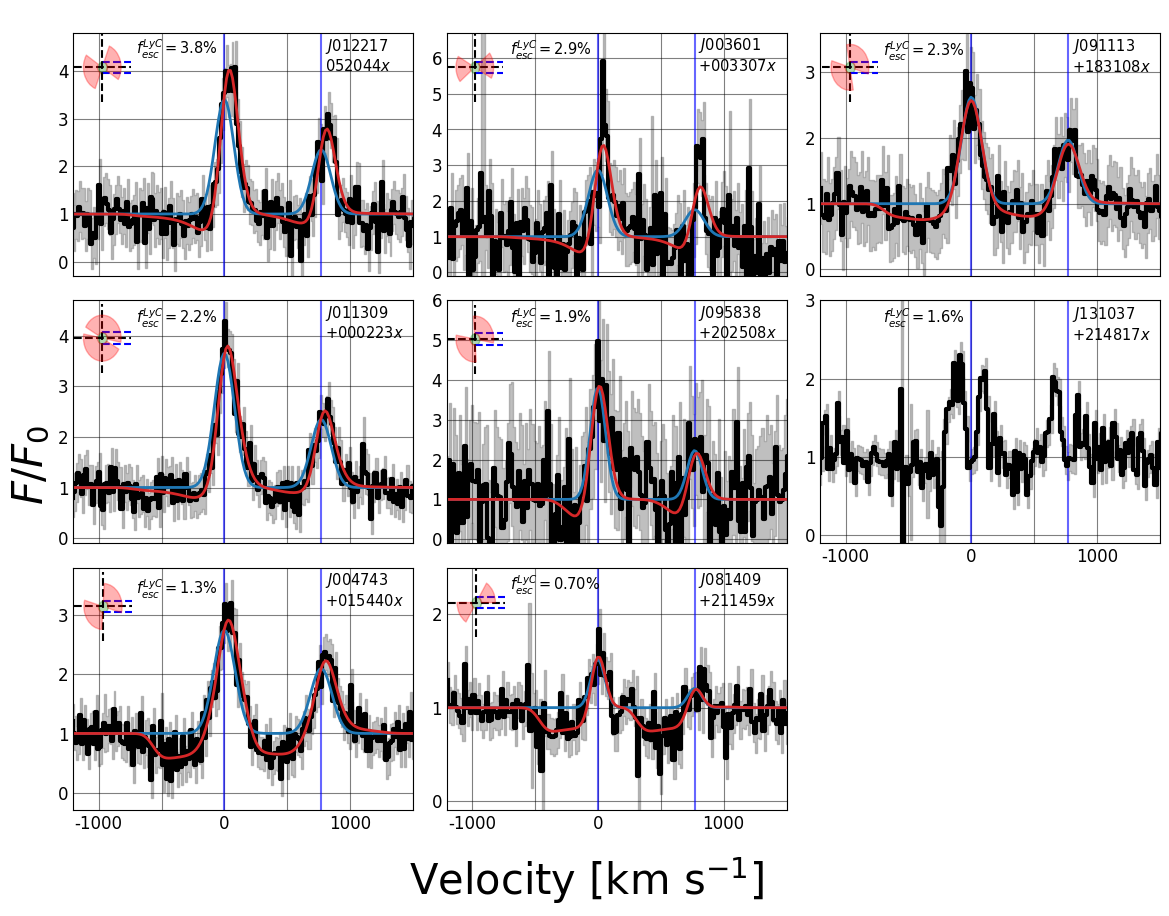}
	\caption{shows the Mg II 2796\AA, 2804\AA\ spectra lines observed with the VLT X-Shooter spectrogprah ($\rm Res \sim 55 \rm \ km \ s^{-1}$).  The overall presentation is the same as in Figure~\ref{fig:MgII_fits_MMT}.  Galaxy $\rm J131037+214817$ was not fit, as it shows double peaked emission lines which are currently not included in the SALT model domain.}
	\label{fig:MgII_fits_Xshooter}
\end{figure*} 

\subsection{Limitations}

By fitting SALT to mock spectra derived from its own parameter space, \cite{Carr2018,Carr2023} were able to study its degeneracy.  A few cases are worth mentioning.

A limiting observing aperture, a small opening angle, and a large amount of dust can all act to reduce the emission profile, while causing negligible changes to the absorption profile.  Although our constraints on $\kappa$ suggest that dust extinction is not a concern, the other factors—particularly the aperture and opening angle—remain unconstrained and may bias SALT towards preferring bi-conical outflow geometries aligned with the line of sight.  Further complicating this matter is the intrinsic emission emanating from the H II regions which may be indistinguishable from the scattered re-emission occurring in the outflow.  For these reasons, we avoid commenting on parameters that depend solely on outflow geometry, such as the global escape fraction, and instead focus on line-of-sight-dependent quantities, like the observed escape fraction, which are more influenced by the absorption component of the line profile.  We suspect this degeneracy to impact the outflow rates by only a factor of a few and at most $4\pi$.     

In certain instances, the parameter $v_{\infty}$ can go unconstrained from above.  This occurs when the density field decays to an undetectable level before the velocity field reaches its terminal value (i.e., $v_{\infty}$).  For this reason, we find it convenient to replace $v_{\infty}$ with $v_{90}$, or the value of $v$ once the EW reaches 90\% of its total value, in our analysis.  

All sources of degeneracy are accounted for in the \texttt{emcee} fitting process and are reflected in our error bars.

\cite{Carr2023} and Carr et al., in prep. also discovered systematic biases in the recovery of certain quantities (most notably the mass outflow rates and column densities) when testing SALT on mock spectra drawn from simulations of turbulent flows.  When assuming a uniform turbulence parameter of $v_{b} = 10\ \rm km\ s^{-1}$, \cite{Carr2023} found that SALT roughly overestimated the mass outflow rates and column densities by a factor of 1.38 and 1.65, respectively, in idealized outflows drawn from a column density range, $13<\log{N}\ [\rm cm^{-2}] <18$.  This value of $v_b$ is consistent with the empirical-based methods of \cite{Chen2023} to measure the non-thermal velocity dispersion of cool clouds tracing the CGM in redshift $z < 1$ galaxies.  Thus we suspect our values to also be biased.  This should not have much impact on our analysis regarding $f_{esc}^{LyC}$ since most of our $\rm H^0$ column densities are well in excess of $10^{17}\rm\ cm^{-2}$ (i.e., are very optically thick).  This caveat must be considered for the outflow rates, however.

\subsection{Identifying Outflows}
To identify the presence of an outflow, we conduct an F-test.  We consider two models: a pure ISM component where the line profile is determined entirely by the two Gaussian profiles and, alternatively, the SALT model.  Note that the ISM model is “nested” in the SALT model (i.e., if $\tau=0$, we recover only the ISM component).  We compute the F-value, $F$, as 
\begin{eqnarray}
    F = \frac{\chi_{\rm ISM}^2-\chi_{\rm outflow}^2}{\chi_{\rm outflow}^2}\frac{n-p_{\rm SALT}}{p_{\rm SALT}-p_{\rm ISM}},
\end{eqnarray}
where $\chi_{i}^2$ is the chi-squared value for the appropriate model, $p_{\rm SALT}=12$ and $p_{\rm ISM}=3$ are the number of free parameters for the SALT and ISM models, respectively, and $n$ is the number of observed velocity bins used to calculate the line profile.  All galaxies with $F>1.7$ ($\rm significance\ level = 0.05$) were said to have outflows.  

We find that 20 out of a possible 26 galaxies have outflows.  Those without outflows include J0047+0154, J0113+00020, J0826+1820, J1033+63530, J1133+45140, J1158+31250, J1235+0635, and J1410+43450 from the MMT sample and J131037+214817x from the X-Shooter sample.  We note that galaxies J0047+0154 and J0113+00020, which were observed with both spectrographs, did not register an outflow in the MMT sample.  This is likely due to the $\sim 2$ increase in resolution between the different instruments.  

We have chosen to remove galaxies J0957+2357, J124835+123403, and J131037+214817 from further analysis, each for separate reasons as we now describe.  We removed J0957+2357 because its spectrum shows strong redshifted absorption which is an indication of galactic inflows \citep{Carr2022a}.  Carr et al., in prep, showed that a failure to model both inflows and outflows in spectral lines can lead to severe biases in SALT parameter estimates.  We removed J131037+214817 because its spectrum shows double peaked emission lines.  We do not account for this feature in the SALT domain.  Lastly, we removed J124835+123403 because we were not able to achieve a satisfactory fit to its spectrum.  We suspect that the fitting procedure failed because the emission lines show a relatively large red-shift from line center, and SALT is unable to reproduce this feature at such high levels of emission.  Since this is the only galaxy with this feature, we have decided to exclude it and suspect the feature could be the result of a poor redshift estimate.  Alternatively, such a feature could be the result of high velocity turbulence.  Since SALT does not account for turbulent line broadening \citep{Carr2023}, this alternative explanation further justifies the omission.     

Figures~\ref{fig:MgII_fits_MMT} and \ref{fig:MgII_fits_Xshooter} display normalized spectral cutouts of the Mg II 2798\AA\ and 2803\AA\ doublet, shown in black, alongside the model fits, shown in red, for the MMT ($\text{res} \approx 90\ \text{km}\ \text{s}^{-1}$) and X-Shooter ($\text{res} \approx 55\ \text{km}\ \text{s}^{-1}$) data, respectively. In addition, the pure ISM fit is shown in light blue.  The panels are arranged from left to right, top to bottom, starting with the strongest LyC emitters as determined by $f_{\text{esc}}^{\text{LyC}}$.  In some instances, e.g., galaxy J072326+414608, the ISM emission cannot be distinguished from the emission component of the outflow, signaling a likely degeneracy.  We discuss this point in the limitations sub section in Section~\ref{sec:model}.  Overall, howerver, our galaxies display a wide ragne of opneing angles and orientations, suggesting a large degree of anisotropy in the cool clouds traced by Mg II. 

We provide the best fit SALT parameters in Table~\ref{tab:best_fits_MgII}.  For galaxies without outflows, only the ISM parameters are provided.  




\section{Modeling LyC Escape} \label{sec:modeling_lyc_escape}

Our ability to predict $f_{\text{esc}}^{\text{LyC}}$ depends on the assumption that low-ionization state (LIS) metals trace $\text{H}^0$. This relationship was empirically confirmed by \cite{Saldana-Lopez2022} for LzLCS galaxies, who found a linear correlation:
\begin{equation}
C_{f,\text{H}^0} = (0.63 \pm 0.19)C_{f,\text{LIS}} + (0.54 \pm 0.09),
\label{eq:Saldana-Lopez}
\end{equation}
between the covering fraction of neutral hydrogen, $C_{f,\text{H}^0}$, and the covering fraction of the LIS gas, $C_{f,\text{LIS}}$, which in their study was inferred from Si II. Other studies, such as \cite{Reddy2016,Gazagnes2018,Reddy2022}, also support the validity of this relation in star-forming galaxies. Physically, the relation stems from the fact that $\text{H}^0$ (13.6 eV) and $\text{Si}^+$ (16.3 eV) have similar ionization potentials.  Given that the ionization potential of $\text{Mg}^+$ is 15.0 eV, we extend this relation to $\text{Mg II}$ as well \citep{Xu2023}.

We note, however, that Equation~\ref{eq:Saldana-Lopez} was derived using low-resolution data (COS G140L, $\sim 250\ \rm km\ s^{-1}$) and may therefore be subject to systematic uncertainties (Jennings et al., in prep). While we continue to use this equation, even if primarily as a proof of concept, future higher-resolution studies (e.g., \citealt{Carr2023HST}) will be necessary for further validation. Reassuringly, Equation~\ref{eq:Saldana-Lopez} predicts that most of our galaxies have  $C_{f,\rm H^0}$ values close to unity, which aligns with direct medium-resolution observations in similar galaxies \citep{Henry2015}.

 \begin{figure}
	\centering
	\includegraphics[width=\columnwidth]{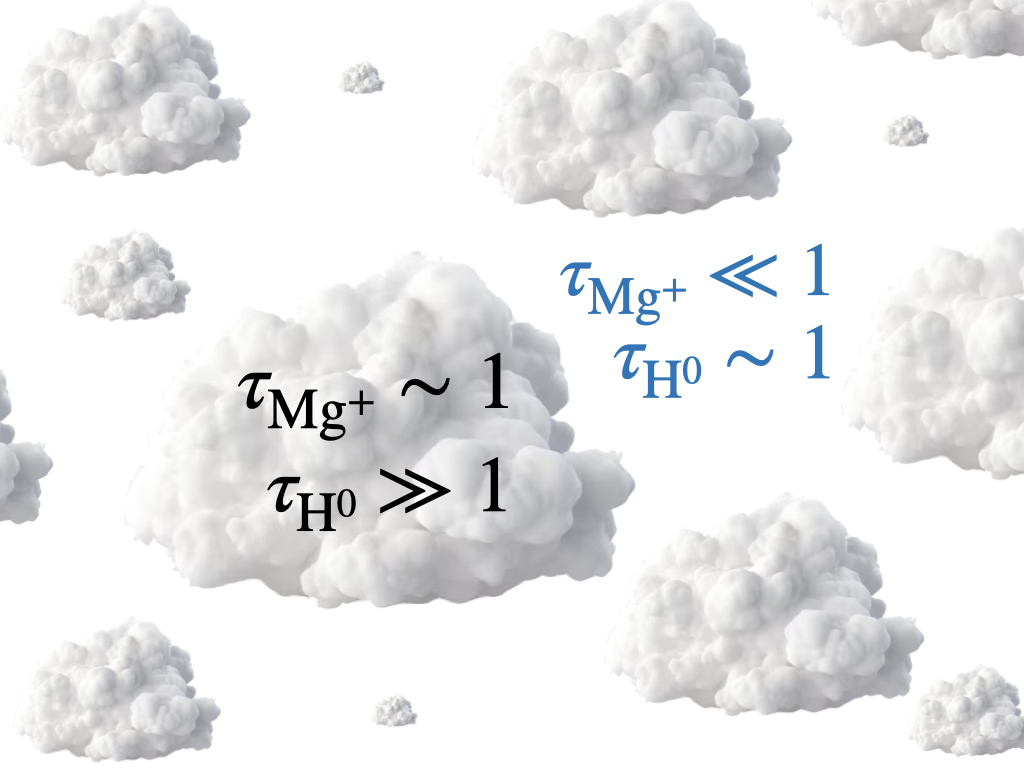}
	\caption{illustrates a multi-phase CGM consisting of cool, dense clouds of $\rm Mg^+$ and $\rm H^0$, surrounded by a more highly ionized ambient medium. Mg II photons only probe the cool clouds in absorption, where the optical depth $\tau_{\rm Mg^+}$ is significant or on the order of unity, while it is ineffective in the ambient medium where $\tau_{\rm Mg^+}$ is much less than one. Conversely, LyC radiation is attenuated both by the cool clouds, where $\tau_{\rm H^0}$ is much greater than one, and by the ambient medium, where $\tau_{\rm H^0}$ is still significant or on the order of unity. This difference in attenuation reflects the higher abundance of $\rm H^0$ compared to $\rm Mg^+$. }
 \label{fig:clouds}
\end{figure} 

Equation~\ref{eq:Saldana-Lopez} demonstrates that while LIS metals trace $\text{H}^0$, they are not evenly distributed: $C_{f,\text{H}^0}$ is systematically larger than $C_{f,\text{LIS}}$ \citep{Mauerhofer2021}.  We hypothesize that this is the case, in large part, because $\text{Mg}^+$ (or any LIS metal tracer in general) is only sensitive to the densest regions of $\text{H}^0$ in absorption.  The underlying geometry is illustrated in Figure~\ref{fig:clouds}.  This assertion is supported by the work of \cite{Chang2024} who compared simulations of Mg II and Ly$\alpha$ radiation transfer in idealized scattering mediums assuming realistic abundance ratios of $\rm H^0$ to $\rm Mg^+$.  If our interpretation is correct, then we can use the SALT constraints on $\text{Mg}^+$ to map the higher density channels of $\text{H}^0$ in the LzLCS galaxies (i.e., the optically thick lines of sight to LyC escape). From this information, we can place an upper estimate on the LyC escape fraction - that is, $f_{esc}^{LyC} \leq 1.0 - C_{f,\rm Mg^+}+C_{f,\rm Mg^+}e^{-\tau_{\rm thick}}$, where $\tau_{\rm thick}$ represents the optical depth of channels optically thick to LyC escape in $\rm H^0$.       



We use photoionization modeling to estimate the abundance ratio of $\rm H^0$ to $\rm Mg^+$, $\alpha_{\rm H^0/Mg^+}$, in each galaxy with an outflow detected in Mg II.  The calculations were conducted using \texttt{Cloudy version c23.01} \citep{Chatzikos2023}, while incorporating the outflow parameters derived from SALT.  Specifically, each outflow was modeled as an expanding spherical cloud with an inner radius, $r = R_{\rm SF}$, and an outer radius, $r = R_{\rm W}$.  We set the outflow porosity to $f_c$ and the turbulence to $10\ \rm km\ s^{-1}$.  The hydrogen distribution was assumed to follow $\rm Mg^+$ as a power law with index $\delta$.  Additionally, we adopted the gas metallicity values from \cite{Flury2022b}. 

The sources are characterized by spectral energy distributions (SEDs) from the Binary Population and Stellar Synthesis (BPASS) models (version 2; \citealt{Stanway2018}). Specifically, we use the \texttt{bin-imf135\_100} model with binary stars, which assumes continuous star formation and a Salpeter initial mass function (IMF; \citealt{Salpeter1955}) with lower and upper mass cutoffs of 0.1 and 100 $\text{M}_{\odot}$, respectively, and a slope $\alpha_s = -2.35$ \citep{Xu2022a}. The age and stellar metallicity were sourced from \cite{Saldana-Lopez2022}. Additionally, we include the \texttt{Cosmic rays background [1.2]} module, which assumes a mean $\rm H^0$ cosmic ray ionization rate of $2 \times 10^{-16}\ \text{s}^{-1}$ \citep{Indriolo2007}.

Following \cite{Xu2022b}, we constructed grids of Cloudy models varying the ionization parameter $U$ over a range of [-4,0] in steps of 0.25, and hydrogen densities over a range of [0,3] in steps of 0.1 dex in $\log{n_0}\ [\text{g}\ \text{cm}^{-3}]$. We then used a standard chi-squared minimization technique to determine the abundance ratio, $\alpha_{\rm H^0/Mg^+} = N_{\rm H^0}/N_{\rm Mg^+}$ \citep{Xu2022b}, by searching the grid for matching $\rm O_{32}$ and $N_{\rm Mg^+}$ values.  We used the $\rm O_{32}$ values measured by \cite{Flury2022a} and the SALT $N_{\rm Mg^+}$ estimates.

\begin{table*}
\caption{Properties Used to Compute LyC Escape Fractions}
\begin{tabular}{ccccccc}
\hline\hline
Galaxy&$\log{N_{\rm Mg^+}}$ & $C_{f,\rm Mg^+}$ & $f_{esc}^{LyC}(\rm Mg\ II)$ & $\log{N_{\rm H^0}}$ & $C_{f,\rm H^0}$ & $f_{esc}^{LyC}(\rm Mg\ II)\ w/\rm corr.$ \\[.5 ex]
&$M_{\odot}\ yr^{-1}$ &$\%$ &$\%$&$M_{\odot}\ yr^{-1}$&$\%$ &$\%$\\[.5 ex]
\hline
J012910+145935m &   $14.8\substack{1.41\\1.22}$ &  $67.7\substack{14.7\\29.5}$ &     $4.96\substack{44.8\\1.83}$ &   $19.1\substack{1.46\\1.24}$ & $96.7\substack{3.35\\22.2}$ &     $1.92\substack{43.9\\1.18}$ \\
J072326+414608m &   $14.0\substack{1.83\\1.62}$ &  $84.4\substack{7.32\\43.4}$ &      $5.95\substack{53.7\\2.2}$ &   $17.8\substack{2.15\\1.65}$ &  $99.0\substack{1.0\\25.3}$ &     $4.75\substack{49.3\\1.92}$ \\
J081112+414146m & $15.1\substack{0.897\\0.929}$ &  $59.9\substack{13.4\\20.5}$ &     $4.62\substack{41.9\\1.36}$ &  $19.3\substack{0.976\\1.06}$ & $91.8\substack{8.25\\15.7}$ &     $2.11\substack{41.7\\1.22}$ \\
J091208+505009m &   $15.3\substack{0.8\\0.572}$ & $99.1\substack{0.434\\10.4}$ &   $0.14\substack{10.7\\0.0921}$ &  $19.2\substack{0.911\\0.64}$ &  $99.0\substack{1.0\\10.8}$ & $0.0443\substack{10.7\\0.0333}$ \\
J091703+315221m &   $15.1\substack{1.17\\1.16}$ &  $53.3\substack{16.7\\15.0}$ &     $18.8\substack{36.8\\5.21}$ &   $19.1\substack{1.28\\1.19}$ & $87.5\substack{12.5\\13.3}$ &     $9.48\substack{42.6\\3.42}$ \\
J092552+395714m &    $15.0\substack{1.2\\0.95}$ &  $79.2\substack{10.6\\23.3}$ &    $2.58\substack{34.1\\0.917}$ &   $19.3\substack{1.27\\1.02}$ &  $99.0\substack{1.0\\16.6}$ &   $0.816\substack{33.7\\0.488}$ \\
J103816+452718m &   $13.6\substack{1.73\\1.66}$ &  $52.6\substack{23.8\\32.8}$ &      $4.91\substack{71.3\\0.0}$ &   $17.9\substack{1.81\\1.67}$ & $87.1\substack{12.9\\21.1}$ &    $2.86\substack{69.1\\0.504}$ \\
J112933+493525m &  $15.0\substack{1.02\\0.673}$ &   $86.0\substack{6.4\\25.9}$ &     $9.04\substack{31.3\\4.34}$ &  $19.1\substack{1.07\\0.741}$ &  $99.0\substack{1.0\\15.6}$ &     $2.86\substack{30.7\\1.42}$ \\
J124619+444902m &   $13.1\substack{2.34\\1.61}$ &  $92.1\substack{6.55\\55.1}$ &     $6.34\substack{57.8\\4.97}$ &   $16.7\substack{2.47\\1.64}$ &  $99.0\substack{1.0\\26.4}$ &     $5.85\substack{53.4\\1.37}$ \\
J130128+510451m &   $13.9\substack{1.29\\1.43}$ &  $73.6\substack{15.2\\42.2}$ &     $20.2\substack{44.6\\19.7}$ &   $18.8\substack{1.52\\1.52}$ &  $99.0\substack{1.0\\26.4}$ &     $14.5\substack{36.1\\5.08}$ \\
J131419+104739m & $16.1\substack{0.684\\0.732}$ &  $76.7\substack{8.56\\6.81}$ &    $1.07\substack{20.0\\0.544}$ &  $20.5\substack{0.76\\0.883}$ &  $99.0\substack{1.0\\11.2}$ &    $0.34\substack{19.8\\0.117}$ \\
J132633+421824m &   $14.3\substack{1.41\\1.33}$ &  $78.0\substack{10.5\\36.6}$ &     $11.4\substack{49.5\\4.87}$ &   $18.5\substack{1.45\\1.37}$ &  $99.0\substack{1.0\\24.0}$ &     $3.63\substack{56.5\\1.76}$ \\
J134559+112848m & $16.3\substack{0.459\\0.675}$ &  $97.2\substack{1.64\\5.72}$ & $0.0778\substack{5.06\\0.0398}$ & $20.3\substack{0.621\\0.758}$ &  $99.0\substack{1.0\\11.5}$ & $0.0247\substack{4.96\\0.0164}$ \\
J003601+003307x &   $15.4\substack{0.782\\0.7}$ &  $96.3\substack{1.99\\11.5}$ &     $3.18\substack{12.0\\1.61}$ &  $19.3\substack{1.02\\0.887}$ &  $99.0\substack{1.0\\11.6}$ &    $1.01\substack{12.1\\0.532}$ \\
J004743+015440x & $15.8\substack{0.581\\0.553}$ &  $46.3\substack{9.52\\3.24}$ &      $11.0\substack{35.2\\0.0}$ & $20.3\substack{0.718\\0.621}$ & $83.2\substack{12.7\\7.94}$ &     $5.83\substack{30.6\\1.97}$ \\
J011309+000223x & $14.7\substack{0.866\\0.746}$ &  $81.5\substack{6.94\\14.1}$ &     $3.68\substack{23.4\\1.68}$ & $19.3\substack{0.946\\0.927}$ &  $99.0\substack{1.0\\14.2}$ &    $1.17\substack{24.2\\0.604}$ \\
J012217+052044x & $15.1\substack{0.705\\0.693}$ &   $93.6\substack{3.6\\7.31}$ &     $2.64\substack{9.56\\1.31}$ &  $19.2\substack{1.02\\0.711}$ &  $99.0\substack{1.0\\11.4}$ &    $0.836\substack{9.7\\0.401}$ \\
J081409+211459x &  $14.8\substack{1.05\\0.451}$ &  $66.9\substack{18.1\\30.0}$ &    $3.26\substack{51.6\\0.182}$ &  $19.1\substack{1.28\\0.565}$ & $96.2\substack{3.84\\19.4}$ &     $2.34\substack{50.6\\1.01}$ \\
J091113+183108x &   $14.4\substack{1.22\\1.26}$ &  $30.6\substack{28.7\\10.8}$ &    $4.15\substack{55.7\\0.903}$ &   $19.3\substack{1.29\\1.46}$ & $73.3\substack{17.6\\9.52}$ &      $2.4\substack{55.5\\1.03}$ \\
J095838+202508x &   $14.0\substack{1.23\\1.66}$ &  $96.5\substack{1.91\\46.1}$ &     $2.48\substack{49.5\\1.02}$ &   $17.7\substack{1.51\\1.66}$ &  $99.0\substack{1.0\\23.2}$ &     $2.09\substack{49.3\\1.11}$ \\
\hline
\end{tabular} 
\justifying
{\\ \\ From left to right: (1) Object (2) column density of ionized magnesium  (3) covering fraction of ionized magnesium (4) predicted LyC escape fraction from Mg II emission only (5) column density of neutral hydrogen (6) covering fraction of neutral hydrogen from Equation~\ref{eq:Saldana-Lopez} (7) LyC escape fraction with correction for optically thin gas.  Only galaxies with outflows (i.e., available data) are shown.}
\label{tab:lyc}
\end{table*} 

To estimate $f_{esc}^{LyC}$ from the inferred distribution of $\rm H^0$, we adapt our method for computing the column density, $N$, in Equation~\ref{eq:average_N}.  Assuming a porosity $f_c = 1$, the fraction of photons escaping through the bi-cone is given by:

\begin{eqnarray}\label{eq:fesc_fc=1}
    f_{esc}^{LyC}|_{f_c=1} = \sum_i^I \frac{e^{-\sigma_{\rm UV} N_{\rm{H^0},i}}}{I} \times 10^{-0.4 \rm{E(B-V)}_{\rm{UV}}k},
\end{eqnarray}
where $\sigma_{\text{UV}} = 6.3 \times 10^{-18} \, \text{cm}^2$ is the photoionization crosssection for $\text{H}^0$, $N_{\rm{H^0},i} = \alpha_{\rm H^0/Mg^+}N_{\rm{Mg}^+,i}$ is the column density of neutral hydrogen along the $i$'th line of sight, $\rm E(B-V)_{UV}$ represents the UV dust extinction, and $k(912) = 12.87$ is derived from the FUV dust attenuation law of \cite{Reddy2016}. The sum is taken over $I$ randomly sampled sight lines. We find that our results converge to two significant figures by $I = 10^4$, consistent with the column density calculations.  

In the case of a spatially constant column density, such as in a uniform slab, Equation~\ref{eq:f_c_1} simplifies to:
\begin{eqnarray}
f_{\text{esc}}^{\text{LyC}}|_{f_c=1}^{\text{Uniform}} &=& \left(C_{f,\text{Mg}^+} e^{-\sigma_{\text{UV}} N_{\text{H}^0}} + 1 - C_{f,\text{Mg}^+}\right)\\ \nonumber
&&\times 10^{-0.4 \text{E(B-V})_{\text{UV}} k}.
\label{eq:f_c_1}
\end{eqnarray}
In this scenario, the meaning of Equation~\ref{eq:f_c_1} becomes more apparent. The term $C_{f,\text{Mg}^+} e^{-\sigma_{\text{UV}} N_{\text{H}^0}}$ represents the fraction of photons that escape through paths intersecting the cool, dense clouds shown in Figure~\ref{fig:clouds}. The remaining term, $1 - C_{f,\text{Mg}^+}$, accounts for photons traveling through paths that do not intersect these clouds, where we assume no attenuation by the ambient medium illustrated in Figure~\ref{fig:clouds}.       

In the calculation of Equation~\ref{eq:f_c_1}, the Monte Carlo sampling takes into account the portion of the covering fraction, $C_f$, attributed to the geometry and orientation of the bi-cone.  We can generalize Equation~\ref{eq:f_c_1} to allow for holes in the bi-cone (i.e., $f_c <1$) as follows:
\begin{eqnarray}
f_{esc}^{LyC} &=& \resizebox{0.3\textwidth}{!}{$\left[\Sigma_i^I \frac{e^{-\sigma N_{\rm{H^0},i}}}{I} + (1+f_c)\left(1- \Sigma_i^I \frac{e^{-\sigma N_{\rm{H^0},i}}}{I}\right)\right]$}\\ \nonumber
&& \times 10^{-0.4 \text{E(B-V)}_{\text{UV}}k}\\ \label{eq:f_c<1}
    &=& \Sigma_i^I \frac{1.0-fc(1.0-e^{-\sigma N_{H^0,i}})}{I}e^{-\tau_{\rm dust}}\\ 
    && \times 10^{-0.4 \text{E(B-V)}_{\text{UV}}k}. \nonumber
    \label{eq:fesc_mg}
\end{eqnarray}

\begin{figure}
	\centering
	\includegraphics[width=\columnwidth]{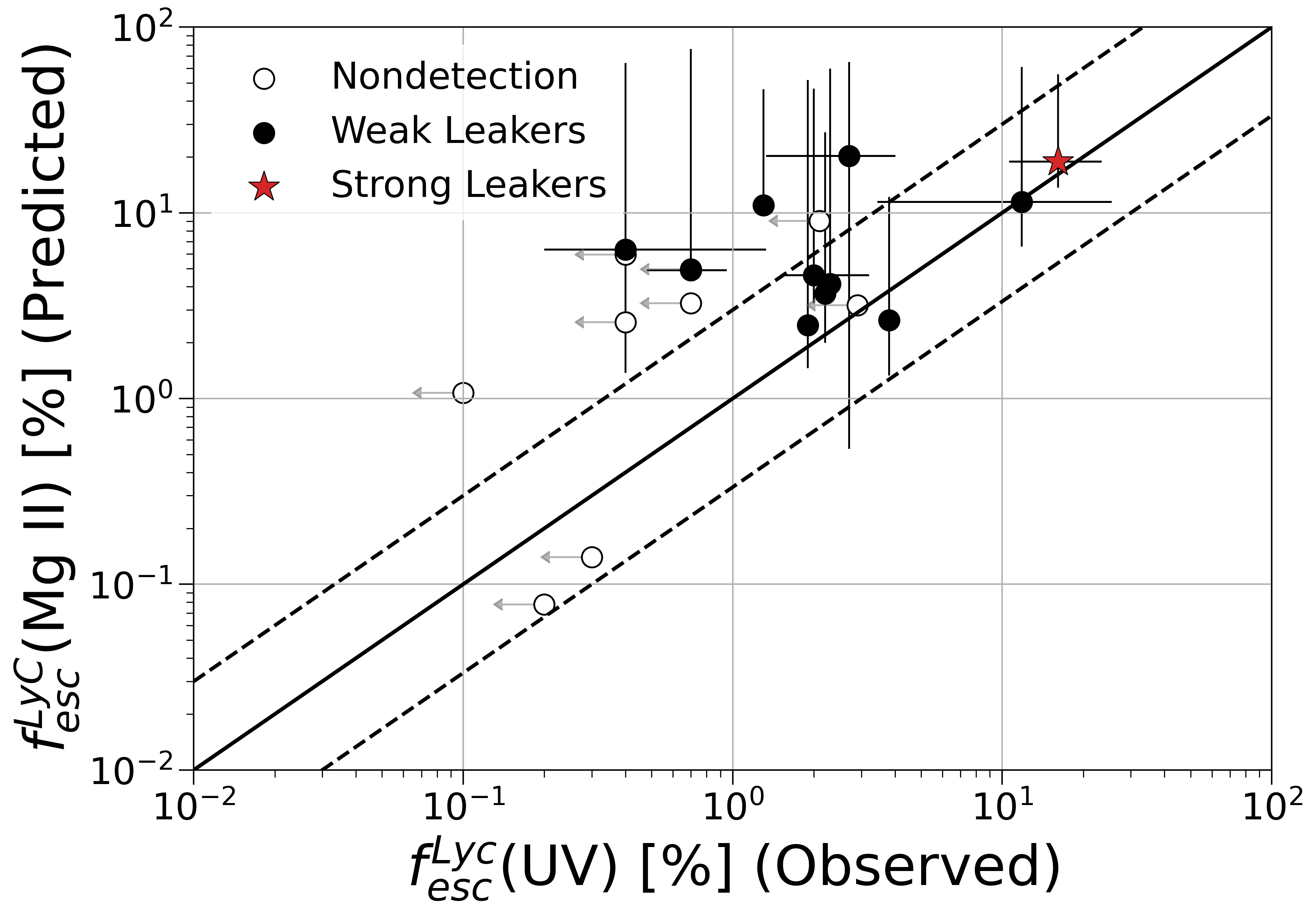}
	\caption{compares the Mg II-based predictions of the Lyman Continuum (LyC) escape fraction, $f_{\text{esc}}^{\text{LyC}}(\text{Mg II})$ (vertical axis), defined by Equation~\ref{eq:f_c<1}, with the observed values, $f_{\text{esc}}^{\text{LyC}}(\text{UV})$ (horizontal axis). Hollow circles denote non-detections, serving as upper limits; black circles indicate weak leakers; and red stars represent strong leakers. The dashed black lines illustrate a factor of three deviation from the one-to-one line (solid black line). Generally, the predictions tend to overestimate the observed values. This discrepancy is attributed to low column density regions of $\text{H}^0$ that are not detected by Mg II in absorption.}
	\label{fig:LyC_predictions}
\end{figure}

Figure~\ref{fig:LyC_predictions} compares the LyC escape fraction estimates, derived using Equation~\ref{eq:fesc_mg}, $f_{\text{esc}}^{\text{LyC}}(\text{Mg II})$, with the observed values reported by \cite{Flury2022a}, $f_{\text{esc}}^{\text{LyC}}(\text{UV})$. These values are also recorded in Table~\ref{tab:lyc}. We opted to use the UV-derived  $f_{\text{esc}}^{\text{LyC}}$ values because the $\text{H}\beta$-derived estimates may be biased by older stellar populations ($>10$ Myr) that emit optical emission \citep{Flury2022a}.  Leakers are categorized following the criteria of \cite{Flury2022a, Flury2022b}: strong leakers ($>5\sigma$ LyC detection and $f_{\text{esc}}^{\text{LyC}} > 5\%$), weak leakers ($>2\sigma$ LyC detection, but not strong), and nondetections ($<2\sigma$ LyC detection). The results reveal a clear bias, with $f_{\text{esc}}^{\text{LyC}}(\text{Mg II})$  generally overestimating $f_{\text{esc}}^{\text{LyC}}(\text{UV})$.   

The overestimate of $f_{esc}^{LyC}$ was expected since we assumed that $C_{f,\rm{H}^0} = C_{f,\rm{Mg}^+}$, and neglected the larger covering fractions of $\rm H^0$ reported by \cite{Saldana-Lopez2022} in Equation~\ref{eq:Saldana-Lopez}.  To test our hypothesis that the difference between the covering fractions of $\rm Mg^+$ and $\rm H^0$ is due primarily to optically thin gas, still sensitive to LyC absorption, but not Mg II, we develop an empirical based correction to Equation~\ref{eq:fesc_mg}, where we measure the optical depth, $\tau_{\rm thin}$, of the gas required to recover the observed escape fraction.  

We consider a three component model of the ISM/CGM consisting of optically thick, thin, and transparent pathways to LyC emission - that is,
\begin{equation}
\begin{split}
    f_{\text{esc}}^{\text{LyC}} &= \left( C_{f,\text{thick}}e^{\tau_{\text{thick}}} + C_{f,\text{thin}}e^{\tau_{\text{thin}}}  + C_{f,\text{transparent}}\right)\\
    & \quad \times 10^{-0.4\text{E(B-V)}_{\text{UV}}k},
\end{split}
\label{eq:general_fesc}
\end{equation}
where $C_{f,\text{thick}} = C_{f,\text{Mg}^+}$, $C_{f,\text{thin}} = C_{f,\text{H}^0}-C_{f,\text{Mg}^+}$, and $C_{f,\text{transparent}} = 1-C_{f,\text{H}^0}$ in our interpretation.  We recover $C_{f,\text{H}^0}$ from Equation~\ref{eq:Saldana-Lopez}, and set all values which exceed unity to one.   We can rewrite Equation~\ref{eq:general_fesc} in terms of Equation~\ref{fig:LyC_predictions} as
\begin{eqnarray}
\resizebox{0.4\textwidth}{!}{$
\begin{aligned}
    f_{\text{esc}}^{\text{LyC}} &= f_{\text{esc}}^{\text{LyC}}(\text{MgII})\\
    &+\left[\left(C_{f,\text{H}^0}-C_{f,\text{Mg}^+}\right) e^{-\tau_{\text{thin}}}-\left(C_{f,\text{H}^0}-C_{f,\text{Mg}^+}\right)\right]\\
    &\times 10^{-0.4 \rm{E(B-V)}_{\rm UV}k}.
\end{aligned}
$}
\label{eq:fesc_correction}
\end{eqnarray}

When possible\footnote{In some instances, a solution could not be found.  This occurs either because $C_{f,\text{transparent}}$ is too large given our method for dust attenuation or $f_{\text{esc}}^{\text{LyC}}$ was under predicted.  In regards to the former, we suspect that this is due to a lack of precision in Equation~\ref{eq:Saldana-Lopez}.}, we solve for $\tau_{\text{thin}}$ by setting Equation~\ref{eq:fesc_correction} equal to $f_{esc}^{LyC}$.  We find an average/median value, $\bar{\tau}_{\text{thin}} = 1.3/0.4$, over a range $0.093<\tau_{\text{thin}}<3.1$.  This corresponds to a column density of roughly $\bar{N}_{\rm H^0} = \bar{\tau}/\sigma = 2.1/0.63\times 10^{17}\rm\ cm^{-2}$, or, after scaling by ${\bar{\alpha}_{\rm H^0/Mg^+}}^{-1} = 4.6\times10^{-5}$, $\bar{N}_{\rm Mg^+} = 9.5/2.8\times 10^{12}\rm\ cm^{-2}$.  According to mock spectra drawn from the idealized outflow models studied by \cite{Huberty2024} (see their Figure 8), this is just below the detection limit of the COS spectrograph ($\rm Res \sim 20\ km\ s^{-1},\ S/N = 10$) for Si II 1190\AA, 1193\AA\ in absorption.  Given that the ratio of oscillator strength-wavelength products is only a factor of a few between line comparisons and the overall lower resolution of our survey, we can safely assume that pathways at this column density will also go undetected in Mg II in absorption, in support of our hypothesis.     



\begin{figure}
	\centering
	\includegraphics[width=\columnwidth]{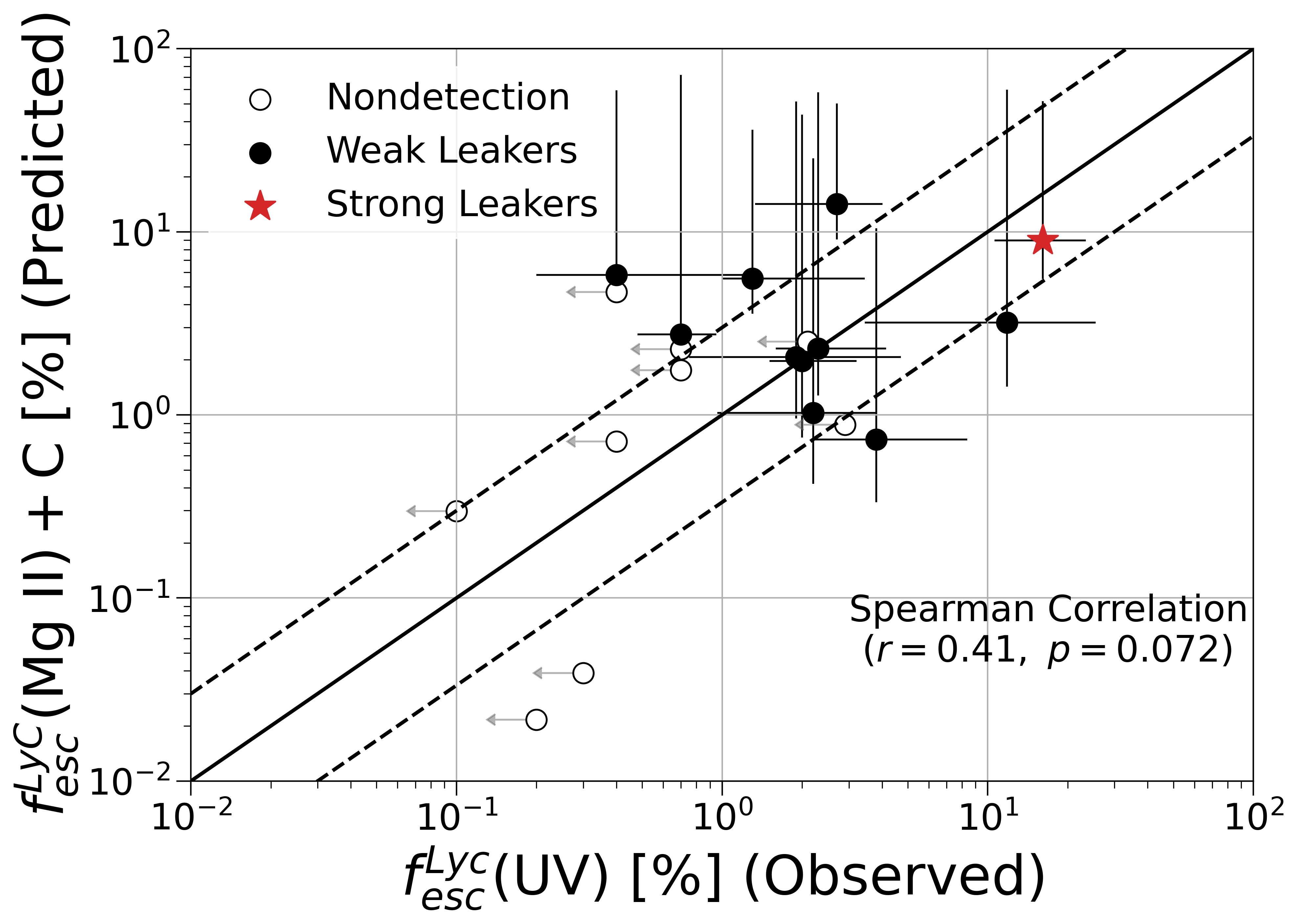}
	\caption{compares the corrected Mg II-based predictions of the Lyman Continuum (LyC) escape fraction, $f_{\text{esc}}^{\text{LyC}}(\text{Mg II})$, defined in Equation~\ref{eq:fesc_correction}, with the observed values, $f_{\text{esc}}^{\text{LyC}}(\text{UV})$. The corrections account for low column density channels of $\text{H}^0$. The presentation follows the same format as Figure~\ref{fig:LyC_predictions}. The updated predictions show improved agreement, and the bias observed in Figure~\ref{fig:LyC_predictions} is no longer present.}
 \label{fig:LyC_predictions_corrected}
\end{figure}

Given the tight range in $\tau_{\text{thin}}$, we can plug the median value back into Equation~\ref{eq:fesc_correction} to compute an empirically corrected escape fraction for each galaxy.  The results are shown in Figure~\ref{fig:LyC_predictions_corrected}.  We recover a Spearman Correlation coefficient of $r = 0.41$ and a p-value of $p = 0.072$ suggesting a moderate correlation.  These results are similar to those of \cite{Xu2023} who predicted LyC escape fractions for many of these same galaxies, while excluding the strongest absorbers and keeping the most optically thin cases which we excluded.

We mention that we also tested the two component partial covering model, $f_{\text{esc}}^{\text{LyC}} = C_{f,\text{thick}}e^{\tau_{\text{thick}}} + C_{f,\text{thin}}e^{\tau_{\text{thin}}}$ (e.g., \citealt{Zackrisson2013,Reddy2016,Chisholm2020,Saldana-Lopez2022,Xu2022a,Xu2023}), and found similar results.  This suggests that the boost in generality offered by the three component model may not be needed given the current precision and resolution of our data set.  Still, we prefer the three component model, as it offers a simple explanation for scenarios where $C_{f,\rm H^0}<1$.

In summary, we consider an ISM/CGM which has both optically thick, thin, and transparent pathways to LyC escape along the line of sight.  We argue that Mg II absorption lines are dominated by optically thick pathways.  We attribute the lack of sensitivity to the roughly $10^{4}$ factor difference in abundance between $\rm Mg^+$ and $\rm H^0$.  Naively, this picture might appear to be in contrast with the works of \cite{Chisholm2020,Xu2022a,Xu2023} who posit that Mg II escapes along optically thin pathways.  The key difference between these studies and our own is that they are focusing on Mg II (escape) in emission, while we are focusing on Mg II in absorption.  While SALT also expects Mg II emission to escape along low density channels, its primary purpose is to model the resonant scattering of photons in galactic outflows.  With this in mind, it should be noted that Equation~\ref{eq:fesc_correction} can be used to predict $f_{\text{esc}}^{\text{LyC}}$ in galaxies which show Mg II scattering in absorption, but not otherwise - for example, in density bounded scenarios.   In these cases, the methods of  \cite{Chisholm2020,Xu2022a,Xu2023} are better suited.

\section{The Cool Outflows of LzLCS}\label{sec:outflows}

\begin{table*}
\caption{Neutral Hydrogen Outflow Rates and Loading Factors}
\centering
\begin{tabular}{cccccccc}
\hline\hline
Galaxy&$\log{\dot{M}_{\rm H^0}}$ &$\log{\dot{P}_{\rm H^0}}$&$\log{\dot{E}_{\rm H^0}}$&$\log{\eta_{\rm M,H^0}}$&$\log{\eta_{\rm P,H^0}}$&$\log{\eta_{\rm E,H^0}}$\\
&$\rm M_{\odot}\ yr^{-1}$&$\rm M_{\odot}\ km\ s^{-1}\ yr^{-1}$&$\rm M_{\odot}\ km^{2}\  s^{-2}\ yr^{-1}$&&\\[.5 ex]
\hline
J012910+145935m &   $0.104\substack{1.54\\1.29}$ &   $2.03\substack{1.46\\1.45}$ &  $3.67\substack{1.49\\2.09}$ &    $-1.01\substack{1.54\\1.29}$ &   $-1.59\substack{1.46\\1.45}$ &  $-3.13\substack{1.49\\2.09}$ \\
J072326+414608m &   $-1.21\substack{2.21\\2.04}$ &  $0.485\substack{2.52\\1.88}$ &  $1.92\substack{2.79\\2.12}$ &    $-2.13\substack{2.21\\2.04}$ &   $-2.94\substack{2.52\\1.88}$ &  $-4.68\substack{2.79\\2.12}$ \\
J081112+414146m &  $0.645\substack{0.897\\1.38}$ &  $2.54\substack{0.985\\1.52}$ &  $4.11\substack{1.11\\1.69}$ &  $0.0895\substack{0.897\\1.38}$ & $-0.519\substack{0.985\\1.52}$ &  $-2.12\substack{1.11\\1.69}$ \\
J091208+505009m &   $0.21\substack{0.935\\1.04}$ &   $1.61\substack{1.01\\1.29}$ &  $2.69\substack{1.32\\1.58}$ &  $-0.929\substack{0.935\\1.04}$ &   $-2.03\substack{1.01\\1.29}$ &  $-4.13\substack{1.32\\1.58}$ \\
J091703+315221m &   $0.115\substack{1.34\\1.33}$ &    $2.1\substack{1.28\\1.46}$ &  $3.78\substack{1.19\\1.61}$ &    $-1.18\substack{1.34\\1.33}$ &    $-1.7\substack{1.28\\1.46}$ &  $-3.19\substack{1.19\\1.61}$ \\
J092552+395714m &    $0.68\substack{1.35\\1.13}$ &     $2.6\substack{1.45\\1.2}$ &  $4.24\substack{1.56\\1.34}$ &  $-0.0856\substack{1.35\\1.13}$ &   $-0.664\substack{1.45\\1.2}$ &   $-2.2\substack{1.56\\1.34}$ \\
J103816+452718m &  $-0.858\substack{1.88\\1.96}$ &   $1.04\substack{1.97\\2.41}$ &  $2.64\substack{1.96\\2.86}$ &    $-2.46\substack{1.88\\1.96}$ &   $-3.06\substack{1.97\\2.41}$ &  $-4.64\substack{1.96\\2.86}$ \\
J112933+493525m &    $0.262\substack{1.05\\1.0}$ &   $2.24\substack{1.04\\1.04}$ &  $3.93\substack{1.05\\1.12}$ &    $-0.433\substack{1.05\\1.0}$ &  $-0.954\substack{1.04\\1.04}$ &  $-2.44\substack{1.05\\1.12}$ \\
J124619+444902m &    $-1.7\substack{2.49\\1.67}$ & $0.0422\substack{2.55\\1.99}$ &  $1.51\substack{2.75\\2.43}$ &    $-3.17\substack{2.49\\1.67}$ &   $-3.93\substack{2.55\\1.99}$ &  $-5.63\substack{2.75\\2.43}$ \\
J130128+510451m &  $-0.436\substack{1.72\\1.57}$ &   $1.25\substack{1.79\\1.93}$ &  $2.68\substack{1.92\\2.49}$ &    $-1.82\substack{1.72\\1.57}$ &   $-2.63\substack{1.79\\1.93}$ &  $-4.37\substack{1.92\\2.49}$ \\
J131419+104739m &  $1.82\substack{0.824\\0.982}$ &   $3.79\substack{0.88\\1.09}$ & $5.48\substack{0.925\\1.23}$ &  $0.466\substack{0.824\\0.982}$ & $-0.0653\substack{0.88\\1.09}$ & $-1.55\substack{0.925\\1.23}$ \\
J132633+421824m &  $-0.293\substack{1.48\\1.33}$ &   $1.53\substack{1.54\\1.48}$ &   $3.09\substack{1.6\\1.74}$ &    $-1.62\substack{1.48\\1.33}$ &    $-2.3\substack{1.54\\1.48}$ &   $-3.92\substack{1.6\\1.74}$ \\
J134559+112848m &  $1.04\substack{0.881\\0.927}$ &  $2.76\substack{0.918\\1.39}$ &   $4.14\substack{1.01\\1.9}$ &  $-0.19\substack{0.881\\0.927}$ & $-0.968\substack{0.918\\1.39}$ &   $-2.76\substack{1.01\\1.9}$ \\
J003601+003307x &  $-0.429\substack{1.34\\1.04}$ &  $0.883\substack{1.56\\1.32}$ &  $1.91\substack{1.79\\1.62}$ &    $-1.61\substack{1.34\\1.04}$ &    $-2.8\substack{1.56\\1.32}$ &  $-4.95\substack{1.79\\1.62}$ \\
J004743+015440x &   $1.01\substack{0.868\\1.07}$ &     $2.5\substack{1.1\\1.28}$ &   $4.19\substack{1.25\\1.53}$ &  $-0.055\substack{0.868\\1.07}$ &  $-1.3\substack{1.1\\1.28}$ &   $-2.7\substack{1.25\\1.53}$ \\
\hline
\end{tabular}
\begin{minipage}{\textwidth}
\justifying
\small
{\noindent From left to right: (1) Object (2) mass outflow rate of neutral hydrogen (3) momentum outflow rate of neutral hydrogen (4) energy outflow rate of neutral hydrogen (5) mass loading factor of neutral hydrogen (6) momentum loading factor of neutral hydrogen (7) energy loading factor of neutral hydrogen}
\end{minipage}
\label{tab:mor}
\end{table*}


Galactic outflows are multi-phase phenomena and can span a temperature range from $10^2$ to $10^7\ \text{K}$. To fully understand their impact on LyC escape, a multi-wavelength approach is warranted \citep{Gonzalez-Alfonso2012, Falstad2015, Herrera-Camus2020, Chisholm2018c, Carr2021a, Laha2018}.  Indeed, the phases do not necessarily share the same kinematics (see \citealt{Chisholm2016a,Chisholm2018c, Carr2021a}) preventing a single tracer such as Mg II and photoionization model such as \texttt{Cloudy} from capturing the full impact of the outflows.  We will see that this is generally the case for our galaxies as well, when we compare the velocities of the outflows traced by Mg II to those traced by the broad emission components observed in the [O III] 5007\AA\ lines at the end of the next section.

Ultimately, for a galaxy to be a LyC emitter, it must successfully clear pathways through the ISM/CGM of $\rm H^0$ and dust.  With this in mind, we focus on the outflows traced by $\rm H^0$ while adopting the kinematics from Mg II.  As we saw in Section~\ref{sec:modeling_lyc_escape}, the Mg II clouds appear to account for the majority of $\rm H^0$ in mass.  Kinematically, we associate these outflows with the cool phase ($10^4\rm\ K$) of the wind, but acknowledge that they represent a lower estimate of the total mass in this phase, which can be predominantly ionized \citep{Werk2013}.  Indeed, Mg II exists at a ionization potential ranging from 7.6 to 15 eV, and traces both neutral and partially ionized gas.  In this section, we focus on the general properties of the neutral outflows in LzLCS and study their impact on LyC escape in the next section.

\begin{figure}
	\centering
 \includegraphics[width=\columnwidth]{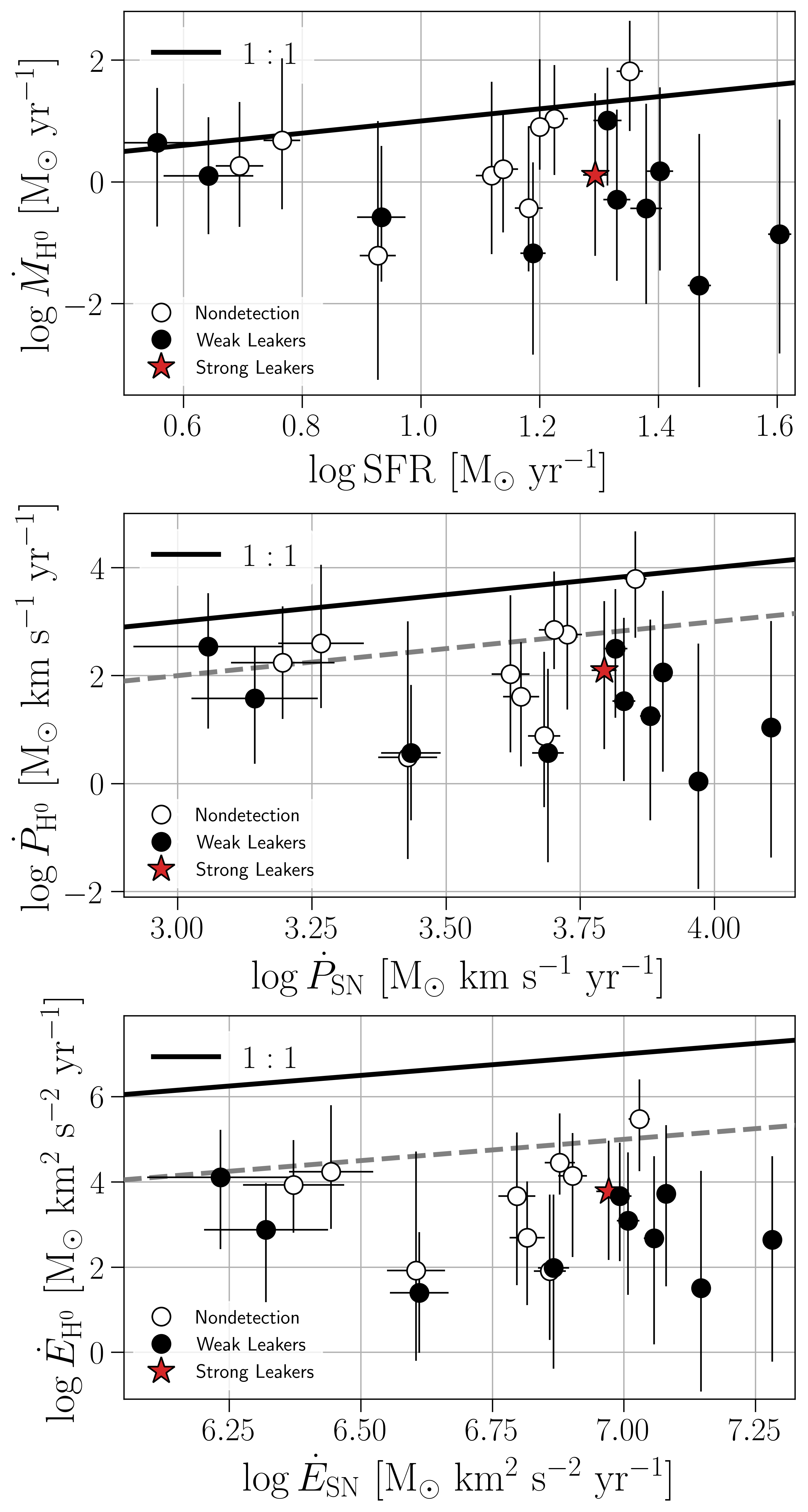}
	\caption{presents various relationships concerning the neutral outflow rates in LzLCS.  \emph{\textbf{Top Panel}}: The mass outflow rate is plotted against the star formation rate (SFR), with the one-to-one line shown in black.  \emph{\textbf{Middle Panel}}: The momentum outflow rate is compared to the momentum deposition rate predicted from supernovae. The one-to-one line is depicted in black, while 10\% of the supernovae-predicted momentum is shown by the dashed grey line.  \emph{\textbf{Bottom Panel}}:  The energy outflow rate is plotted against the energy deposition rate predicted from supernovae. Here, the one-to-one line is in black, and 1\% of the supernovae-predicted energy is indicated by the dashed grey line.  The wide range of values in these plots illustrates the diverse characteristics of outflows in the LzLCS galaxies which likely reflect the time line relative to the starburst and the primary driving mechanism of the outflows.}
	\label{fig:rate_vs_dep}
\end{figure} 

We begin by showing the mass ($\dot{M}_{\rm H^0}$), momentum ($\dot{P}_{\rm H^0}$), and energy ($\dot{E}_{\rm H^0}$) outflow rates versus the SFR, momentum deposition rate from supernovae ($\dot{P}_{\rm SN}$), and energy deposition rate from supernovae ($\dot{E}_{\rm SN}$), in the top, middle, and bottom panels of Figure~\ref{fig:rate_vs_dep}, respectively.  The outflow rates were obtained by scaling the $\rm Mg^+$ outflow rates, defined in Equations~\ref{eq:SALT_MOR}, \ref{eq:SALT_POR}, and \ref{eq:SALT_EOR}, by the scale factor, $\alpha_{\rm H^0/Mg^+} = N_{\rm H^0}/N_{\rm Mg^+}$, obtained with \texttt{Cloudy} models in Section~\ref{sec:modeling_lyc_escape}.  We chose to compute the outflow rates at the UV half-light radius to better align with the SFR \citep{Huberty2024}.  Furthermore, evaluating our outflow rates at small radii should aid in our later comparisons to the winds probed by [O III] 5007\AA\ which is expected to occur on relatively small scales on the order of super star clusters (\citealt{Komarova2021}, Komarova et al., in prep).  We adopt the relation, $\dot{P}_{\rm SN} = 3.17\times 10^2\rm\ M_{\odot}\ km\ s^{-1}\ yr^{-1} \left(\frac{\rm SFR}{M_{\odot} \ \rm yr^{-1}}\right)$ from \cite{Murray2005}, and the relation, $\dot{E}_{\rm SN} = 4.76\times 10^5\ \rm M_{\odot}\ km^{2}\ s^{-2}\ yr^{-1}\left(\frac{\rm SFR}{M_{\odot} \ \rm yr^{-1}}\right)$ from \cite{Leitherer1999}.  Both predictions assume one supernova per $100 M_{\odot}$ of stars formed, Salpeter IMFs \citep{Salpeter1955}, and do not include other forms of feedback such as stellar winds. 

Overall, we observe a significant variation in the values of the three quantities relative to the measured error bars. This variability likely reflects the broad range of outflow velocities detected in the Mg II absorption lines, which span from nearly zero to several hundred kilometers per second which may be linked to different system ages. Indeed, \cite{Hayes2023a} observed an increase of several orders of magnitude in energy loading with stellar age in a study of 87 local starbursts.  The highest recorded values of $\dot{M}_{\text{H}^0}$ are on the order of the star formation rate (SFR), while the maximum values for $\dot{P}_{\text{H}^0}$ and $\dot{E}_{\text{H}^0}$ are below 10\% and 1\% of the total energy output expected from supernovae, respectively, in broad agreement with \cite{Hayes2023a}. 

The lower momentum and energy loading factors are expected, in part because we only account for the mass and energy in the $\text{H}^0$ phase of the outflows, but also because supernovae at these metallicities are expected to be delayed.  \cite{Jecmen2023} demonstrate the consequences for mechanical feedback, and show that kinetic energy is expected to be suppressed more strongly than momentum, consistent with our results.  This is consistent with a scenario where, as we will discuss, a substantial fraction of the galaxies are experiencing predominantly radiation feedback whereas supernovae are expected to deliver significantly more mechanical energy to the outflows.  The matter is further complicated by the fact that we used SFR derived from H$\beta$ which is likely biased to younger stars and might possibly miss some energy from supernovae.  Later, we will examine how the metallicities and conditions of our galaxies make them prime candidates for delayed supernovae feedback \citep{Jecmen2023}.

We observe a potential trend where non-leakers exhibit higher energy and momentum loading, and possibly higher mass loading as well. This is consistent with the expectation that outflows with high density and large covering fractions of neutral hydrogen should favor high outflow rates and lower $f_{\text{esc}}^{\text{LyC}}$. Again, the age of the starburst may play a role. Multiple supernovae are required to drive a wind, and in systems with delayed supernovae, this process can take up to 10 million years or longer \citep{Jecmen2023}. As a result, lower metallicity systems with strong supernova-driven outflows may host fewer massive stars, leading to reduced LyC production and, consequently, lower $f_{\text{esc}}^{\text{LyC}}$.  We will examine the relationship between LyC escape and outflow rates in more detail in the next section.

\begin{figure}
	\centering
	\includegraphics[width=\columnwidth]{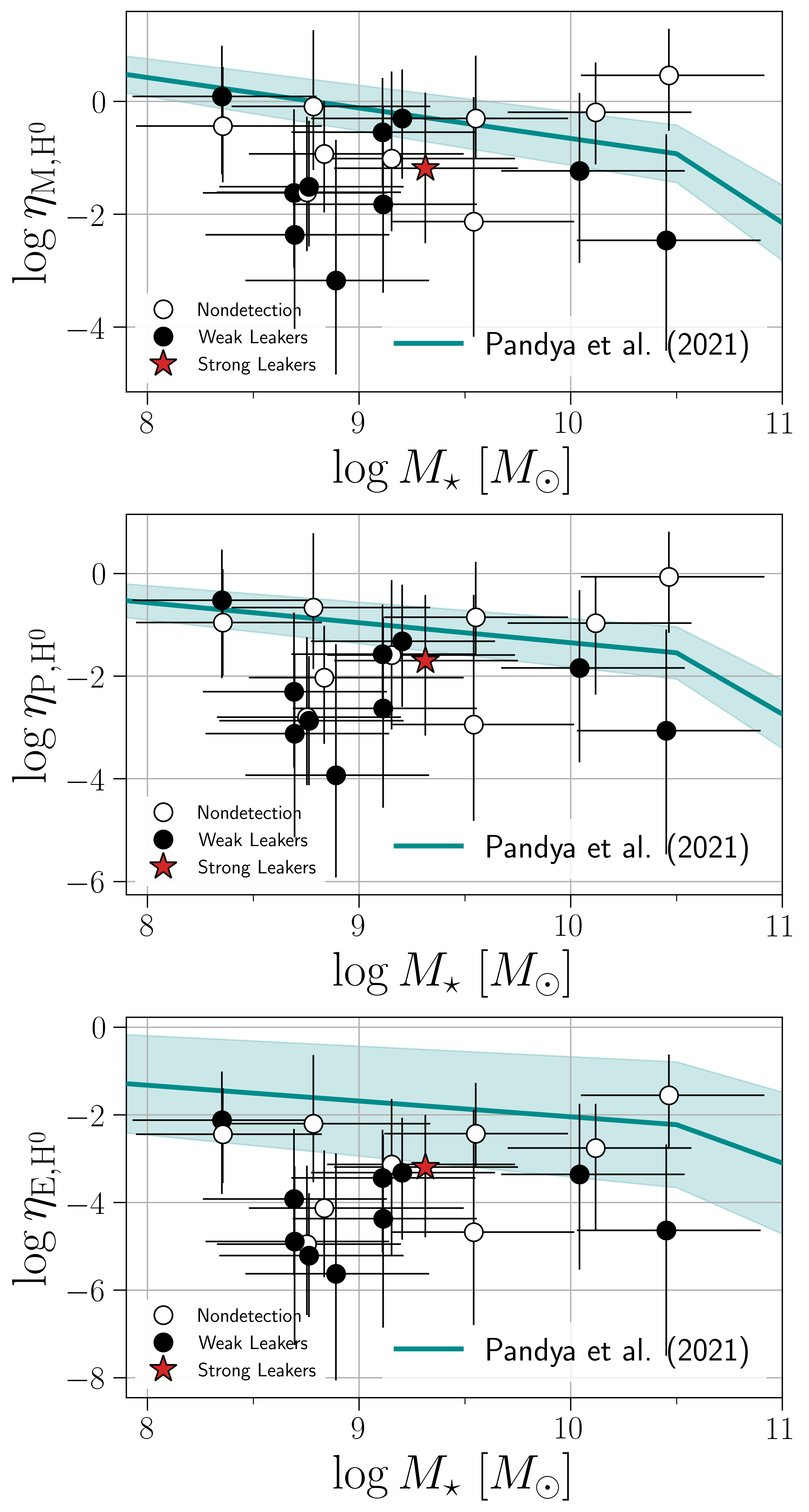}
	\caption{compares the mass, momentum, and energy loading factors of neutral hydrogen measured in LzLCS to the stellar mass in the \emph{\textbf{Top}}, \emph{\textbf{Middle}}, and \emph{\textbf{Bottom Panels}}, respectively. Theoretical predictions from the FIRE-2 simulations, as presented by \cite{Pandya2021}, are overlaid in turquoise. These predictions correspond to the portion of the loading factors associated with the cool phase of the outflows. Although we only consider the neutral component of the cool phase in LzLCS, several galaxies still achieve loading factors that align with the theoretical predictions.}
	\label{fig:eta_vs_mstar}
\end{figure} 

We compare the mass ($\eta_{\rm M,H^0} = \dot{M}_{\rm H^0}/\rm{SFR}$), momentum ($\eta_{\rm P,H^0} = \dot{P}_{\rm H^0}/\dot{P}_{\rm SN}$), and energy ($\eta_{\rm E,H^0} = \dot{E}_{\rm H^0}/\dot{E}_{\rm SN}$) loading factors against stellar mass ($M_{\star}$) in the top, middle, and right panels of Figure~\ref{fig:eta_vs_mstar}, respectively.  We have overlaid theoretical predictions from the FIRE-2 simulations \citep{Hopkins2018} as presented by \cite{Pandya2021}. These predictions correspond to the portion of each loading factor associated with the cool phase of the outflows, defined by a temperature range of $10^3$ to $10^5$ K.

Once again, we observe a substantial spread in the measured values relative to the error bars. While the upper envelope of the data aligns with the FIRE-2 predictions, the majority of the data points fall two to three orders of magnitude lower.  The spread and differences with FIRE-2 could depend on a number of factors.  The FIRE-2 loading factors taken from \cite{Pandya2021} represent burst averaged quantities with outflow rates measured at 0.15 $R_{\text{vir}}$, or about 35 kpc from galactic center.  Hence, they avoid scatter by definition.  In contrast, our measurements are instantaneous and measured at the UV half-light radius, which is typically only a few kpc from galactic center.  Outflow rates can vary significantly over this distance, with values at 0.15 $R_{\text{vir}}$ generally being higher in the FIRE-2 simulations.  Lastly, the differences proposed to explain the lower outflow rates carry over.  We note that the lower energy and momentum loading factors are also indications of systems experiencing delayed supernovae and prolonged periods of radiation dominant feedback \citep{Jecmen2023}

The same pattern regarding LyC observed in Figure~\ref{fig:rate_vs_dep} emerges in Figure~\ref{fig:eta_vs_mstar}.  As before, we deem these results to coincide with our intuition that outflows with higher densities and covering fractions of $\rm H^0$ will generally have higher mass loading and lower $f_{esc}^{LyC}$.  These relationships are complex, however, and likely depend on the source of feedback and various galaxies properties such as compactness which we will explore later on.       

Ultimately, we detect substantial differences between the FIRE-2 predictions and our estimates of the loading factors.  While this is likely due in large part to the differences in technique and the explicit quantities measured, we emphasize the importance of incorporating delayed supernovae in low metallicity systems and accounting for the extended periods of radiation-dominant feedback.


\begin{figure*}
	\centering
	\includegraphics[width=\textwidth] {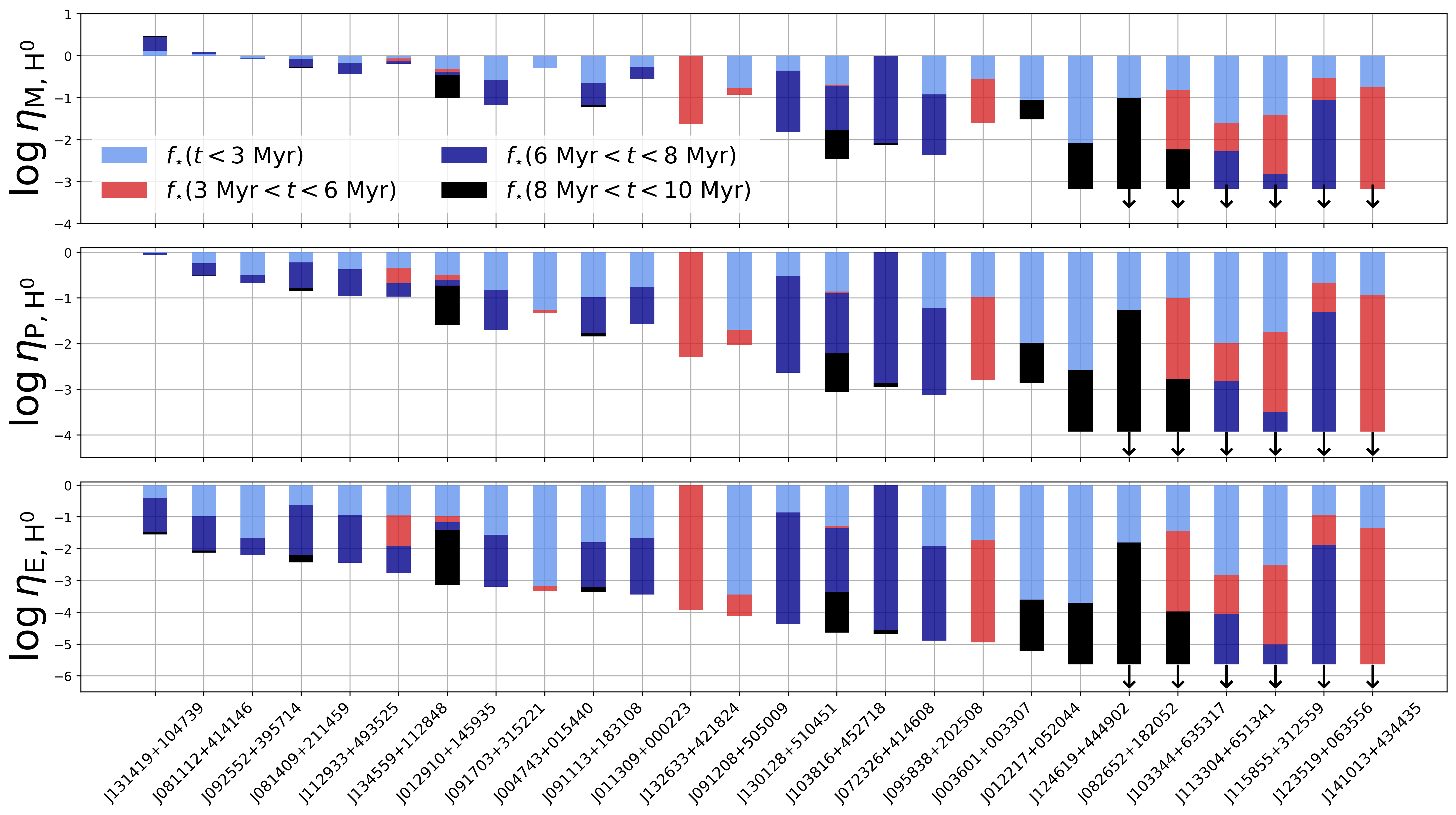}
	\caption{illustrates the light fractions associated with different loading factors for each galaxy. The \textbf{\emph{Top Panel}} shows the light fractions according to the mass loading factor, the \textbf{\emph{Middle Panel}} displays them according to the momentum loading factor, and the \textbf{\emph{Bottom Panel}} plots them according to the energy loading factor.  The loading factors are shown on a logarithmic scale, with values less than zero displayed from the bottom of each bar and values greater than zero from the top. The sub-bars, colored differently, represent the light fractions and are visually scaled. For example, if 75\% of a bar is light blue, this indicates that 75\% of the light in that galaxy originates from stellar populations younger than 3 Myr.  Objects which failed to shows signs of galactic winds were assigned to the lowest measured values and represent upper estimates.  All galaxies are arranged from left to right with decreasing energy loading.  Most galaxies, except for J072326+414608, contain significant contributions from young stellar populations ($t < 3\ \text{Myr}$). Generally, galaxies exhibiting the highest mass, momentum, and energy loading factors also show substantial fractions of both young and middle-aged to old stellar populations ($6\ \text{Myr} < t < 8\ \text{Myr}$). Conversely, galaxies with the lowest energy and momentum loading factors often have notable amounts of young to middle-aged stellar populations ($3\ \text{Myr} < t < 6\ \text{Myr}$).}
	\label{fig:lightfraction_eta}
\end{figure*} 

To reveal the possible sources of feedback driving the outflows, we use light fractions, $f_{\star}$, which represent the fraction of total starlight contributed by stellar populations within a specified age range.  Our $f*$ values were taken from \cite{Saldana-Lopez2022} and were derived from \texttt{STARBURST99} \citep{Leitherer1999} template fits to the COS UV spectra, considering stellar populations ranging up to 10 Myr.  We consider the same age bins as Flury et al. (in prep): Young stars, $t < 3\ \rm Myr$, contain the most massive stars and are responsible for producing copious amounts of LyC radiation.  Since these populations are too young for supernovae, the outflows in galaxies dominated by these populations are likely driven by radiation or stellar winds.  We consider two middle age ranges for post starbursts, $3\ \rm Myr < t < 6\ \rm Myr$ and $6\ \rm Myr < t < 8\ \rm Myr$.  The former range corresponds to the Wolf-Rayet stage \citep{Crowther2007,Ekstrom2012}.  Wolf-Rayet stars are very massive and capable of producing copious amounts of LyC emission and high-speed stellar winds \citep{Senchyna2021,Martins2022,delValleEspinosa2023,Rivera-Thorsen2024}.  Feedback in the latter stage should become increasingly more supernovae dominant.  Finally, we consider older stellar populations, $8\ \rm Myr<t<10\ \rm Myr$.  At this stage, mechanical feedback should be dominated by supernovae \citep{Leitherer2014,Stanway2018}.

We note that stellar winds are anticipated to contribute substantially less mechanical feedback at the low-metallicities of our sample \citep{Vink2022}.  This claim is understood both theoretically and with observations (see \citealt{Jecmen2023} for a discussion and references within).  Thus, throughout this paper we will often refer to galaxies with young stellar populations as being radiation feedback dominant.

While linear SED fitting provides a robust indication of the light-weighted age and metallicity of the LzLCS galaxies, the stellar population fractions within individual age bins may be quite uncertain. This occurs because, in the LzLCS COS wavelength range (950-1250\AA\ rest-frame), two stellar-wind features—OVI 1037 Å and NV 1240 Å—are particularly sensitive to the age of the underlying stellar population. These features, combined with the UV continuum slope, help to break the age-dust degeneracy, which is a key advantage of the UV regime over optical continuum fitting. Additionally, the strength of weaker photospheric lines, such as C III 1175\AA, provides insights into metallicity. However, determining the age and metallicity for individual stellar components (i.e., individual light fractions) can still be challenging, especially in noisy spectra. Consequently, Flury et al. (in prep) used light fractions in stacked LzLCS galaxies.  Therefore, while we continue to conduct our analysis on an individual galaxy basis, we refrain from making a deep quantitative analysis which relies on individual age bins, and instead focus on general trends.     

We show the light fractions reported for each galaxy according to the different loading factors in Figure~\ref{fig:lightfraction_eta}.  The galaxies are arranged from left to right with decreasing energy loading.  The loading factors are positioned according to a log scale, but the light fractions are shown according to the visual scale (i.e., they are not on a log scale).  For example, if 75\% of a bar is colored in light blue, then 75\% of that galaxy's stellar populations are less than $3\ \rm Myr$ old.  Galaxies which failed the F-test are included, and are assigned the lowest measured values which are meant to represent upper estimates.  

The distributions in Figure~\ref{fig:lightfraction_eta} reveal that young stellar populations are present in all but one galaxy, J072326+414146, which is characterized by a significant presence of young-to-middle-aged (3-6 Myr) stellar populations.  This suggests that most galaxies have the potential to produce substantial amounts of LyC radiation. Despite considerable variation, some general trends emerge:  The five galaxies with the highest energy loading factors contain substantial fractions of middle-to-old-aged (6-8 Myr) stellar populations.  Conversely, seven of the nine galaxies with the lowest energy loading factors are predominantly composed of young ($ < 3\ \text{Myr}$) and young-to-middle-aged (3-6 Myr) stellar populations \citep{Hayes2023a}.  This dichotomy implies that the energy and momentum carried by neutral outflows in galaxies with radiation-driven winds are lower compared to those with supernova-driven winds.  This trend likely demonstrates that the neutral gas kinematics are more closely linked to the SN-driven feedback, whereas the radiation-driven wind has a very low filling factor and cannot effectively move dense gas \citep{Komarova2021}.  The collapse in wind velocity in these conditions may be associated with rapid or catastrophic cooling in the radiation driven winds \citep{Silich2003,Silich2004,Silich2018}.  We explore these possibilities in greater depth in the next sections.   

As noted, there is significant variation in the distributions shown in Figure~\ref{fig:lightfraction_eta}, especially in the middle of the diagram or away from the extremes. Ultimately, the loading factors will be heavily influenced by galaxy properties such as compactness and star formation surface density, which will lead to variation. These effects and their relationship to LyC escape will be further explored in the next section. 

In summary, our analysis of the LzLCS galaxies reveals considerable variation in their mass, momentum, and energy outflow rates of $\rm H^0$ or the neutral component of their galactic winds. The mass outflow rates ($\dot{M}_{\rm H^0}$) can reach values comparable to the SFR, while the momentum outflow rates ($\dot{P}_{\rm H^0}$) can be up to 10\% of the total supernovae momentum deposition rate, and the energy outflow rates ($\dot{E}_{\rm H^0}$) can be up to 1\% of the total supernovae energy deposition rate.  The loading factors show a similar range of variation, with the highest values aligning with predictions for the cool phase of outflows in the FIRE-2 simulations. Stellar population synthesis models reveal that our galaxies are undergoing both radiation and supernovae feedback, with some cases showing predominately radiation driven winds. These factors, along with intrinsic galaxy properties, likely contribute to the observed deviations in the measured values.  Overall, we find that galaxies with supernovae-driven winds tend to have higher mass, momentum, and energy loading factors related to the cool phase of the outflows compared to those driven primarily by radiation. 

\section{Cool Outflows and LyC Escape}\label{sec:lyc_escape}

Galactic outflows have long been considered a necessary\footnote{See \citep{Marques-Chaves2022b} for an interesting case of a LyC leaker undergoing galactic inflows.} condition for LyC escape \citep{Heckman2001}. Outflow signatures, such as blueshifted absorption lines \citep{Chisholm2017a,Marques-Chaves2021}, broad emission line components \citep{Amorin2024}, and double-peaked Lyman alpha emission profiles \citep{Izotov2021,Izotov2022}, are commonly observed in LyC leakers. However, not all galaxies with outflows exhibit LyC escape.  The relationship between outflows and LyC escape is complex and influenced by the timing and specific details of how feedback affects the neutral gas and dust distribution in the ISM and CGM. In this section, we aim to unravel this relationship by examining the connection between the neutral outflows, quantified in the previous section, and LyC escape observed in the LzLCS sample.

\begin{figure}
	\centering
	\includegraphics[width=\columnwidth]{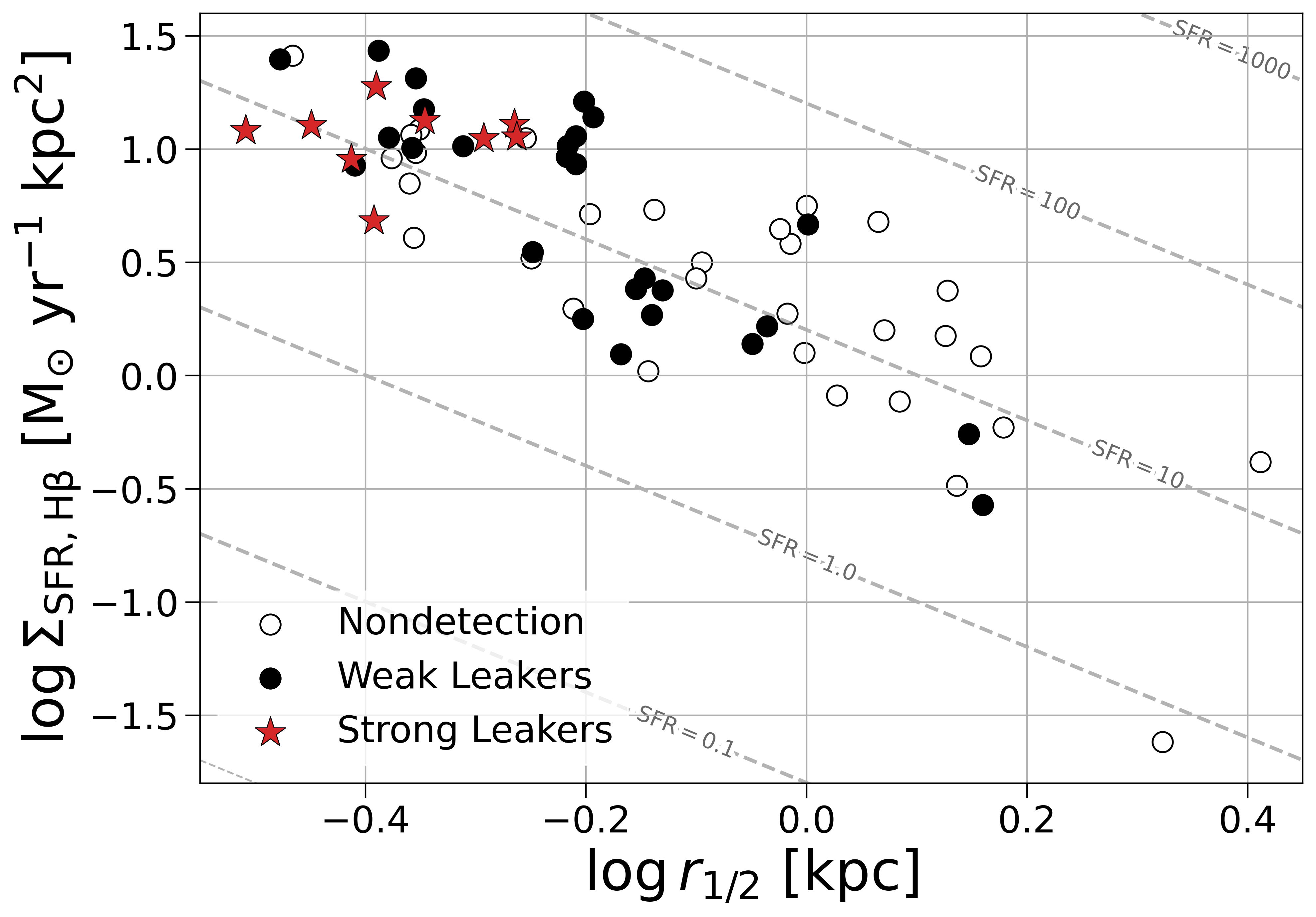}
	\caption{presents the LyC escape fraction for the entire LzLCS galaxy sample \citep{Flury2022a} plotted against star formation surface density, $\Sigma_{\rm SFR}$, and UV half-light radius, $r_{1/2}\ [\rm kpc]$. Non-detections are indicated by hollow circles and are considered upper estimates; black circles represent weak leakers, while red stars denote strong leakers. Most strong leakers are concentrated in the upper left corner of the plot. According to \cite{Cen2020}, galaxies in this region of the $\Sigma_{\rm SFR}-r_{1/2}$ parameter space are expected to have optimal conditions for the formation of galactic winds capable of clearing the neutral ISM/CGM, thereby facilitating LyC escape.}
 \label{fig:podigous_leakers}
\end{figure}

In star-forming galaxies, it is well established that LyC leakers often exhibit compact morphology and exceptionally high star formation surface densities, $\Sigma_{SFR}$ (see \citealt{Naidu2020} for a list of early references, \citealt{Flury2022b}).  It was proposed by \cite{Heckman2001,Heckman2011} and later justified by \cite{Cen2020}, using simple analytical arguments, that these conditions are ideal for the production of galactic winds, or outflows capable of clearing pathways through the neutral ISM/CGM for LyC escape.  To demonstrate the significance of the relation, we show $f_{esc}^{Lyc}$ as a function of $\Sigma_{SFR}$ and the UV half-light radius, $r_{1/2}$, for 66 LzLCS galaxies in Figure~\ref{fig:podigous_leakers}.  While \cite{Flury2022b} initially presented these results separately, this figure combines them to reveal the interaction between the parameters. The trend shows that as $\Sigma_{SFR}$ increases at a constant SFR, the frequency of LyC leakers also rises, with the most efficient leakers appearing in the upper left corner of the parameter space.

We choose to work in the $\Sigma_{SFR} -r_{1/2}$ parameter space for the remainder of this section.  This will allow us to focus our study on galaxies with the conditions necessary for feedback to effectively alter the geometry of the ISM/CGM to enable LyC escape.

\subsection{Evidence for Feedback}

\begin{figure}
	\centering
	\includegraphics[width=\columnwidth]{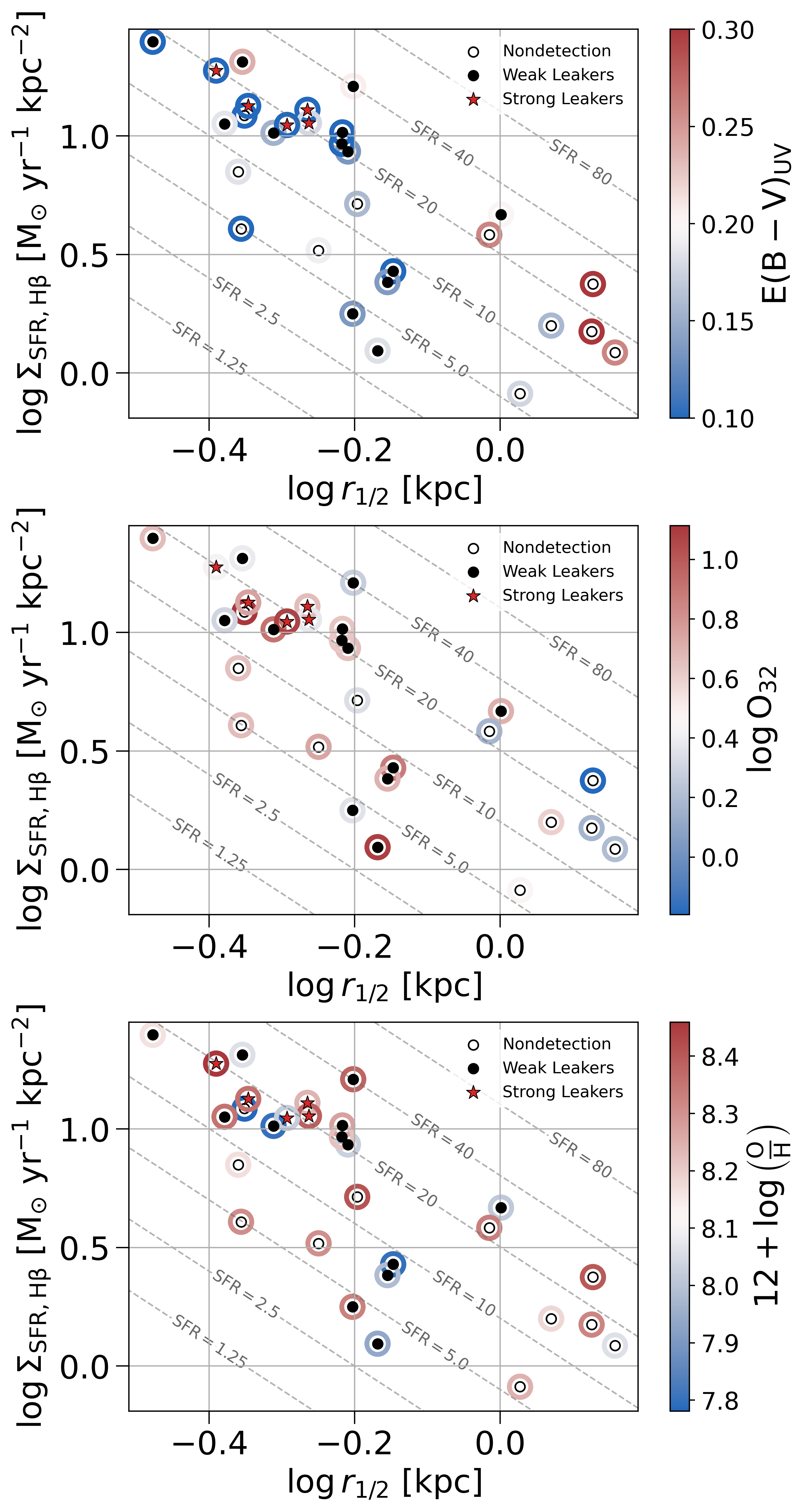}	\caption{displays various quantities relevant to LyC escape plotted against the star formation surface density, $\Sigma_{\rm SFR}$, and UV half-light radius, $r_{1/2}$.  The \textbf{\emph{top panel}} shows the UV dust extinction, the \textbf{\emph{middle panel}} presents the $O_{32}$ ratio, and the \textbf{\emph{bottom panel}} displays the gas metallicity fraction relative to solar. These panels follow the same format as Figure~\ref{fig:podigous_leakers}. In summary, the region characterized by the highest $\Sigma_{\rm SFR}$ and lowest $r_{1/2}$, which corresponds to the strongest LyC leakers, exhibits lower dust extinction, higher $O_{32}$ ratios, and a range of gas metallicities.   }
	\label{fig:dust_Sigma_RUV}
\end{figure}

\begin{figure*}
	\centering
	\includegraphics[width=\textwidth]{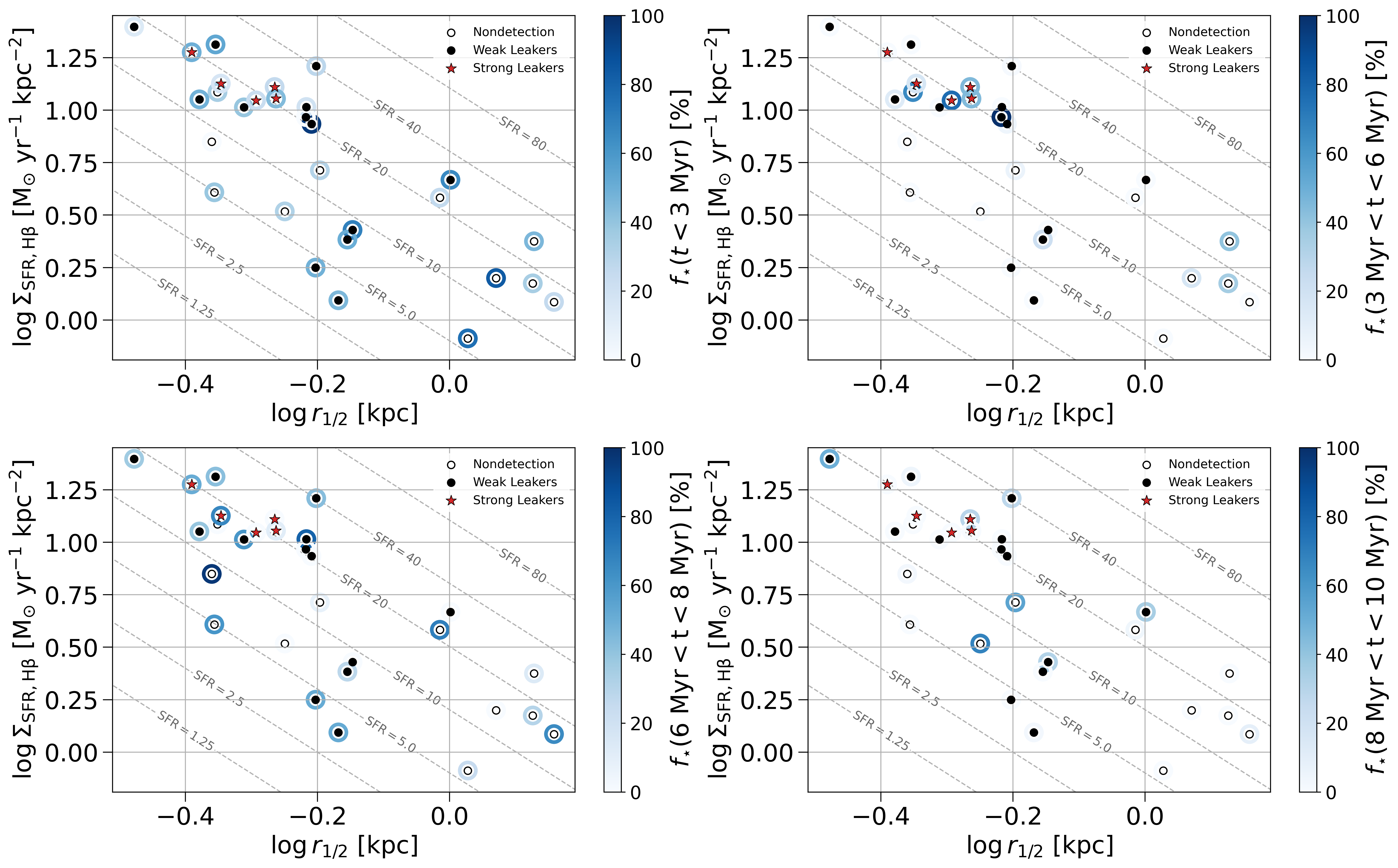}	\caption{displays the light fractions for each galaxy are plotted as a function of the star formation surface density, $\Sigma_{\rm SFR}$, and UV half-light radius, $r_{1/2}$. The stellar population ages increase from left to right and top to bottom. For each galaxy, the light fractions sum to 100\% across all panels. The presentation follows the format of Figure~\ref{fig:podigous_leakers}. It is noteworthy that the strong LyC emitters predominantly exhibit both young and intermediate-age stellar populations. }  
 \label{fig:lightfrac_vs_sigma_ruv}
\end{figure*} 

We plot UV dust extinction, $\rm E(B-V)_{UV}$, $\rm O_{32}$, and gas phase metallicity, in the $\Sigma_{\rm SFR} - r_{1/2}$ parameter space in the top, middle, and bottom panels of Figure~\ref{fig:dust_Sigma_RUV}, respectively. This analysis encompasses all 29 galaxies in our sample.  Higher values are shown in red and lower values in blue.

Strong LyC leakers, concentrated in the upper left corner of the $\Sigma_{\rm SFR} - r_{1/2}$ space, exhibit lower dust extinction, higher $\rm O_{32}$ ratios, and a range of gas metallicities reaching up to about 0.5 $Z_{\odot}$ \citep{Saldana-Lopez2022,Flury2022b,Chisholm2022}. While it is possible that the galaxies in the upper left corner are simply producing less dust (unlikely given the range in metallicity and stellar age, see below), there are plenty of reasons to anticipate the trend. UV radiation can destroy dust grains through processes such as photo-dissociation, heating, and sputtering near hot stars \citep{Salpeter1977}. Destruction can be difficult, however, for fully formed grains, even in H II regions \citep{Salpeter1977}.  Indeed, the dust survival time in the ISM is typically determined by destruction processes occurring in supernova-generated shock waves \citep{Jones2004}.  An alternative explanation for the reduced extinction is the removal of dust grains from the ISM through radiation-driven winds \citep{Zhang2017,Komarova2021}. Photons can accelerate dust grains to large distances from the galaxy, effectively reducing the dust optical depth and/or covering fraction. Moreover, intense UV radiation fields can cause the destruction of cool clouds and their associated dust by crushing the clouds which then mix with the hotter ambient medium \citep{Huang2020}.    

To better understand the role of feedback, we plot the light fractions, $f_{\star}$, for the same stellar ages specified in section~\ref{sec:outflows}, in the $\Sigma_{\rm SFR}-r_{1/2}$ parameter space in Figure~\ref{fig:lightfrac_vs_sigma_ruv}.  The panels are organized from left to right, top to bottom with increasing stellar age.  Note that $f_{\star}$ should sum to 100\% when adding values from matching galaxies across the four panels.  

The strongest LyC leakers appear to predominantly have both young ($<3\ \rm Myr$) and young-to-middle-aged ($3-6\ \rm Myr$) stellar populations. For reference, the strong LyC leakers include galaxies J091703+315221, J103344+635317, J115855+312559, J123519+063556, and J141013+434435. As seen in Figure~\ref{fig:lightfraction_eta}, galaxies J103344+635317, J115855+312559, J123519+063556, and J141013+434435 exhibit no signs of outflows in Mg II absorption. All but J123519+063556--the weakest leaker of the five--have at least 70\% of their stellar populations younger than six Myr. In contrast, J091703+315221 displays evidence of galactic winds in Mg II. Specifically, J091703+315221 lacks young-to-middle-aged populations, but has nearly 50\% of its stars in the middle-to-old age range ($6 - 8\ \text{Myr}$). Looking further, many of the relatively weaker leakers neighboring the strong leakers have winds and middle-to-old-aged stellar populations.

These findings support a general picture where the strongest leakers have young stellar populations and are primarily experiencing radiation feedback, absent supernovae.  In contrast, the relatively weaker leakers have older stellar populations and are likely experiencing supernovae feedback \citep{Kimm2019}.  Under specific conditions—such as high $\Sigma_{\rm SFR}$ and low $r_{1/2}$, possibly enhanced by Wolf-Rayet stars—the strongest leakers seem to lack cool clouds in their winds.  The two exceptions to this trend likely represent strong leakers which are transitioning to primarily supernovae dominant feedback based on the ages of their stellar populations.  These galaxies may be experiencing a two-staged burst scenario, where both mechanical feedback and radiation feedback act to clear channels for optimal LyC escape (\citealt{Micheva2017,Enders2023}, Flury et al. in prep).  As first suggested in Section~\ref{sec:outflows}, these findings could be indicative of catastrophic cooling.  We will explore this possibility in more detail in the next section. 

These results are consistent with \cite{Bait2023} who studied the radio properties of the LzLCS galaxies.  They found that the strongest leakers ($f_{esc}^{LyC} > 10\%$), had flat radio spectra which are consistent with young stellar populations and a lack of supernovae.  Additionally, they found steeper spectra in the relatively weaker leakers ($1\%<f_{esc}^{LyC}<10\%$) indicating the presence of relativistic cosmic rays accelerated by supernovae.  Further evidence has been found in stacks of the LzLCS galaxies by Flury et al. (in prep) and in an independent study focusing on the broad wing morphology of high ionization state (HIS) lines, such as the [O III] 5007\AA\ line, by Komarova et al. (in prep).  We will discuss the latter in more detail in the next half of this section.  


\begin{figure}
	\centering
    \includegraphics[width=\columnwidth]{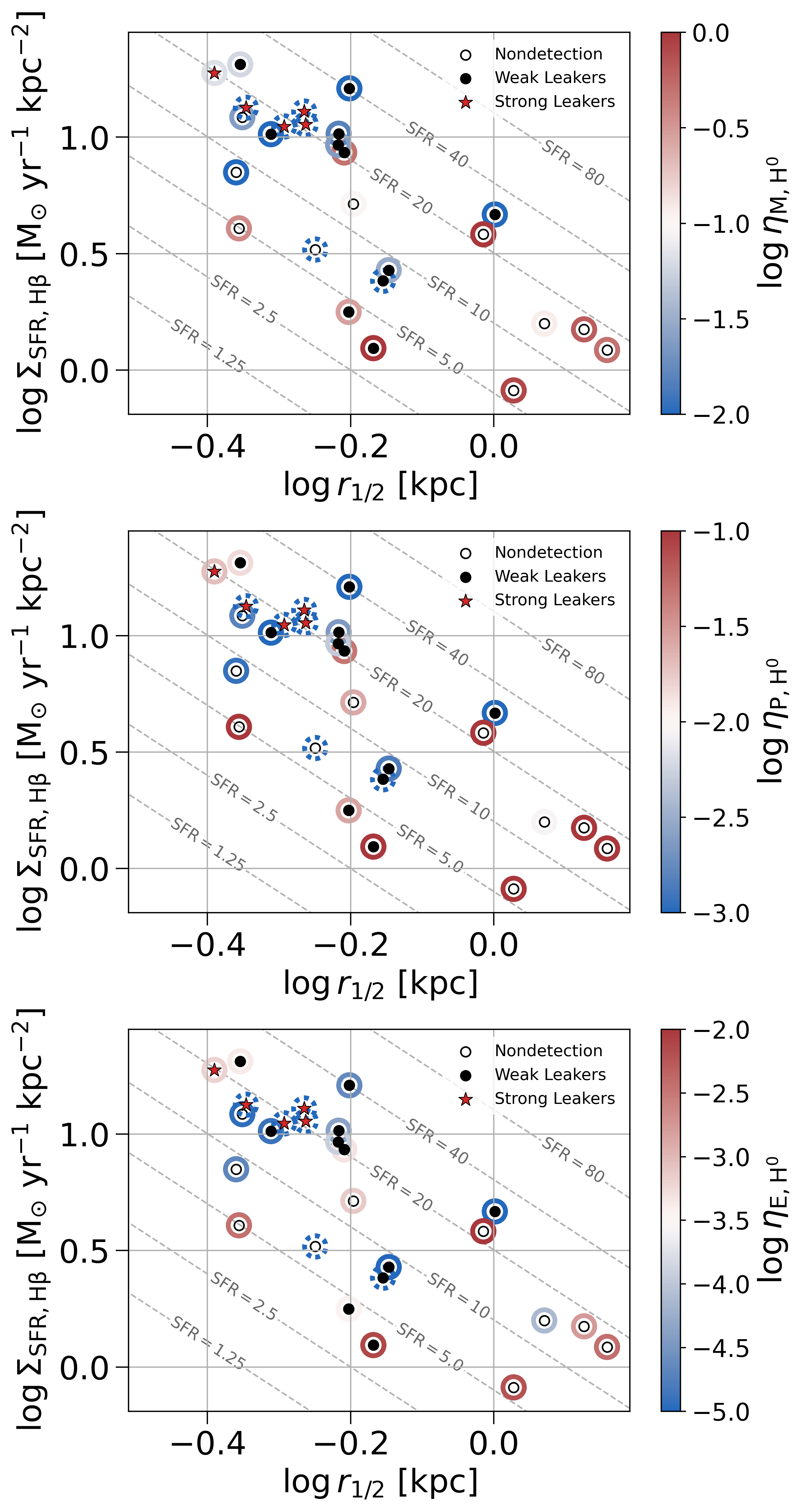}
	\caption{displays the loading factors plotted against the star formation surface density, $\Sigma_{\rm SFR}$, and UV half-light radius, $r_{1/2}$.  The overall format follows that of Figure~\ref{fig:podigous_leakers}. The \textbf{\emph{top panel}} displays the mass loading factor, the \textbf{\emph{middle panel}} presents the momentum loading factor, and the \textbf{\emph{bottom panel}} shows the energy loading factor. In summary, the neutral hydrogen outflows ($\rm H^0$) in the region characterized by high $\Sigma_{\rm SFR}$ and low $r_{1/2}$, which includes the strongest LyC leakers, exhibit lower mass, momentum, and energy loading.}
	\label{fig:eta_Sigma_RUV}
\end{figure} 

\begin{figure*}
    \centering
    \subfigure{
        \includegraphics[width=0.485\textwidth]{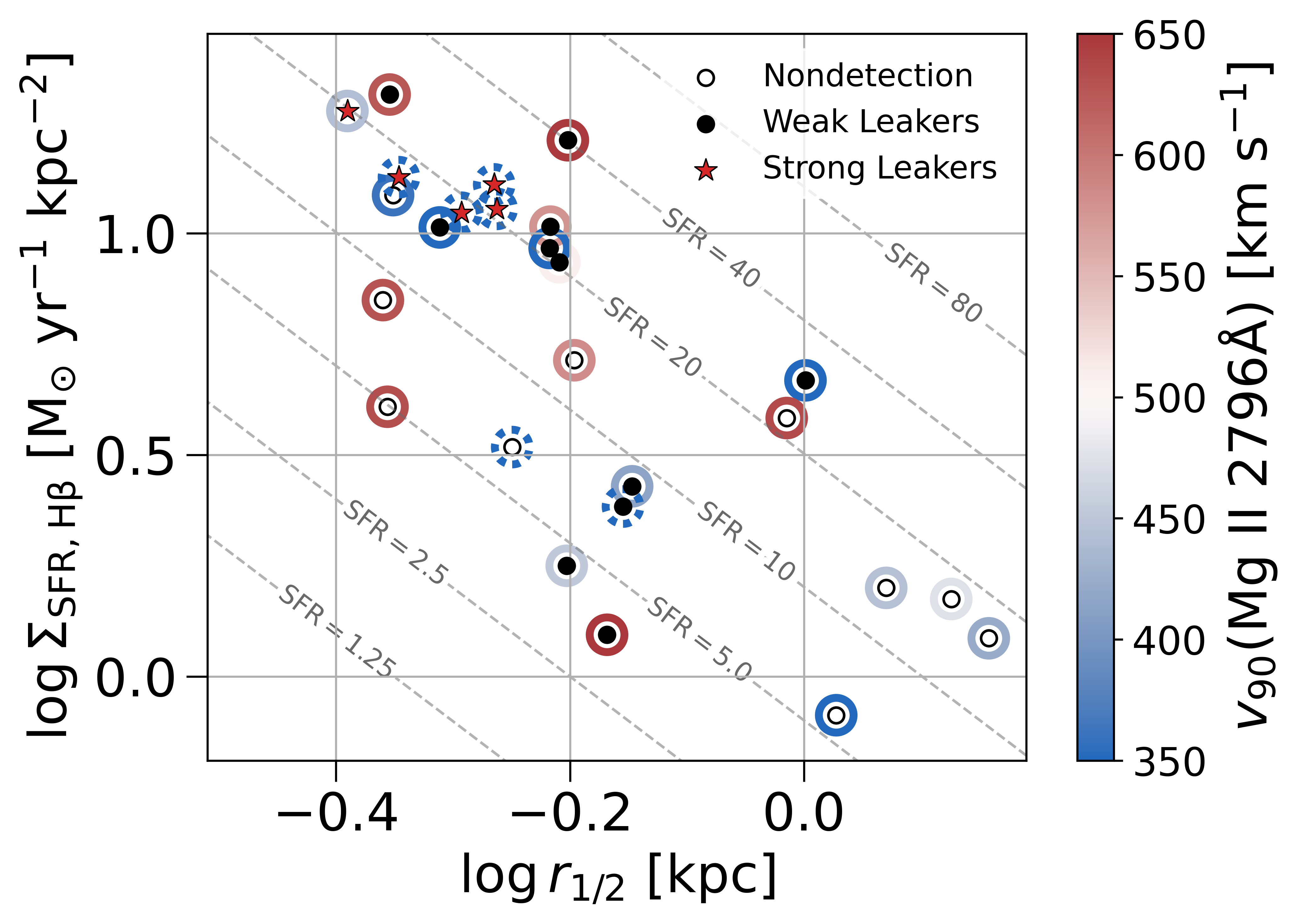}
    }
    \hfill
    \subfigure{
        \includegraphics[width=0.485\textwidth]{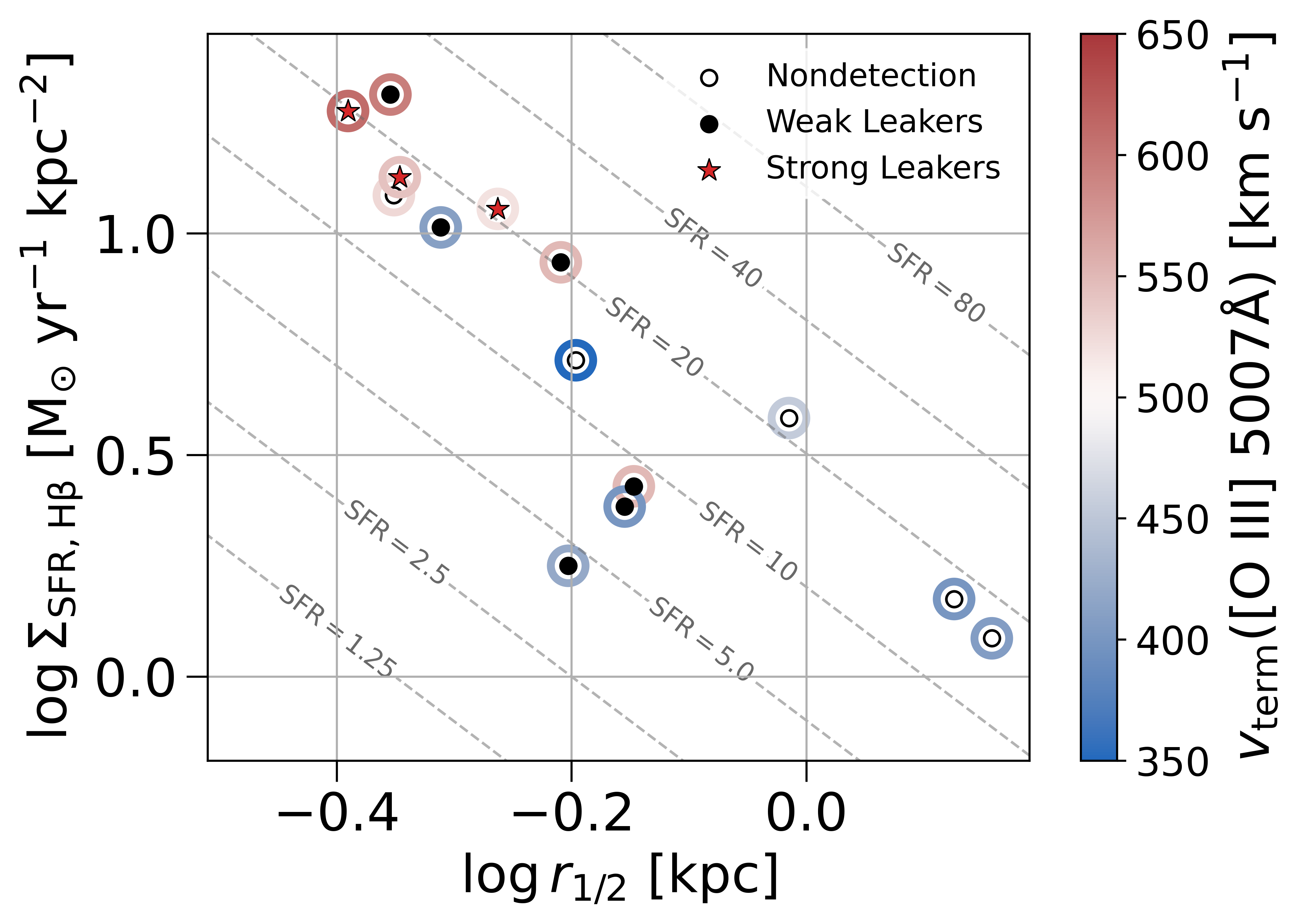}
    }
    \caption{compares the speeds of the cool and warm phases of the outflows in LzLCS. Color bars are on the same scale.  \textbf{\emph{Left Panel}}: The velocity at 90\% of the EW, $v_{90}$, measured from the Mg II 2796\AA\ line, probing the cool phase of the outflow, is plotted against star formation surface density, $\Sigma_{\rm SFR}$, and UV half-light radius, $r_{1/2}$. Blue circles indicate low $v_{90}$ values, while red circles indicate high values. Dashed blue circles represent galaxies without blue-shifted absorption in the Mg II 2796\AA\ line (i.e., $v_{90} = 0\rm \ km\ s^{-1}$). Generally, $v_{90}$ increases with compactness and higher $\Sigma_{\rm SFR}$, but some variation is observed. Notably, four out of the five strongest LyC leakers show no signs of absorption in Mg II. \textbf{\emph{Right Panel}}: The terminal velocity, $v_{\rm term}$, of the warm outflow, derived from power-law fits to the [O III] 5007\AA\ line by Komarova et al. (in prep), is plotted against $\Sigma_{\rm SFR}$ and $r_{1/2}$. $v_{\rm term}$ consistently increases with compactness and higher $\Sigma_{\rm SFR}$, showing less deviation than $v_{90}$ values. All three strong leakers with available data exhibit fast winds ($> 500 \ \rm km\ s^{-1}$) according to the [O III] 5007\AA\ line.}
    \label{fig:v90}
\end{figure*}

\subsection{Cool Winds and LyC Escape}

We now investigate the relationship between neutral outflows, as defined in Section~\ref{sec:outflows}, and LyC escape. In the top, middle, and bottom panels of Figure~\ref{fig:eta_Sigma_RUV}, we plot the mass, momentum, and energy loading factors within the $\Sigma_{\rm SFR}-r_{1/2}$ parameter space. For this analysis, we focus on the 26 galaxies that were successfully fit with the SALT model. Objects with high loading factors are depicted in red, while those with low loading factors are shown in blue. Galaxies that failed the F-test are assumed to have outflow rates below the minimum detected level and are indicated by dashed blue lines.  

In general, as one moves diagonally along the paths of constant SFR towards increasing $\Sigma_{\rm SFR}$, the loading factors tend to decrease, albeit with some scatter.  A “hole” appears corresponding to four of the five strong leakers, where the loading factors suddenly collapse to zero or drop below the detection limit.   From a strictly observational perspective, this may be expected: LyC leakers are anticipated to have lower covering fractions and column densities of neutral hydrogen ($N_{\rm H^0}$).  In fact, the cool phase of outflows may influence LyC escape by simply getting in the way \citep{Carr2021a}.  Still, these results are rather surprising, since the analytical arguments of \cite{Cen2020} predicted that high $\Sigma_{\rm SFR}$ in compact galaxies would generate higher velocity winds and other observations of LIS lines have generally shown this (e.g., \citealt{Heckman2016}).  Moreover, \cite{Amorin2024} found a significant correlation between $f_{esc}^{LyC}$ and the velocity dispersion in the broad line component of the [O III] 5007\AA\ line in the LzLCS galaxies.  O III (35-55 eV) probes a higher ionization potential than Mg II (7.6-15 eV), however, indicating a potential phase dependence.  These observed trends are consistent with the scenario that supernovae are delayed at lower metallicities, which substantially prolongs the period for radiation-dominated feedback \citep{Jecmen2023}, of which strong [O III] is an indicator.  

To investigate the phase dependence of the relations in Figure~\ref{fig:eta_Sigma_RUV}, we plot measures of the maximum outflow velocities corresponding to the phases probed by the Mg II 2796\AA\ (cool phase) and [O III] 5007\AA\ (warm phase) lines in the $\Sigma_{\rm SFR}-r_{1/2}$ parameter space in the left and right panels of Figure~\ref{fig:v90}, respectively. The outflow velocities for Mg II are determined using the $v_{90}$ values (Section~\ref{sec:model}) obtained from SALT model fits to the 2796\AA\ line. For this calculation, we only consider the absorption component of the continuum line profile without smoothing to the instrumental resolution, effectively removing the effects of emission infilling \citep{Prochaska2011,Scarlata2015,Mauerhofer2021}, instrumental smoothing, and nebular emission. The O III outflow velocities were provided by Komarova et al. (in prep) and correspond to terminal velocities derived from power-law fits to the broad-line components.  The measured values are provided in Table~\ref{tab:wind_speeds}.  

The wind velocities generally increase along paths of constant SFR towards regions of high $\Sigma_{\rm SFR}$, albeit with some scatter. Once again, a “hole” appears, corresponding to four of the five strongest LyC leakers, where velocities drop to zero or fall below the detection threshold. These may be objects where no SNe have yet occurred.  In contrast, the outflow velocities traced by O III consistently increase along these same paths towards higher $\Sigma_{\rm SFR}$. Even in the strongest leakers, winds are still present, with velocities exceeding 500 km s$^{-1}$ in the three strong leakers with available data.

\begin{figure}
	\centering
	\includegraphics[width=\columnwidth]{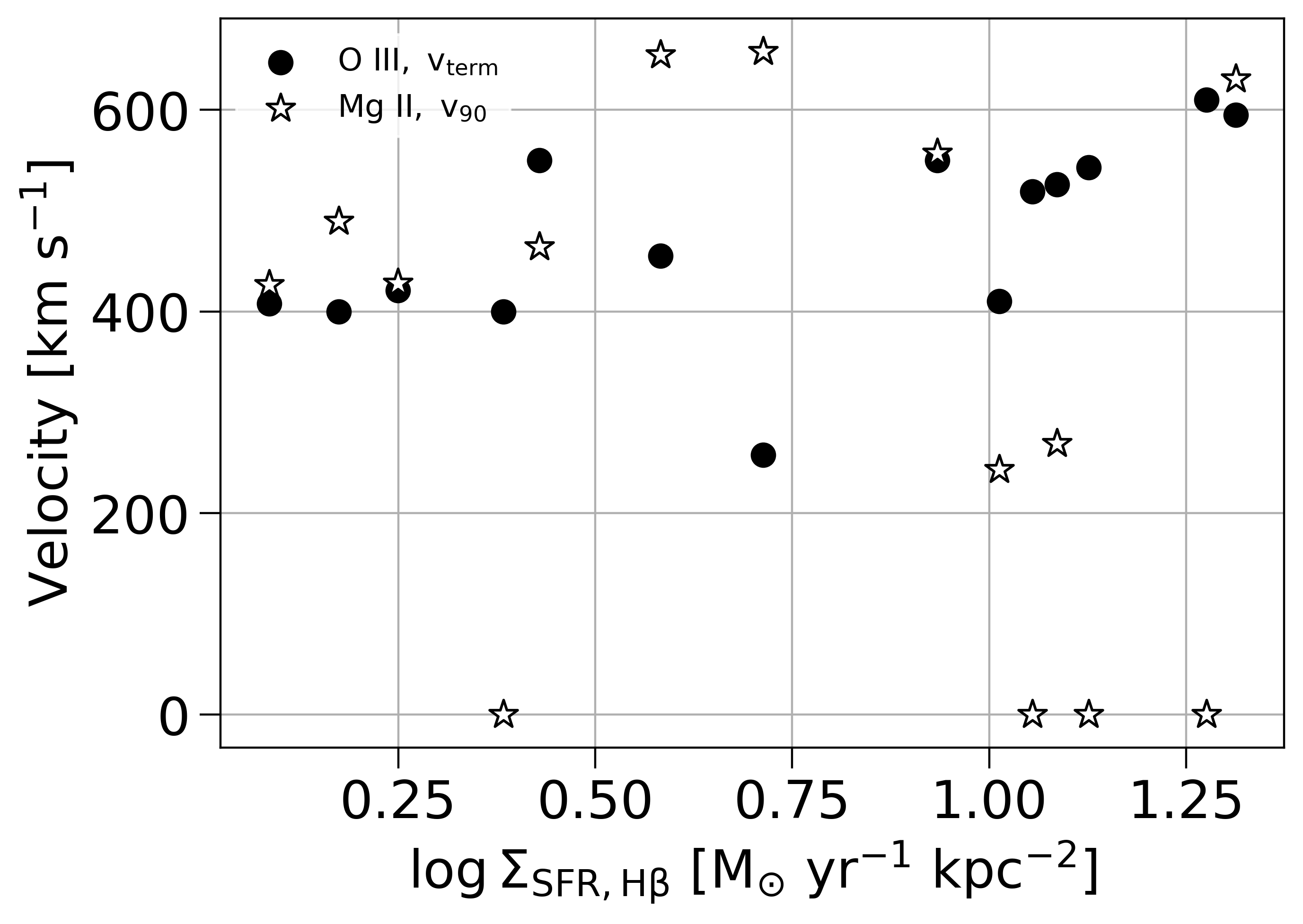}
	\caption{compares wind speeds corresponding to the Mg II 2796\AA\ $v_{90}$ values and the [O III] 5007\AA\ $v_{\rm term}$ values measured by Komarova et al. (in prep) against star formation surface density ($\Sigma_{\rm SFR}$).  In general, both speeds tend to increase with increasing $\Sigma_{\rm SFR}$.  However, at high $\Sigma_{\rm SFR}$ the wind speeds traced by Mg II appear to collapse and fall to low velocities or below the detection threshold.  We attribute the kinematic decoupling of the winds at high $\Sigma_{\rm SFR}$ to rapid or “catastrophic” cooling.  }
	\label{fig:decoupled_winds}
\end{figure}

We more directly compare the velocities of the different phases in Figure~\ref{fig:decoupled_winds} where we display only the galaxies with velocity measurements for each phase according to $\Sigma_{\rm SFR}$.  In general, both $v_{90}$ and $v_{\rm term}$ increase with increasing $\Sigma_{\rm SFR}$.  However, the Mg II values begin to drop below the O III values at high $\Sigma_{\rm SFR}$.  The galaxies with $v_{90} \ll v_{\rm term}$ could be the result of low densities in highly ionized winds, or a geometric effect where the wind geometry is a narrow bi-cone oriented away from the line of sight.  The cool clouds traced by Mg II in absorption have become kinematically decoupled from the warmer phase of the wind traced by O III due to rapid cooling.  Understanding this behavior will be our primary focus in the next section.  

\begin{table}
\centering
\caption{Cool and warm wind max velocities.}
\resizebox{\columnwidth}{!}{$\begin{tabular}{cccc}
\hline\hline
Galaxy& Leaker &$\rm{Mg\ II}\ 2796\text{\AA}\ v_{90}$ & $[\rm{O\ III}]\ 5007\text{\AA}\ v_{\rm term}$ \\[.5 ex]
&&$\rm km\ s^{-1}$ &$\rm km\ s^{-1}$\\[.5 ex]
\hline
\hline
J012910+145935m & Nondetection & 584.4 & 257.5 \\
J072326+414608m & Nondetection & 627.9 &    -- \\
J081112+414146m &         Weak & 730.4 &    -- \\
J082652+182052m & Nondetection &   -- &    -- \\
J091208+505009m & Nondetection & 444.3 &    -- \\
J091703+315221m &       Strong & 442.3 &   610 \\
J092552+395714m & Nondetection & 303.3 &    -- \\
J103344+635317m &       Strong &   -- &    -- \\
J103816+452718m &         Weak & 647.4 &    -- \\
J112933+493525m & Nondetection & 632.4 &    -- \\
J113304+651341m &         Weak &   -- &   400 \\
J115855+312559m &       Strong &   -- &   519 \\
J123519+063556m &       Strong &   -- &   543 \\
J124619+444902m &         Weak &   8.67 &    -- \\
J130128+510451m &         Weak & 578.4 &    -- \\
J131419+104739m & Nondetection & 635.9 &   455 \\
J132633+421824m &         Weak & 346.3 &    -- \\
J134559+112848m & Nondetection & 475.3 &   400 \\
J141013+434435m &       Strong &   -- &    -- \\
J003601+003307x & Nondetection & 360.8 &   526 \\
J004743+015440x &         Weak & 508.8 &   550 \\
J011309+000223x &         Weak & 452.3 &   421 \\
J012217+052044x &         Weak & 415.3 &   550 \\
J081409+211459x & Nondetection & 423.3 &   408 \\
J091113+183108x &         Weak & 625.9 &   595 \\
J095838+202508x &         Weak & 245.2 &   410 \\
\hline
\end{tabular} $}
\justifying
{\\ \\ From left to right: (1) Galaxy ID (2) Leaker Status (3) $v_{90}$ measured from Mg II 2796\AA\ line (4) $v_{\rm term}$ measured from [O III] 5007\AA\ line by Komarova et al. (in prep).}
\label{tab:wind_speeds}
\end{table}

In summary, we observe that the strongest LyC leakers in our sample generally exhibit high $\Sigma_{\rm SFR}$, compact morphology, young stellar populations, low dust extinction, elevated $\rm O_{32}$ ratios, and lower mass, momentum, and energy outflow rates associated with the neutral hydrogen content of the winds. Interestingly, while four out of five of the strongest leakers in our sample show no absorption features in Mg II lines, they still possess high-velocity winds (over 500 km/s) evident in the broad line components of higher ionization lines.  While many of these trends were previously noted by \cite{Flury2022a}, our findings regarding the absence or reduction of cool-phase outflow signatures and the associated lower outflow rates are new and somewhat unexpected. Simulation informed predictions from \cite{Cen2020} and other observations, such as by \cite{Heckman2016}, suggest that compact galaxies with high $\Sigma_{\rm SFR}$ should generate faster winds. However, in the case of the strong leakers, the cool clouds probed by Mg II seem to be kinematically decoupled from the hotter, more highly ionized phase of the winds.

\section{Discussion}\label{sec:discussion}


In Section~\ref{sec:modeling_lyc_escape}, we estimated $f_{esc}^{LyC}$ assuming a perfect correspondence between $\rm Mg^+$ and $\rm H^0$. Our predictions, however, systematically overestimated the observed values. Given that \cite{Saldana-Lopez2022} measured the covering fractions of $\rm H^0$ to be greater than that of the LIS metals in LzLCS, we hypothesized that the overestimation stemmed from missing low-density gas—too optically thin to be detected by Mg II in absorption, yet still capable of absorbing significant amounts of LyC radiation. To test this, we calculated the average absorption across the residual covering fraction necessary to match the observed $f_{esc}^{LyC}$ values. The correction terms yielded an average optical depth close to one, supporting our hypothesis.

These results can be understood within the framework of a multi-phase ISM/CGM, consisting of cool, high-density clouds rich in $\rm Mg^+$ and other LIS metals, embedded in a hotter, lower-density ambient medium (see Figure~\ref{fig:clouds}). In this scenario, the clouds are optically thick to LyC radiation, while the surrounding hotter medium may be optically thin or even transparent. Cool clouds therefore act to block LyC escape.  This emerging model of the ISM/CGM is gaining traction in the literature to explain a wide range of phenomena related to the structure of the ISM/CGM. For instance, \cite{Erb2023} used a variation of this model to simultaneously account for the Ly$\alpha$ profiles and the high-velocity range of LIS metal lines in a sample of $z\sim 2$ galaxies.

Thus, our results align with the paradigm of a multi-phase medium, where the removal of cool clouds is crucial for enabling LyC escape. The primary goal of this paper is to explore the mechanisms by which galactic feedback facilitates this process.

Both radiation and mechanical feedback are widely recognized as necessary—though not always sufficient—mechanisms for clearing the ISM/CGM of cool clouds to enable LyC escape \citep{Heckman2011,Trebitsch2017,Kimm2019,Cen2020,Amorin2024}. The question of sufficiency was addressed by \cite{Cen2020}, who showed that feedback can only generate the necessary pressure to lift cool clouds and facilitate LyC escape under specific conditions, particularly high $\Sigma_{\rm SFR}$ and compact morphologies. While Figure~\ref{fig:podigous_leakers} generally supports these claims, the precise mechanisms by which feedback-efficient environments remove cool clouds remain unclear. For instance, although Figure~\ref{fig:v90} suggests that all galaxies\footnote{More precisely, we mean all galaxies with available data.} in our sample exhibit outflows, the cool phase of the outflows traced by Mg II does not trace the warm phase traced by O III. In the most extreme scenario, four of the five strongest LyC leakers show no evidence of cool gas in absorption lines. 

To understand how feedback clears the ISM/CGM of cool clouds in local star forming galaxies, we consider three potential scenarios, focusing mainly on the strongest Leakers in LzLCS: 

\begin{itemize}
    \item \textbf{Full Ionization of Clouds:} There is sufficient LyC radiation to completely ionize the clouds, resulting in a fully density-bounded ISM/CGM.
    \item \textbf{Cloud Fragmentation:} In the absence of strong mechanical feedback, clouds are subject to dynamical and gravitational instabilities that cause clumping.  This in turn creates a “picket fence” geometry conducive to LyC escape.
    \item \textbf{Ram Pressure Acceleration:} The outflows exert enough ram pressure on the cool clouds to accelerate them to distances where they can no longer impede LyC escape.
\end{itemize}


Figures~\ref{fig:lightfraction_eta} and \ref{fig:lightfrac_vs_sigma_ruv} demonstrate that the majority of strong LyC leakers—particularly those lacking signatures of cool winds in Mg II absorption—are characterized by young stellar populations and, consequently, radiation dominant feedback (Flury et al. in prep). The presence of young-to-middle-aged (3–6 Myr) stellar populations suggests that Wolf-Rayet stars may also be amplifying the LyC emission in the strongest leakers. These conditions create an environment conducive to higher ionization parameters and the formation of density-bounded regions. This is further supported by the elevated $\rm O_{32}$ ratios observed in the strong leakers shown in Figure~\ref{fig:dust_Sigma_RUV}.

The simulations of \cite{Kakiichi2021} provide a relevant context, as they show that turbulence in H II regions leads to the formation of low column density channels, which are eventually evacuated by thermal pressure generated by the ionizing front.  These simulations suggest a combination of low- and high-density channels within the ISM/CGM, which likely correspond to the conditions in our LyC leakers.  Indeed, the escape fractions of the strong leakers span a range of only $31-5\%$.  This fact along with the overall lower dust extinction in the strong leakers (see Figure~\ref{fig:dust_Sigma_RUV}), suggests that there is still a considerable amount of absorption of the LyC by neutral gas in our galaxies.  Fully density bounded geometries may be more appropriate for objects with exceptionally high escape fractions (e.g., \citealt{Nakajima2014,Izotov2018b,Gazagnes2020,Flury2022a,Marques-Chaves2022b}).  Hence, we find this to be an unlikely explanation for the absence of cool clouds in the strongest leakers of our sample, which do not have exceptionally high $f_{esc}^{LyC}$.  However, we suspect ionized, or density bounded channels to be the primary means of LyC escape in the strong leakers with radiation dominant feedback. 

As hinted at in previous sections, we favor suppressed supernovae feedback as the primary cause for the absence of Mg II winds observed in the majority of strong LyC leakers in our sample.  This is due to delayed supernovae at low metallicity \citep{Jecmen2023}, which promotes catalstrophic cooling.  Analytical models by \citet{Silich2003, Silich2004} suggest that rapid cooling within ionized winds from compact star clusters can lead to the collapse and accumulation of cool clouds at small radii.  This offers a plausible explanation for the lack of absorption in the Mg II 2796\AA, 2804\AA\ lines, which still exhibit narrow emission.  In the strongest leakers, this is observed alongside the broad line components observed in the [O III] 5007\AA\ lines, which is a signature of radiation-dominated winds (\citealt{Komarova2021}, Komarova et al, in prep).  Specifically, the four strong leakers without observed winds show an average width of $\bar{\sigma}_{2796,2804} = 25\ \rm km\ s^{-1}$ in the Mg II lines and average terminal velocities in the [O III] line exceeding 500 $\rm km\ s^{-1}$. In contrast, the remaining galaxies with outflows have an average width of $\bar{\sigma}_{2796,2804} = 59\ \rm km\ s^{-1}$. 

Recent hydrodynamic simulations by \citet{Gray2019, Danehkar2021, Danehkar2022} further support these findings, demonstrating that the relatively high metallicities (right panel, Figure~\ref{fig:dust_Sigma_RUV}), low outflow speeds (Figure~\ref{fig:v90}), and low energy loading factors (bottom panel, Figure~\ref{fig:lightfraction_eta}) in our sample create the conditions most conducive to catastrophic cooling and the suppression of galactic winds.

In this scenario, rapid cooling leads to the fragmentation and clumping of cool clouds, resulting in “picket fence” geometries where LyC radiation escapes through ionized channels \citep{Jaskot2017, Jaskot2019}. This is consistent with the radiation feedback and elevated $\rm O_{32}$ ratios observed in the strongest LyC leakers in our sample.

Finally, we consider the possibility that supernova blast waves could accelerate cool clouds to sufficient distances from their host galaxies, thereby facilitating LyC escape, as proposed by \citet{Cen2020}. Figures~\ref{fig:lightfrac_vs_sigma_ruv} and \ref{fig:v90} show that most galaxies exhibiting winds in Mg II absorption lines tend to have older (6–10 Myr) stellar populations. These galaxies generally exhibit weaker LyC leakage, with two exceptions: one strong leaker with a significant fraction of older stellar populations and another with only 3–6 Myr stellar populations. This suggests that many of the galaxies in our sample with cool winds are likely experiencing mechanical feedback from supernovae, unlike the strong leakers with younger stellar populations. The two strong leakers with winds may be in the early stages of mechanical feedback, potentially entering a two-stage burst scenario where supernova driven winds clear low-density channels of cool clouds, which are then further evacuated by the ionization front (Flury et al., in prep). Therefore, we suspect supernovae to play a distinct, but adjacent role to radiation feedback in facilitating LyC escape in our galaxies.

Our model constraints for the outflows traced by Mg II favor bi-conical geometries with Mg II covering fractions less than unity, indicating significant anisotropy in the $\rm H^0$ content of winds in galaxies undergoing supernovae feedback.  Conversely, because we do not detect winds in Mg II in the strong leakers experiencing radiation dominant feedback, we do not detect anisotropy in the $\rm H^0$ content of their winds.  These results align with Flury et al. (in prep) who found evidence suggesting supernovae feedback is critical for the neutral gas anisotropy of LyC leakers purported by simulations \citep{Cen2015} and observed in local galaxies \citep{Zastrow2011,Komarova2024}.

\begin{figure*}
	\centering 
	\includegraphics[width=\textwidth]{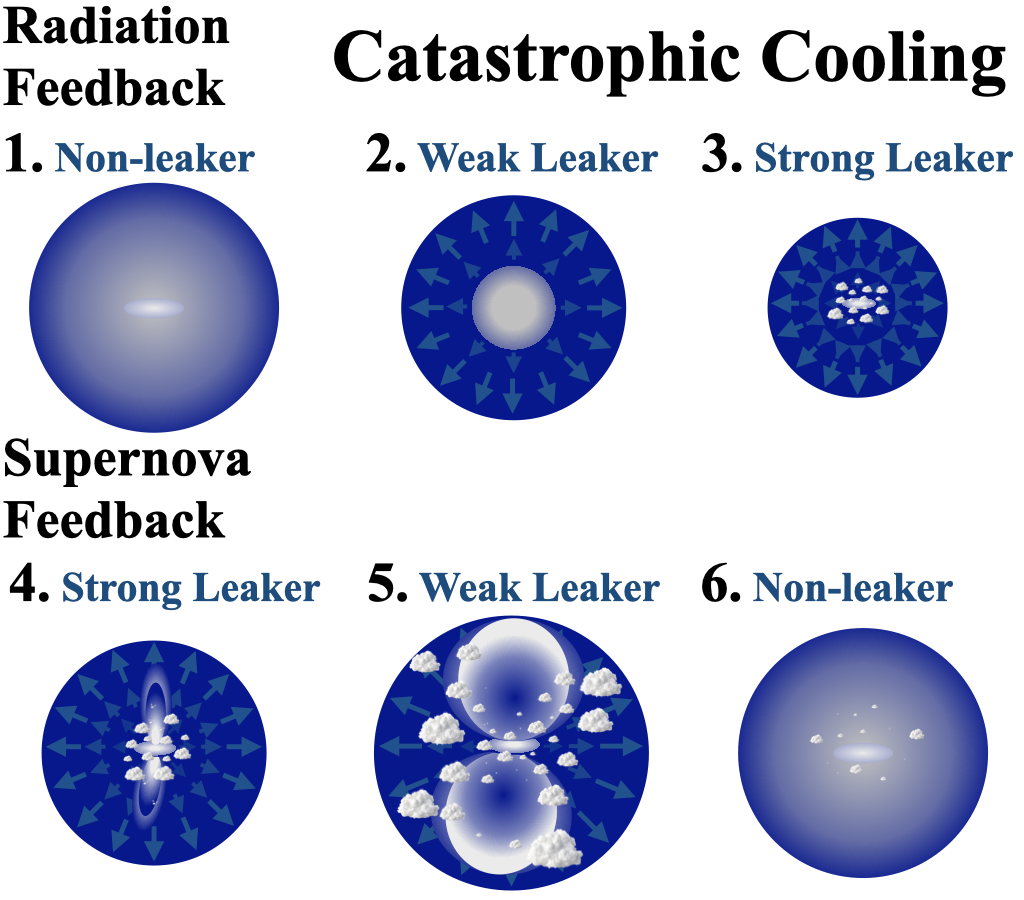}
	\caption{illustrates a proposed sequence of events linking star formation to LyC escape with time evolving from left to right, top to bottom. The complete sequence, steps 1-6, is likely relevant only in compact galaxies with high star formation surface densities ($\Sigma_{\rm SFR}$) and sub-solar metallicities; in other cases, steps 2-3 might be bypassed.  The \emph{\textbf{Top Row}} shows a radiation dominant feedback sequence. \textbf{1.} The initial stage of the star burst is marked by young stellar populations ($<3$ Myr) producing copious amounts of LyC radiation.  The ISM/CGM is filled with neutral gas and potentially dust.  Feedback has not had time to clear the ISM/CGM of neutral material and $f_{esc}^{LyC}$ is low.  \textbf{2.} Radiation drives a fast very diffuse ionized wind from the galaxy.  Cooling occurs and dense clouds remain close to the ionizing cluster. $f_{esc}^{LyC}$ is still low, but leakage can occur.  \textbf{3.} Under conditions such as low energy loading and relatively high metallicity, rapid or catastrophic cooling occurs causing a build up of cool clouds at small radii.  The clouds fragment creating low density channels.  The intense ionization front, potentially amplified by Wolf-Rayet stars, evacuates the low density channels facilitating LyC escape. $f_{esc}^{LyC}$ is high. The \emph{\textbf{Bottom Row}} depicts the onset of supernovae or a supernova-dominant feedback sequence.  \textbf{4.} Supernovae release mechanical energy into the surrounding environment and a super bubble begins to form.  Supernovae and radiation facilitate LyC escape in a two-stage burst scenario, where mechanical feedback clears low-density channels of cool clouds while radiation further evacuates the channel of neutral gas.  $f_{esc}^{LyC}$ is high, but asymmetries appear.  \textbf{5.} Mechanical feedback continues to drive the wind, accelerating cool clouds up to 1000 $\rm km\ s^{-1}$.  The lifting of cool clouds from the ISM and their potential expansion act to increase the covering fraction of neutral gas. Ionizing stars have expired and $f_{esc}^{LyC}$ begins to wane and is highly anisotropic. \textbf{6.}  The majority of massive stars have reached the end of their life cycles.  $f_{esc}^{LyC}$ is low.  Recycled gas and the accretion of pristine gas from the IGM fuel the next generation of star formation.  The local environment becomes more neutral and dusty, as star formation ramps up and the next cycle begins.}
	\label{fig:timeline}
\end{figure*}

In summary, our results suggest the following sequence of events linking star formation to LyC escape.  Initially, radiation feedback dominates in galaxies with young star clusters. In compact galaxies with high $\Sigma_{\rm SFR}$, this can lead to catastrophic cooling, where cool clouds condense at small radii and eventually fragment, creating low-density channels \citep{Jaskot2019,Jecmen2023}. These channels are then evacuated by a strong ionizing front, potentially amplified by Wolf-Rayet stars, enabling LyC escape. This period likely accounts for many of the strongest leakers in our sample, characterized by narrow emission features in LIS lines, weak or absent absorption features, and extended broad line components in HIS lines. As supernovae begin to occur, their mechanical energy can further accelerate the cool clouds. Initially, this might enhance LyC escape, as in a two-stage burst scenario. However, over time, the blast waves may lift enough cool clouds to increase the covering fraction of neutral gas and dust, thereby reducing $f_{esc}^{LyC}$.  Moreover, the top of the IMF will have collapsed into black holes at this stage, eliminating sources of LyC.  The supernova-dominated stages are marked by strong winds in both LIS and HIS lines.  The proposed sequence of events is illustrated in Figure~\ref{fig:timeline}.

An evolutionary sequence in LyC escape was initially proposed by \cite{Flury2022b}, distinguishing between radiation-dominant feedback periods and those with supernovae, while highlighting the importance of a two-stage burst scenario. Our work extends this timeline, with a particular emphasis on the role of delayed supernovae in low-metallicity systems and catastrophic cooling.  Additional evidence of an evolutionary sequence was put forth by \cite{Hayes2023b} who studied the impact of stellar age and the Lyman alpha (Ly$\alpha$) escape fraction, $f_{esc}^{Ly\alpha}$.  They observed an anti-correlation between $f_{esc}^{Ly\alpha}$ and the age of the starburst which can be explained by the increase of cool clouds (or dust) as the supernovae blastwave lifts cool clouds, or with the increase of mass loading as suggested by \citep{Hayes2023a} and observed in Figure~\ref{fig:lightfraction_eta}.     

The exact timeline of the events described above remains uncertain. The time spans associated with the light fractions in our study depend on the assumptions made in \texttt{STARBURST99} \citep{Saldana-Lopez2022}, such as the fraction of stars that undergo supernovae at a given stellar mass, making them highly model-dependent. Crucially, \citet{Jecmen2023} linked the metallicities of our galaxies to delayed mechanical feedback, which occurs when higher-mass stars collapse directly into black holes without a supernova explosion \citep{OConnor2011}, thereby prolonging conditions favorable for catastrophic cooling. As a result, radiation-dominated periods may be more prevalent in low-metallicity, compact galaxies, such as Green Peas. Notably, similar conditions have been proposed as a natural explanation for the early UV-bright galaxies observed by JWST \citep{Dekel2023,Menon2024}. Therefore, the collection of radiation-dominated strong leakers in our study could serve as critical examples of the early star forming galaxies thought to have reionized the Universe. We emphasize their importance in studies like \citet{Jaskot2024a, Jaskot2024b} for estimating $f_{esc}^{LyC}$ at high redshifts.

\section{Conclusions}\label{sec:conclusion}

In this study, we used semi-analytical line transfer (SALT) models \citep{Carr2023} to map the $\rm H^0$ content of galactic winds in 26 out of a potential 29 galaxies taken from the Low-z Lyman Continuum Survey (LzLCS) \citep{Flury2022a}. Specifically, we modeled the Mg II 2798\AA, 2804\AA\ emission lines and converted the inferred $\rm Mg^+$ abundance to $\rm H^0$ using Cloudy photoionization models \citep{Chatzikos2023}.  We detected outflows in 20 of the 26 galaxies. Using the model inferred column densities of $\rm H^0$, we estimated the LyC escape fractions through the winds and compared to observations.  In addition, we computed the mass, momentum, and energy outflow rates/loading factors and compared them to SFR-based deposition rates and predictions from the FIRE-2 simulations \citep{Pandya2021}.

We also analyzed various feedback-related properties, such as dust extinction, the $\rm O_{32}$ ratio, gas metallicity, and loading factors, within the context of galaxy conditions believed to optimize feedback efficiency—specifically, high star formation surface density ($\Sigma_{\rm SFR}$) and compact morphology. By utilizing light fractions \citep{Saldana-Lopez2022}, which represent the fraction of a galaxy’s starlight emanating from stellar populations within a specified age range, we identified the dominant feedback mechanisms at play. In particular, this approach allowed us to determine whether radiation or supernova feedback was responsible for driving the observed outflows by correlating these light fractions with the age of the star formation episode.

Below are our main conclusions and findings.



\begin{enumerate}
  \item We considered a three component model of the ISM/CGM containing optically thick, thin, and transparent pathways to LyC escape.  After proposing that Mg II absorption traces only the optically thick pathways to LyC escape, we then demonstrated that a strictly Mg II inferred $f_{esc}^{LyC}$ will typically overestimate the observed value.  We attributed this fact to the large $\rm H^0$ to $\rm Mg^+$ abundance ratios in the LzLCS sample, which we found to be equal to about $2\times 10^4$ on average.  We inferred the covering fraction for optically thin sight lines by subtracting the optically thick or Mg II derived covering fraction from the total covering fraction of $\rm H^0$ measured by \cite{Saldana-Lopez2022}.  We found that additional absorption along these pathways, with an average optical depth of $\bar{\tau}_{\rm thin} = 1.3$, was necessary to match the observed $f_{\rm esc}^{LyC}$ values, further supporting our interpretation.  These results support a multi-phase model of the ISM/CGM, where low-ionization state (LIS) metals trace cool, dense clouds embedded within a hotter, lower-density ambient medium.

  \item We proposed a method (Equation~\ref{eq:fesc_correction}) for predicting $f_{\rm esc}^{LyC}$ from Mg II absorption, incorporating an empirical correction term for the additional absorption occurring along optically thin pathways, derived from our galaxy sample. This method applies to systems where Mg II appears in absorption or as a combination of absorption and emission. Pure emission systems, however, should be addressed using different approaches (e.g., \citealt{Chisholm2020,Xu2022b,Xu2023}).
       
  \item While focusing solely on the $\rm H^0$ content of the winds, we measured mass outflow rates approaching the SFR, momentum outflow rates reaching up to 10\% of the expected supernova deposition rate, and energy outflow rates approaching 1\% of the expected supernova deposition rate (assuming 1.0 SN per 100 $\rm M_{\odot}$). All measurements showed significant variation, spanning 2, 2.5, and 3 dex, respectively. We found that the upper limits of the corresponding loading factors align well with FIRE-2 predictions \citep{Pandya2021} when considering only the cool phase ($10^3-10^5\ \rm K$) of the winds. However, the majority of the data points fall two to three orders of magnitude lower.  Despite the scatter, we observed that galaxies with supernova feedback tend to exhibit the highest loading factors, while those dominated by radiation feedback show the lowest. The reduction in energy and momentum is qualitatively consistent with predictions by \cite{Jecmen2023}, which are based on delayed supernova feedback in low-metallicity environments.  These results underscore the importance of including delayed supernovae in low-metallicity systems in simulations of galaxy formation.       

  \item The majority (4/5) of the strong LyC leakers in our sample are characterized by younger ($<6\ \rm Myr$) stellar populations and lack detectable galactic outflows in the Mg II 2796\AA, 2804\AA\ absorption lines. Nevertheless, these galaxies still exhibit signs of ionized galactic winds, evident through a broad line component in the [O III] 5007\AA\ emission line, as shown by \cite{Amorin2024}. In contrast, the relatively weaker leakers tend to have older ($>6\ \rm Myr$) stellar populations and display outflows in both the Mg II absorption lines and the broad component of the [O III] emission line. These observations suggest that the strongest leakers are predominantly influenced by radiation feedback, while the weaker leakers are more heavily influenced by supernova-based feedback. These findings are consistent with \cite{Bait2023} and the near future papers by Flury et al. (in prep) and Komarova et al. (in prep).

  \item We found that compact galaxies with high $\Sigma_{\rm SFR}$ generally exhibit higher $f_{esc}^{LyC}$, lower dust extinctions, higher $O_{32}$ ratios, and lower mass, momentum, and energy loading in $\rm H^0$. The wind speeds traced by the [O III] 5007\AA\ emission line tend to increase monotonically with rising $\Sigma_{\rm SFR}$ and compactness. In contrast, the wind speeds measured from the Mg II 2796\AA\ line show much greater variation and are not kinematically coupled to the more highly ionized winds traced by the [O III] 5007\AA\ line.  In the most extreme case, a “hole” appears in the parameter space, where the wind speeds inferred from low-ionization state (LIS) lines collapse to zero or fall below the detection limit. This “hole” corresponds to the 4 out of 5 strong leakers that do not show signs of galactic outflows in Mg II lines, but still have winds exceeding 500 $\rm km\ s^{-1}$ in the [O III] 5007\AA\ line. We attribute this phenomenon to the suppression of mechanical feedback caused by the low-metallicity delay in the onset of supernovae and subsequent catastrophic cooling \citep{Jecmen2023}.
  
\item We proposed a general sequence of events for LyC escape in feedback-efficient environments, particularly in compact galaxies with high star formation rate surface densities and sub-solar metallicities. During the early stages of a starburst, radiation serves as the primary feedback mechanism, which can lead to catastrophic cooling and the formation of cool clouds at small radii. This cooling causes fragmentation, creating low-density, ionized channels—resulting in a “picket fence” geometry that provides optimal conditions for LyC escape, producing the strongest leakers \citep{Jaskot2019}. This stage may be prolonged in low-metallicity environments due to delayed mechanical feedback \citep{Jecmen2023}.  It is characterized by narrow emission lines from low-ionization metals that lack absorption features, along with broad line components in higher-ionization metal emission lines, indicating fast very diffuse ionized winds.  As supernovae begin to occur, they release mechanical feedback that acts to lift and accelerate the cool clouds to greater distances. Initially, this can enhance LyC escape, as in a two-stage burst scenario, where mechanical feedback removes cool clouds from low density channels which are further evacuated by the ionization front (Flury et al. in prep). However, over time, the blast waves may lift enough cool clouds to increase the covering fraction of neutral gas and dust, thereby reducing $f_{esc}^{LyC}$ along with the death of the most massive stars.  At this stage, galaxies exhibit signs of outflows in both low and high-ionization state lines.  Note that catastrophic cooling may be bypassed in diffuse galaxies with low $\Sigma_{\rm SFR}$ and/or high metallicity.  
  

  \item Our constraints favor bi-conical outflow geometries for the winds traced by Mg II, as opposed to spherical, in the majority of weak leakers which typically have older stellar populations.  This suggests a significant anisotropy in the neutral gas content of galaxies undergoing supernova-based feedback in agreement with Flury et al. (in prep). These results further support the idea that whether a galaxy is a LyC leaker is highly sensitive to the viewing angle, in agreement with simulations \citep{Cen2015} and other observations \citep{Zastrow2011,Komarova2024}.

\item The evidence observed in our data for delayed supernovae \citep{Jecmen2023}, such as winds detected in the [O III] 5007Å line but absent in Mg II lines (Komarova et al. in prep), suggests that radiation-dominated feedback periods may be common in low-metallicity compact galaxies with high star formation surface densities, like Green Peas and primordial galaxies \citep{Menon2024}. Indeed, similar conditions have been shown to explain the excess of UV bright galaxies at high redshfits \citep{Dekel2023}.  We speculate that such galaxies may have played a critical role in Reionization and should be emphasized in studies aimed at predicting $f_{esc}^{LyC}$ at higher redshifts (e.g., \citealt{Jaskot2024a,Jaskot2024b}).

\end{enumerate}

\begin{acknowledgments}
We acknowledge Nordita for sponsoring the Cosmic Dawn at High Latitudes Conference where invaluable discussions regarding this work occurred.  We acknowledge the Minnesota Supercomputing Institute (MSI) at the University of Minnesota for providing the majority of computational support for this project.  Support for program number HST-AR-16606 was provided by NASA through a grant from the Space Telescope Science Institute, which is operated by the Association of Universities for Research in Astronomy, Incorporated, under NASA contract NAS5-26555.  M.S.O. and L. K. are supported by NASA grant HST-GO-16261.  
\end{acknowledgments}

\bibliography{sample631}{}
\bibliographystyle{aasjournal}



\end{document}